\documentclass[9pt]{article}
\usepackage{times}
\usepackage[affil-it]{authblk}
\usepackage[english]{babel}
\usepackage{blindtext}
\usepackage{sidecap}
\usepackage[utf8]{inputenc}
\usepackage{geometry}
\usepackage{caption}

\usepackage{subfig}
\usepackage{longtable}
\usepackage{bm}
\usepackage{mathtools}
\usepackage{amsmath}
\usepackage{slashed}
\usepackage{amsfonts} 
\usepackage{MnSymbol}
\usepackage{sidecap}
\usepackage{xcolor}
\usepackage{hyperref}
\usepackage{url}
\usepackage{geometry} 
\usepackage{floatrow}
\usepackage{graphicx}
\makeatletter 

\title{\textbf{{Relativistic Calculations of Energies, Transition Parameters, Hyperfine Structures, Land\'e g$_J$ factors and Isotope Shifts for Na-like Ar$^{7+}$, Kr$^{25+}$ and Xe$^{43+}$ ions}}}

\author[1]{Shikha Rathi}
\author[1,*]{Lalita Sharma}

\affil[1]{%
  Indian Institute of Technology Roorkee, Roorkee 247667, India }
 
\affil[*]{Corresponding author: lalita.sharma@ph.iitr.ac.in}
\DeclareUnicodeCharacter{2212}{-}
\DeclareUnicodeCharacter{2264}{+}
\begin{document}





\maketitle
\section*{Abstract}
In this study, comprehensive calculations of energies, hyperfine structure constants, Land\'e g$_J$ factors and isotope shifts have been performed for the lowest 71 states of Na-like Ar$^{7+}$, Kr$^{25+}$ and Xe$^{43+}$ ions. Radiative parameters viz., wavelengths, transition rates, oscillator strengths and lifetimes are also estimated for the electric and magnetic dipole (E1,M1) and quadrupole (E2,M2) transitions among these levels. The states under consideration include $1s^22s^22p^6nl$ for $n = 3-9$, $l = 0-6$, and the fully relativistic multiconfiguration Dirac–Fock (MCDF) method integrated in the latest version of the general-purpose relativistic atomic structure package (GRASP) is used for the calculations. The relativistic corrections, viz., Breit interactions and quantum electrodynamics effects are also included in the present work and their effects on energies and other parameters are analysed. We also examined the impact of including the core-core and core valence correlations on level energies. Further, to inspect the reliability of our MCDF results, we performed another set of calculations using the many-body perturbation theory inbuilt into the Flexible Atomic Code (FAC). A thorough comparison between the two obtained calculations and with the previous theoretical and experimental results, wherever available, is carried out and a good agreement is observed. The present calculations will considerably extend the already existing results for these Na-like ions as most of the results are reported for the first time. Further, we believe that our results are of high accuracy and will be beneficial for spectral analysis and diagnosis in plasma and astrophysics.

\section{Introduction}
Spectroscopy of highly charged ions (HCI) has become an exquisite research topic as knowledge of their atomic structures and other parameters plays a crucial role in various fields of science and technology such as astrophysics, lasers, plasma and collision physics \cite{gillaspy2001highly}. For instance,  HCI may serve as the source of extreme ultraviolet (EUV) radiation for lithography. Also, they are being used as added impurities for fusion plasma diagnosing, e.g., ET-EFDA. In addition, precise atomic results such as fine-structure levels, transition parameters, hyperfine structure constants, etc., are vital for developing sophisticated stellar atmosphere models, calculating the opacities and other atmospheric parameters \cite{doi:10.1139/p10-105} and designing new atomic clocks. Sodium-like ions are of particular interest as their emission lines are the sources of prominent features in the spectra of sun \cite{mohan1996forbidden}, high-temperature tokamak plasma and thus have diagnostic applications \cite{hinnov1976highly,keenan1986relative}. 
Moreover, these ions are appropriate for examining the electron-correlation and quantum electrodynamics (QED) effects on atomic properties \cite{konan2018energies}.\\
However, due to the challenges in experimental techniques for HCI systems, estimation of their atomic parameters relies primarily on precise theoretical studies. Thus, driven by the vast applications of HCI and the upsurge in demand for their electronic structure data in astrophysics and projects like ITER, we purpose to carry out a comprehensive study of atomic parameters for sodium-like Ar$^{7+}$, Kr$^{25+}$ and Xe$^{43+}$ ions.
%
Their chemical inertness makes them befitting candidates for use in tokamaks, fusion plasma research and diagnostics. 
For example, these elements are typically used as trace gases in reactive plasma to examine the density variation of the radicals formed \cite{coburn1980optical}. Xe and Kr have a rich emission spectrum and are an indispensable element in developing lasers, laser techniques, and light sources. Further, xenon ions are likely to be employed as an edge plasma coolant in the ITER fusion reactor and are also being used in gas-discharge lamps, plasma contactors, spacecraft propulsion \cite{beattie1990xenon} and a potential EUV source for next-generation lithography \cite{PhysRevApplied.10.034065}. Krypton has excellent implications in plasma physics as it is usually employed in ECR ion sources and controlled fusion \cite{guerra2013electron}. It is proposed that injection of Kr in the ITER will assist in cooling the outermost plasma region and reduce erosion by producing a peripheral radiating mantle. Further, lines of HCIs of Kr and Xe have been observed in tokamak plasma \cite{hinnov1976highly}.
Moreover, xenon and krypton have been identified in planetary nebula NGC 7027 \cite{pequignot1994identification}, hot DO-type white dwarf RE 0503−289 \cite{werner2012first}. The excessive abundance of Xe ions is observed in almost all mercury–manganese stars \cite{10.1111/j.1365-2966.2008.12937.x}. Furthermore, Xe and Kr are neutron(n)-capture elements. Understanding their properties play a crucial role in enhancing the understanding of nucleosynthesis and mixing processes in the stars, e.g., AGB stars \cite{busso2001nucleosynthesis,doi:10.1146/annurev.astro.43.072103.150600} and can provide a way to map the universe's chemical evolution \cite{doi:10.1146/annurev.astro.46.060407.145207}. However, a scarcity of reliable atomic results for these trans-iron elements has hindered their accurate spectra analysis and precise abundance predictions in astrophysical objects \cite{sterling2015abundances, zhang2006abundances}. Likewise, argon has been detected in interstellar clouds \cite{morton1974interstellar}, stellar atmosphere, solar flares and is one the most abundant elements in the universe \cite{werner2007discovery}. Lines of Na-like Ar$^{7+}$ have been detected in radio-loud quasar\cite{finn2014compact} and in laboratory plasma \cite{bartnik2015photoionized}.\\
Further, various isotopes of Xe, Ar and Kr have been detected in the atmosphere of Sun, Earth, Mars \cite{owen1977composition}, martian meteoroids \cite{eugster2002ejection} and coma of comet 67P/Churyumov-Gerasimenko \cite{rubin2018krypton}. Hyperfine interaction effects also exist in many of these isotopes with non-zero nuclear spin. These effects can lead to broadening and shifting of the atomic lines and, thus, are unavoidable for correctly interpreting spectra. Also, hyperfine interaction can open up many forbidden transitions, which are beneficial for improving the plasma models and diagnosis. Results of isotope shifts (IS) and hyperfine structures (HFS) are also integral to study isotope anomalies and understanding nucleosynthesis processes in astronomical objects \cite{proffitt1999goddard,roederer2012new}. Besides plasma and astrophysical applications, these effects are also significant in nuclear physics, e.g., to estimate the mean-square nuclear charge radii and nuclear quadrupole moment. Further, the strong magnetic field in stellar atmosphere necessitates the knowledge of Land\'e g$_J$ factors. Therefore, for understanding the dynamics and properties of astrophysical objects and laboratory plasma, precise and complete calculations of atomic structures, including the HFS, IS and Land\'e g$_J$ factors, are essential.
%
%
%

To investigate the atomic properties of Na-like Ar$^{7+}$, Kr$^{25+}$ and Xe$^{43+}$, several experimental and theoretical studies have been performed in the recent past.
Theoretical work on Na-like ions includes the study of ionization potential and oscillator strengths ($f$) of electric dipole (E1) transition $nl-n'l'$ ($n, n': 3-4, l, l': s-p$) for Na iso-electronic series by Fischer \cite{doi:10.1139/p76-174} using the multiconfigurational Hartree-Fock (MCHF) method.  To ease the reading we omit closed-shells i.e., $1s^2 2s^2 2p^6$, for representing the states of Na-like ions. 
Bi\'emont \cite{biemont1978theoretical} calculated the $f$ values for $nl-n'l'$ ($n, n' \leq 8; l, l' = s,p,d,f;$ Z$: 11-26$) using the monoconfigurational Hartree-Fock (HF) procedure. Godefroid et al. \cite{godefroid1985forbidden} have estimated the oscillator strengths for $3s ~^2S - 3d ~^2D$ electric quadrupole (E2) lines up to Z = 26 by employing the MCHF technique. Theodosiou and Curtis \cite{theodosiou1988accurate} calculated the lifetime for $3p$ and $3d$ levels of Na-like ions for Z = 11–54, 74, 79, 90, and 92 using the experimental data and MCDF method. Relativistic many-body perturbation theory (RMBPT) was applied by Guet et al. \cite{guet1990relativistic} to determine $f$ of $3p-3s$ E1 lines for Z = $11-29$ Na-like ions.
Johnson el al. \cite{johnson1996transition} used the third-order many-body perturbation theory to evaluate the energy and transition rates ($A$) for $3p-3s$ and $4s-3p$ lines for Na-like ions from Z = 3-100. Semi empirical calculations of energies and electric dipole transition probabilities  for Na-like ions (Z = 30 - 60) have been performed by Matsushima et al. \cite{matsushima1991spectra} using the relativistic parametric potential approach.
%
Siegel et al. \cite{SIEGEL1998303} applied monoconfigurational Dirac-Fock theory to calculate $f$ values between states with $n = 3-5$ and $l = s,p,d,f$ of sodium iso-electronic sequence from Na to Ca. Charro and Mart\'in \cite{charro2002relativistic} used the relativistic quantum defect orbital (RQDO) method to compute  $A$ values between $nl-n'l (n, n'=3-4)$ states for Si IV to Kr XXVI ions . Later, Charro and Mart\'in \cite{charro2002systematic} employed the same method to calculate oscillator strengths for $np ~^2P$ - $n'f ~^2F$ ($n = 3-5, n' = 3-6$) E2 lines of Na-like K$^{8+}$ to Cs$^{44+}$ ions.
%
Applying the C1V3 code of Hibbert, Younis et al. \cite{younis2005energy} determined the energy levels and dipole oscillator strengths between $n = 3-5$ and $l = 0-3$ levels of Na-like ions up to krypton. Fischer et al. \cite{fischer2006relativistic} reported the energy levels, transition probabilities and lifetimes of Ar$^{7+}$ using the MCDF method for $n = 3-4$ and $l = 0-3$. Calculations of energies, E1 transitions rates and lifetimes for 5 lowest levels of Na-like Xe$^{43+}$ have been done by Vilkas et al. \cite{vilkas2008relativistic} using the relativistic many-body Møller–Plesset perturbation theory. Sapirstein and Cheng \cite{sapirstein2015s} implemented the S-matrix-based QED formalism to calculate the $3s-3p$ transition energies of Na isoelectronic sequence ranging from Z = 30-100. 
Apart from the structure calculations, Fontes and Zhang \cite{fontes2017relativistic} used relativistic distorted wave (RDW) theory to obtain the collision strengths for 10 transitions with $n =$ 3 in the 67 Na-like ions (Z = 26-92). Electron impact excitation data belonging to $nl-n'l (n,n' = 3-6$ and $l = 0-5)$ have been estimated for Na-like Mg$^+$ to Kr$^{25+}$ ions using the intermediate coupling frame transformation R-matrix method by Liang et al. \cite{liang2009r}. They have also computed the level energies, $A$ and weighted oscillator strengths ($gf$) among these levels using the AUTOSTRUCTURE \cite{badnell1986dielectronic} program. Sampson et al. \cite{SAMPSON1990209} used the Dirac-Fock Slater potential to determine the electronic radial functions, transition energies and oscillator strengths and further employed the RDW theory to calculate collision strengths for
{$3l-nl$, $n = 4-5$} states of Z = 22-92 Na-like ions. 
Further, neglecting the relativistic effects, large-scale calculations for allowed transition between $n\leq10$ and $l\leq4$ levels of Ar VIII were performed under the Opacity Project \cite{TOPbase}. However, these results are available only for the fine-structure unresolved multiplets.\\
%
%
On experimental front, using the beam-foil technique and ANDC principle, Livingston et al. \cite{Livingston:81} studied the spectra of Ar V - Ar VIII using beam-foil excitation and determined the lifetime of 3p and 3d states. Later, Reistad et al. \cite{reistad1986oscillator} measured the f values for $3s-3p$ transition in Ar VIII and also evaluated the lifetime for $3p ~^2P_{1/2,3/2}$ levels. 
Gillaspy et al. \cite{gillaspy2013transition} measured the wavelength of D lines of Na -like Xe using the NIST electron beam ion trap (EBIT) and also applied the RMBPT procedure to evaluate the wavelengths of D lines for Na-like ions from argon to uranium. 
Kink et al. \cite{kink1997lifetime} carried out lifetime measurement of $3p ~^2P_{3/2}$ level of Kr$^{25+}$ using the beam-foil excitation and cascade-corrected analysis. Lifetime of $3p$  and $3d$ levels of Xe$^{43+}$ have been measured by Tr\"abert et al. \cite{trabert1994experimental} using the time-resolved spectroscopy of foil-excited ion beams. Later, Tr\"abert et al. \cite{trabert2003extreme} measured the wavelengths of Xe$^{43+}$ at the SuperEBIT. Fahy et al. \cite{fahy2007extreme} performed the extreme-ultraviolet  spectroscopy of Xe ions using NIST EBIT and measured the wavelength of lines arising from $3s$–$3p$ transitions in Xe$^{43+}$ and other ionic states. Osin et al. \cite{osin2012extreme} identified $n = 3$ - $n^\prime= 3$ transitions from Ni-like to Na-like xenon at the NIST EBIT and measured the corresponding wavelengths. \\
%
%
Reader et al. \cite{reader19873s} measured the wavelength of $3s-3p$, $3p-3d$ and $3d-4f$ transitions in Kr$^{25+}$ using the photographing laser-produced plasmas and tokamak plasmas with grazing-incidence spectrographs. Further, they extrapolated their results by least square fitting to estimate the wavelengths of Ar$^{7+}$ to Xe$^{43+}$. Wavelengths of $3s-3p$ emission lines of Kr$^{25+}$ have been measured by Kondo et al.\cite{kondo1986precision} in the Alcator C tokamak plasmas. \\
Studies on HFS, IS and Land\'e g$_J$ factors are scarce, particularly for higher excited levels. Dutta and Majumder \cite{dutta2013electron} used the relativistic coupled-cluster (RCC) theory and calculated the HFS constants A$_J$ and B$_J$ of states with configuration $nl (n = 3$ and $l = 0-2)$ for Na-like Si to V ions. They also studied the electron correlation effects on HFS.  Using the RMBPT method Safronova and Johnson \cite{PhysRevA.64.052501} derived the field shift (FS) and specific mass shift (SMS) constants for $3s,3p$ and $3d$ levels in sodium like ions with Z = 12 - 18 and Z = 26. Tupitsyn et al. \cite{tupitsyn2003relativistic} reported the FS and SMS for $3p_{1/2}-3s_{1/2}$ transition of Na-like Ar and Xe ions by performing large-scale configuration interaction Dirac-Fock (CIDF) calculations. Using the EBIT facility at NIST, Silwal et al. \cite{silwal2018measuring,silwal2020determination} measured the mass shifts (MS) and FS for D1 and D2 transitions between sodium like $~^{124}$Xe and $~^{136}$Xe ions.\\
An evident scarcity of the results of the atomic structures for higher excited states of sodium-like ions can be pictured from the studies mentioned above, especially for Xe and Kr; only a few are fully relativistic. Further, there is a lack of results on transition parameters of forbidden spectral lines, HFS, IS and Land\'e g$_J$ factors. However, beside allowed transitions, results on forbidden lines are vital for determining precise lifetimes of energy levels, effective and accurate plasma modeling. Also, due to the non-zero nuclear spin of numerous isotopes of these noble gasses, precise spectral analysis demands hyperfine structures and isotope shift factors. Therefore, we will provide complete and comprehensive results for energy levels, (E1,M1) and (E2,M2) transition probabilities, lifetime, HFS constants A$_J$-B$_J$, Land\'e g$_J$ and IS factors for Na-like Ar$^{7+}$, Kr$^{25+}$ and Xe$^{43+}$ ions in this work.
An evident scarcity of the atomic structures results for higher excited states of sodium-like ions can be pictured from the studies mentioned above, especially for Xe and Kr; only few are fully relativistic. Transitions parameters of a small number of forbidden and allowed spectral lines are available. Further, there is a paucity of results on HFS, IS and Land\'e g$_J$ factors. However, due to the non zero nuclear spin and presence of numerous isotopes of these noble gasses, precise spectral analysis demands for the hyperfine structures and isotope shift factors.  Therefore, we will provide complete and comprehensive results for energy levels, (E1,M1) and (E2,M2) transition probabilities, lifetime, HFS constants A$_J$-B$_J$, Land\'e g$_J$ and IS factors for Na-like Ar$^{7+}$, Kr$^{25+}$ and Xe$^{43+}$ ions in this work.
In this connection, we implement the MCDF procedure to perform a systematic and fully relativistic calculations for the lowest 71 levels of these ions. The Breit interaction, QED effects, normal and specific mass shift corrections are also taken into consideration to enhance the accuracy of calculated parameters. The configuration of our interest includes $1s^22s^22p^6nl$, where $n=3-9$ and $l=0-6$. Energy levels, HFS constants A$_J$-B$_J$ and IS are estimated for all the above mentioned levels. 
 All the parameters mentioned earlier are calculated using GRASP2018 package \cite{fischer2019grasp2018} except for IS, for which RIS4 module \cite{ekman2019ris} is employed. To examine the reliability of our results, we have performed similar calculations using the many body perturbation theory (MBPT) inbuilt in flexible atomic code (FAC)\cite{gu2008flexible}. A good agreement is achieved by comparing the two results and also with the available previous results. We also analyse the effect of correlations  by considering the core-valence and core-core correlations for Na-like Xe$^{43+}$. We believe that our calculations are precise enough to be used in studying the plasma properties, chemical abundance of stars and stellar atmosphere modeling. 
%
%
%
\section{Theory}
\subsection{The MCDF-RCI method}
We have given here only a brief description of the methods applied in the present calculations. Although theoretical details are widely available in the literature, they can also be followed from previous papers \cite{fischer2019grasp2018, rathi2022comprehensive, verdebout2014hyperfine}. \\
The MCDF method is based on the Dirac-Coulomb Hamiltonian, which for an N electron system is written as, 
\begin{eqnarray}
\label{H_DC}
H_{DC} \; = \; \sum_{s=1}^{N} \left[ c\bm{\alpha}_s \cdot \bm{p}_s + (\beta_s-1)c^2  + V_s^{nuc} \right ]  + \sum_{s<t}^{N}\frac{1}{r_{st}}  \; .
\end{eqnarray}
Here, $\bm{\alpha}$ and $\beta$ are the 4 x 4 Dirac matrices and $\bm{ p}_{s}$ is the momentum operator of the $s^{th}$ electron. $V^{nuc}$ and $c$ are the electron-nucleus interaction potential and the speed of light, respectively. The last term denotes the electron-electron Coulomb interaction. \\
The atomic state functions (ASFs), describing the energy states for $N$ electrons, are derived as a linear combination of the configuration state functions (CSFs),
\begin{eqnarray}
\big|\Psi\big> \; = \; \sum_{s=1}^{N_{C S F}}e_{s}\big|\gamma_{s}P J M_J\big> , 
\end{eqnarray}
where e$_s$ are the mixing coeﬃcients. $P$, $J$ and $M_J$ represent the parity, total angular momentum quantum number and its associated magnetic quantum number, respectively. $\gamma$ refers to the occupation number and all other appropriate quantum labels that are required to uniquely specify a CSF. These CSFs are in-turn built from the anti-symmetric products of Dirac's one-electron orbitals.\\
%
%
%
%
Employing the variational principle method and the relativistic self consistent field (RSCF) procedure, the radial components of one electron orbitals and the mixing coefficients are optimised to self consistency. Further, to perform the relativistic configuration interaction (RCI) calculations, the Breit interaction and quantum electrodynamics (QED) effects, viz., self-energy, vacuum polarization, and specific mass shift corrections are also added to the Dirac-Coulomb Hamiltonian (equation(\ref{H_DC})).

All the further atomic structure properties like transition rates, hyperfine structure constants, Landé $g_J$ factors, and isotope shift factors can be estimated from the ASFs and are briefly described in the following sections.
\subsubsection{Transition parameters}
The line parameters for a transition between upper $u$ and lower $l$ states, can be expressed in terms of the reduced matrix elements of operator $(O)$ corresponding to the multipole expansion of electron-photon interaction \cite{grant1974gauge},  
\begin{eqnarray}
\big<u\big|\big|O_\xi^q\big|\big|l\big> .
\end{eqnarray}
Here $\xi$ is the rank of the operator. Its values can be 0 or 1, which defines the electric or magnetic type of multipole. Further, $q$ is 1 and 2, referring to the dipole and quadrupole transitions.
By applying the Racah algebra, these matrices can be further simplified into a weighted sum over radial integrals and the weights can be obtained by performing angular integration. We have studied only electric and magnetic dipole $(q =$1) and quadrupole ($q=$2) transitions in the present work.
\subsubsection{Hyperfine structures and Landé g$_{J}$ factors}
%
%
The hyperfine constants A$_J$ and B$_J$ are defined as,
\begin{eqnarray}
A_J\; = \; \frac{\mu_I}{I}\frac{1}{{\sqrt{(J(J+1))}}}\big<\gamma P J\big|\big|\bm{T}^{(1)}\big|\big|\gamma P J\big>,
\end{eqnarray}
\begin{eqnarray}
B_J\; = \;2Q_I\frac{\sqrt{J(2J-1)}}{\sqrt{(J+1)(2J+3)}}\big< \gamma P J\big|\big|\bm{T}^{(2)}\big|\big|\gamma P J\big>,
\end{eqnarray}
where $\mu_I$ and $Q_I$ are the nuclear magnetic dipole and electric quadrupole moments, respectively. $\bm{T}^{(k)}$ is a spherical tensor operator in the electron space having a rank $k$  \cite{osti_4571333}. In this study, only the diagonal hyperfine components are considered.
Further, the Landé $g_J$ factor determines the splitting of the magnetic levels in the presence of an external magnetic field and can be written as \cite{verdebout2014hyperfine}, 
\begin{eqnarray}
g_J \; = \;
\frac{2}{\sqrt{J(J+1)}} \big< \gamma P J  \big\| \sum_{s=1}^N  \left[ -\iota \frac{\sqrt{2}}{2 \alpha^2} r_s \big( \bm{\alpha}_s\bm{X}_s^{(1)} \big )^{(1)}  + \frac{g_{spin}-2}{2} \beta_s \Gamma_s\right] \big\|\gamma P J\big> \:,
\end{eqnarray}
where $\Gamma$ and $\alpha$ are the relativistic spin matrix and fine-structure constant, respectively. $\bm{X}$ refers to a spherical tensor. The values of the spin g factor, $g_{spin}$, is 2.00232.
\subsubsection{Isotope shift}
The difference in the atomic transition frequencies of the isotopes of an atom is predominantly due to the difference in the masses and nuclear charge distributions of the isotopes and are called mass shift and field shift, respectively.\\
The mass shift correction operator $H_{MS}$ is given as, \cite{shabaev1994relativistic},
  \begin{eqnarray}
 \label{hms}
 H_{MS} = \frac{1}{2M} \sum_{s,t}^N \left (
 \bm{p_s \cdot p_t} - \frac{\alpha Z}{r_s} 
 \left(\bm{\alpha_{s}}+ 
 \frac{(\bm{\alpha_{s}\cdot r_{s}})\bm{r_s}}{r_s^2}
 \right)\cdot \bm{p_t}
 \right) .
 \label{mass_shift}
 \end{eqnarray}
Equation \ref{mass_shift} can be further separated into the one body ($s = t$) and two body $(s \neq t)$ terms, corresponding to the NMS and SMS, respectively. 
For two isotopes $A$ and $A'$ of mass $M$ and $M'$, the level mass shift can be written as, 
\begin{eqnarray}
 \delta E_{i,MS}^{A,A'} = \delta E_{i,NMS}^{A,A'} - 
 \delta E_{i,SMS}^{A,A'} = \frac{M-M'}{MM'}K_{MS}\hspace{3mm} ,
\end{eqnarray}
where $K_{MS}$ is the electronic factor defined as,
\begin{eqnarray}
\frac{K_{MS}}{M} \; = \; \langle \gamma P J M_J\big|H_{MS}\big|\gamma P J M_J \rangle \hspace{3mm}.
 \label{kms}
\end{eqnarray}
The energy shift due to different nuclear charge radii of two isotopes $A$ and $A'$ for a particular atomic state, is, 
\begin{eqnarray}
\delta E_{A,A^\prime}^s \; = \; F^s
 ( \langle r^2 \rangle _A \;  - \; \langle r^2 \rangle _{A^\prime} ) \; ,
\end{eqnarray}
where $F^s$ is the field shift factor given by,
\begin{eqnarray}
F^s \; = \; \frac{2\pi}{3}\left(\frac{Ze^2}{4\pi\epsilon_o}\right)\Delta|\Psi(\bm{O})|^2_s \; . 
\end{eqnarray}
Here $\Delta|\Psi(\bm{O})|^2_s$ is the electron probability density at origin for the state $s$.\\
\subsection{The MBPT approach}
The present work also implements the MBPT method included in the Flexible Atomic Code \cite{gu2008flexible} to calculate the electronic structure parameters. It is based on no-pair Dirac-Coulomb-Breit Hamiltonian, written as,
\begin{eqnarray}
H_{DCB}\:=\: \sum_{s=1}^N \left[ h_d(s) - \frac{Z}{r_s} \right]\:+\: \sum_{s < t} \left( \frac{1}{r_{st}} + B_{st}\right)\:,
 \label{mbpt}
\end{eqnarray}
where h$_d$ is one electron Dirac Hamiltonian and B$_{st}$ is the Breit interaction. Further, the Hamiltonian in equation \ref{mbpt} is divided into a perturbation $V$ and model Hamiltonian $H_o$, given by, 
\begin{eqnarray}
H_{o}\:=\: \sum_{s=1}^N \left[ h_d(s) + U(r_{s}) \right] ,
\end{eqnarray}
and 
\begin{eqnarray}
V\:=\: \sum_{s=1}^N \left[ \frac{Z}{r_s}  + U(r_{s}) \right] \:+\: \sum_{s < t} \left( \frac{1}{r_{st}} + B_{st}\right)\:.
\end{eqnarray}
Here $U(r)$ is a model potential including the screening effect of all the electrons and should be chosen appropriately to make $V$ as small as possible.
The prime feature of the MBPT method is to divide the Hilbert space of the Hamiltonian into M and N orthogonal spaces. Here M represents the model space containing the non-Hermitian effective Hamiltonian and N includes the perturbation expansion. The electron correlation effects are accounted for within the M space, while the perturbation method is applied to work out the interaction between M and N model space. Eigenvalues in second-order are acquired by solving the generalized ﬁrst-order effective Hamiltonian eigenvalue problem.
\section{Computational procedure}
We performed the calculations for the lowest 71 levels having the configurations $1s^22s^22p^6nl$ for $n = 3-9$, $l = 0-6$ of Na-like Ar$^{7+}$, Kr$^{25+}$ and Xe$^{43+}$ ions using the MCDF method inbuilt in GRASP2018 \cite{fischer2019grasp2018}. These states are grouped according to their parity and independent calculations in extended optimal level (EOL) scheme are performed for the even and odd sets. The multi-reference (MR) set for odd states includes 34 levels which are built with the configurations $1s^22s^22p^6nl$ for $n = 3-9$, $l = p,f,h$. Similarly, the even parity states contain 37 levels pertaining to the configurations $1s^22s^22p^6nl$ for $n = 3-9$, $l = s,d,g,i$, making these as the even MR set.  
The initial calculations are performed on the MR sets, which account for the static electron correlations. In the next step, expansion of CSFs is done in a restrictive active space approach, by single and double substitutions of electrons from the MR set orbitals to higher orbitals upto $n=11$ and $l = s - i$ and forms the active space (AS). 
The $1s$ orbital is considered as inactive core, whereas the $2s$ and $2p$ orbitals are taken as the active core. The core-valence and valence-valence correlations are considered due to excitation from $2s^22p^6$ and $nl$ orbitals of the MR set. To monitor the convergence of the computed electronic structure parameters, the active space are divided into layers of the same $n$ and all the corresponding orbital quantum numbers upto $l = 6$. So, we define the AS's as,  AS\{10\} and As\{11\} with ${n = 3-10}$ and  $n = {3-11}$, respectively. Both AS\{10\} and As\{11\} have ${l = s,p,d,f,g,h,i}$. For each AS, only the outer orbitals are optimized while the inner ones are kept frozen. The number of CSFs generated for AS\{10\}/AS\{11\} is 53352/71820 and 47191/63339 for even and odd parities, respectively. Further, we incorporate the transverse photon interaction and QED effects by performing the RCI calculations. Since the calculations are done in the $jj-$coupling scheme, we transform the levels to $LS-$coupling scheme to ease their identification and comparisons with other's results. Also, due to the non-orthogonality of initial and final states, biorthogonal transformations are performed on the odd and even sets to remove the complexities in estimating transition probabilities. \\
The specific mass shift, normal mass shift and field shift are calculated using RIS4. Though it is a sub program written for GRASP2K \cite{jonsson2013new}, we did the required modifications to use it in alliance with GRASP2018 code.\\
To demonstrate the correlation effects, we choose Xe$^{43+}$, which is expected to be the best candidate among the three ions considered in the present work. In this regard, we performed calculations by simultaneously considering the valence-valence, core-valence and core-core correlations with a restriction that correlation orbitals will always have at least two electrons. The CSFs generated for AS\{10\}/AS\{11\} are 933241/1099439 and 726101/853388 for even and odd sets, respectively. These exceedingly large number of CSFs make the computations time-consuming and constrain the computational resources. To mitigate such a situation, we rearranged the CSFs into the zero and first-order spaces. We consider the full interaction of the CSFs in the zero-order space, while only the diagonal interaction due to the CSFs in the first-order space is included.\\
Furthermore, we performed the MBPT calculations using FAC as an additional accuracy test of our results. The $1s^22p^6nl$ configurations for $n = 3$ to $9$ and the corresponding orbital quantum numbers are considered in the model space M. All the possible configurations originated by single and double substitutions from the M space in the N space are considered. Further, the leading order corrections like vacuum polarization, self-energy, etc. are also taken into account in addition to $H_{DCB}$ in equation \ref{mbpt}.
%
%
\section{Results}
\subsection{Energy levels}
We carried out systematic MCDF-RCI calculations for AS\{9\} to AS\{11\} to determine the energies of the lowest 71 levels of Na-like Ar$^{7+}$, Kr$^{25+}$ and Xe$^{43+}$. The corresponding AS label defines the calculated energies for each active space. Further, to examine the importance of relativistic corrections, we performed another set of calculations for AS\{11\} without incorporating the Breit, QED and other corrections (labeled as AS\{11\}\_MCDF). Furthermore, to ascertain the accuracy of our results, independent calculations are performed using the MBPT approach. A thorough comparison with the previously available results and a detailed analysis of each ion are presented in the following subsections. 
\setlength{\tabcolsep}{1pt}
\setlength\LTleft{-35pt}            
\setlength\LTright\fill 

\begin{longtable}{@{\extracolsep{\fill}}|l|l|l|l|l|l|l|l|l|@{}}

\caption{The present MCDF-RCI and MBPT energies as a function of increasing active space for lowest 71 levels of Ar$^{7+}$ and the values from the NIST database \cite{NIST}, Fischer et
al.\cite{fischer2006relativistic} and Liang et al. \cite{liang2009r} .\label{tab argon_energy}} \\
\hline
\multicolumn{1}{|c|}{\textbf{Level  }} & \multicolumn{5}{c|}{\textbf{Present}}  & \multicolumn{3}{c|}{\textbf{Others}}  \\
\cline{2-9}
 & \multicolumn{1}{l|}{\textbf{ AS\{9\}}}&\multicolumn{1}{l|}{\textbf{AS\{10\}}}& \multicolumn{1}{l|}{\textbf{AS\{11\}\_MCDF}}& \multicolumn{1}{l|}{\textbf{AS\{11\}}}& \multicolumn{1}{l|}{\textbf{MBPT }}& \multicolumn{1}{l|}{\textbf{\cite{NIST} }}& \multicolumn{1}{l|}{\textbf{\cite{fischer2006relativistic} }}& \multicolumn{1}{l|}{\textbf{\cite{liang2009r} }}\\ \hline 
\endfirsthead
\multicolumn{9}{l}%
{{\bfseries \tablename\ \thetable{} -- continued from previous page}} \\
\hline
\multicolumn{1}{|c|}{\textbf{Level  }} & \multicolumn{5}{c|}{\textbf{Present}}  & \multicolumn{3}{c|}{\textbf{Others}}  \\
\cline{2-9}
& \multicolumn{1}{l|}{\textbf{AS\{9\} }}& \multicolumn{1}{l|}{\textbf{AS\{10\}}}& \multicolumn{1}{l|}{\textbf{AS\{11\}\_MCDF}}& \multicolumn{1}{l|}{\textbf{AS\{11\}}}& \multicolumn{1}{l|}{\textbf{MBPT }}& \multicolumn{1}{l|}{\textbf{\cite{NIST}}}& \multicolumn{1}{l|}{\textbf{\cite{fischer2006relativistic} }}&\multicolumn{1}{l|}{\textbf{\cite{liang2009r} }}\\ \hline 
\endhead
\hline \multicolumn{9}{|l|}{{Continued on next page}} \\ \hline
\endfoot
\hline \hline
\endlastfoot
$3s  ~^2S _{1/2}$&        0.00 &        0.00 &        0.00 &        0.00 &        0.00 &        0.00 &       0.00 &       0.00 \\
$3p  ~^2P _{1/2}$&   140163.98 &   138562.24 &   139963.65 &   139952.76 &   140274.29 &   140095 &  140354.56 &  139908 \\
$3p  ~^2P _{3/2}$&   142840.25 &   141278.36 &   142792.25 &   142665.77 &   142891.08 &   142808 &  143155.47 &  142372 \\
$3d  ~^2D _{3/2}$&   332531.33 &   332672.19 &   333376.22 &   333132.07 &   332431.55 &   332609 &  334155.28 &  332565 \\
$3d  ~^2D _{5/2}$&   332655.71 &   332237.25 &   333526.52 &   333242.76 &   332517.00 &   332754 &  334322.56 &  332789 \\
$4s  ~^2S _{1/2}$&   573000.41 &   575585.06 &   576912.42 &   576777.26 &   574151.74 &   575958 &  577405.01 &  573583\\
$4p  ~^2P _{1/2}$&   625329.28 &   627236.15 &   629103.42 &   628927.47 &   626667.23 &   628241&  629791.40 &  625834\\
$4p  ~^2P _{3/2}$&   626330.41 &   628248.19 &   630155.18 &   629938.00 &   627633.10 &   629243&  630840.80 &  626741\\
$4d  ~^2D _{3/2}$&   694494.73 &   697073.90 &   698598.47 &   698340.23 &   695383.57 &   697532&  699452.63 &  695061\\
$4d  ~^2D _{5/2}$&   694561.74 &   696861.19 &   698671.44 &   698395.73 &   695430.33 &   697621&  699547.38 &  695162\\
$4f  ~^2F _{5/2}$&   712798.90 &   716021.61 &   718155.32 &   717885.61 &   715162.82 &   716852&  719153.28 &  713644\\
$4f  ~^2F _{7/2}$&   712828.10 &   716051.45 &   718186.06 &   717915.46 &   715188.73 &   716875&  719195.48 &  713676\\
$5s  ~^2S _{1/2}$&   803416.17 &   806675.08 &   808457.39 &   808230.23 &   804477.34 &   807306&            &  804185\\
$5p  ~^2P _{1/2}$&   828458.33 &   831368.80 &   833384.86 &   833158.88 &   829602.51 &   832261&            &  829164\\
$5p  ~^2P _{3/2}$&   828938.80 &   831853.79 &   833890.10 &   833643.06 &   830064.53 &   832749&            &  829598\\
$5d  ~^2D _{3/2}$&   861411.99 &   864656.14 &   866487.80 &   866222.78 &   862257.56 &   865274&            &  862142\\
$5d  ~^2D _{5/2}$&   861449.21 &   864542.75 &   866525.69 &   866251.69 &   862284.47 &   865278&            &  862194\\
$5f  ~^2F _{5/2}$&   870984.01 &   874473.69 &   876614.88 &   876343.30 &   872354.29 &   875329&            &  871840\\
$5f  ~^2F _{7/2}$&   870998.57 &   874488.67 &   876630.53 &   876358.29 &   872366.70 &   875349&            &  871856\\
$5g  ~^2G _{7/2}$&   871468.63 &   875216.75 &   877347.73 &   877077.28 &   873415.41 &   876005&            &  872346\\
$5g  ~^2G _{9/2}$&   871478.18 &   875226.83 &   877357.18 &   877086.71 &   873424.40 &   876019&            &  872356\\
$6s  ~^2S _{1/2}$&   919241.76 &   922756.83 &   924704.39 &   924459.67 &   920092.67 &   923471&            &  920062\\
$6p  ~^2P _{1/2}$&   933102.75 &   936416.30 &   938495.41 &   938248.04 &   934032.59 &            &            &  933888\\
$6p  ~^2P _{3/2}$&   933370.10 &   936685.97 &   938776.38 &   938517.23 &   934288.44 &   937520&            &  934128\\
$6d  ~^2D _{3/2}$&   951407.19 &   954909.60 &   956879.89 &   956611.42 &   952137.44 &   955682&            &  952204\\
$6d  ~^2D _{5/2}$&   951429.52 &   954842.53 &   956900.24 &   956626.66 &   952153.79 &   955618&            &  952234\\
$6f  ~^2F _{5/2}$&   956984.94 &   960604.56 &   962751.77 &   962479.12 &   957975.52 &            &            &  957849\\
$6f  ~^2F _{7/2}$&   956993.26 &   960613.15 &   962760.82 &   962487.72 &   957982.44 &   961492&            &  957858\\
$6g  ~^2G _{7/2}$&   957311.41 &   961087.87 &   963223.16 &   962951.57 &   958541.66 &   961941&            &  958190\\
$6g  ~^2G _{9/2}$&   957316.94 &   961093.95 &   963228.59 &   962956.98 &   958546.88 &   961963&            &  958196\\
$6h  ~^2H _{9/2}$&   957327.47 &   961121.24 &   963278.27 &   963005.75 &   958829.47 &   961923&            &  958204\\
$6h  ~^2H _{11/2}$&   957331.16 &   961124.93 &   963281.97 &   963009.44 &   958832.89 &   961937&            &  958208\\
$7s  ~^2S _{1/2}$&   985637.21 &   989271.57 &   991307.05 &   991045.74 &   986381.71 &   990073&            &           \\
$7p  ~^2P _{1/2}$&   994104.07 &   997611.83 &   999721.98 &   999464.07 &   994885.66 &   998465&            &           \\
$7p  ~^2P _{3/2}$&   994268.09 &   997777.14 &   999894.13 &   999629.08 &   995041.67 &   998582&            &           \\
$7d  ~^2D _{3/2}$&  1005331.99 &  1008899.23 &  1010994.67 &  1010721.19 &  1005973.27 &  1009751&            &           \\
$7d  ~^2D _{5/2}$&  1005344.40 &  1008911.63 &  1011007.08 &  1010733.59 &  1005983.51 &  1009752&            &           \\
$7f  ~^2F _{5/2}$&  1008851.86 &  1012540.63 &  1014691.73 &  1014418.42 &  1009649.34 &            &            &           \\
$7f  ~^2F _{7/2}$&  1008857.06 &  1012546.00 &  1014697.42 &  1014423.81 &  1009653.68 &  1013424 &            &           \\
$7g  ~^2G _{7/2}$&  1009074.15 &  1012864.42 &  1015005.43 &  1014733.15 &  1009972.92 &            &            &           \\
$7g  ~^2G _{9/2}$&  1009077.67 &  1012868.37 &  1015008.88 &  1014736.58 &  1009975.82 &            &            &           \\
$7h  ~^2H _{9/2}$&  1009086.58 &  1012886.05 &  1015043.78 &  1014770.55 &  1010151.10 &  1013780 &            &           \\
$7h  ~^2H _{11/2}$&  1009086.76 &  1012888.36 &  1015046.10 &  1014772.87 &  1010153.31 &  1013780 &            &           \\
$7i  ~^2I _{11/2}$&  1009088.25 &  1012896.74 &  1015053.21 &  1014780.95 &  1010287.11 &            &            &           \\
$7i  ~^2I _{13/2}$&  1009089.07 &  1012898.40 &  1015054.88 &  1014782.61 &  1010288.62 &            &            &           \\
$8s  ~^2S _{1/2}$&  1027229.13 &  1030926.64 &  1033003.69 &  1032739.48 &  1027888.82 &  1031716 &            &           \\
$8p  ~^2P _{1/2}$&  1032770.52 &  1036383.52 &  1038510.78 &  1038247.01 &  1033456.98 &            &            &           \\
$8p  ~^2P _{3/2}$&  1032878.43 &  1036492.18 &  1038623.76 &  1038355.46 &  1033558.85 &  1036968 &            &           \\
$8d  ~^2D _{3/2}$&  1040151.61 &  1043802.58 &  1045919.33 &  1045645.80 &  1040740.23 &  1044625 &            &           \\
$8d  ~^2D _{5/2}$&  1040162.46 &  1043813.58 &  1045930.18 &  1045656.65 &  1040747.11 &  1044645 &            &           \\
$8f  ~^2F _{5/2}$&  1042515.05 &  1046243.63 &  1048397.24 &  1048123.50 &  1043202.94 &            &            &           \\
$8f  ~^2F _{7/2}$&  1042518.53 &  1046247.23 &  1048401.07 &  1048127.11 &  1043205.85 &  1047116 &            &           \\
$8g  ~^2G _{7/2}$&  1042670.94 &  1046468.59 &  1048614.24 &  1048341.50 &  1043400.04 &            &            &           \\
$8g  ~^2G _{9/2}$&  1042673.21 &  1046471.19 &  1048616.45 &  1048343.71 &  1043402.23 &            &            &           \\
$8h  ~^2H _{9/2}$&  1042679.92 &  1046483.68 &  1048641.92 &  1048368.23 &  1043511.79 &            &            &           \\
$8h  ~^2H _{11/2}$&  1042680.81 &  1046485.29 &  1048643.53 &  1048369.84 &  1043513.38 &            &            &           \\
$8i  ~^2I _{11/2}$&  1042681.04 &  1046490.62 &  1048647.61 &  1048374.90 &  1043608.36 &  1047291 &            &           \\
$8i  ~^2I _{13/2}$&  1042682.42 &  1046491.73 &  1048648.73 &  1048376.01 &  1043609.46 &  1047304 &            &           \\
$9s  ~^2S _{1/2}$&  1080418.51 &  1078210.13 &  1072332.40 &  1072058.61 &  1055602.17 &  1059616 &            &           \\
$9p  ~^2P _{1/2}$&  1058817.61 &  1062492.69 &  1064630.11 &  1064362.90 &  1059445.14 &  1063335 &            &           \\
$9p  ~^2P _{3/2}$&  1058892.05 &  1062567.63 &  1064708.10 &  1064437.69 &  1059515.43 &  1063412 &            &           \\
$9d  ~^2D _{3/2}$&  1063925.36 &  1067637.15 &  1069756.52 &  1069484.45 &  1064485.99 &  1068471 &            &           \\
$9d  ~^2D _{5/2}$&  1063936.21 &  1067655.34 &  1069766.64 &  1069493.04 &  1064490.66 &  1068511 &            &           \\
$9f  ~^2F _{5/2}$&  1065590.38 &  1069343.46 &  1071498.73 &  1071224.69 &  1066215.45 &            &            &           \\
$9f  ~^2F _{7/2}$&  1065592.82 &  1069345.98 &  1071501.42 &  1071227.22 &  1066217.53 &  1070277 &            &           \\
$9g  ~^2G _{7/2}$&  1065703.91 &  1069505.87 &  1071654.91 &  1071381.87 &  1066342.42 &  1070380 &            &           \\
$9g  ~^2G _{9/2}$&  1065705.53 &  1069507.74 &  1071656.50 &  1071383.45 &  1066343.87 &  1070380 &            &           \\
$9h  ~^2H _{9/2}$&  1065712.81 &  1069518.05 &  1071676.65 &  1071402.65 &  1066413.56 &            &            &           \\
$9h  ~^2H _{11/2}$&  1065713.85 &  1069519.09 &  1071677.69 &  1071403.69 &  1066414.66 &            &            &           \\
$9i  ~^2I _{11/2}$&  1065711.57 &  1069522.65 &  1071680.00 &  1071406.98 &  1066483.64 &            &            &           \\
$9i  ~^2I _{13/2}$&  1065712.35 &  1069523.43 &  1071680.79 &  1071407.76 &  1066484.24 &            &            &           \\

\end{longtable}
\subsubsection{Ar$^{7+}$}
\begin{figure}[H]
 
    \centering
    \subfloat
    {{\includegraphics[width=8cm,height=5cm]{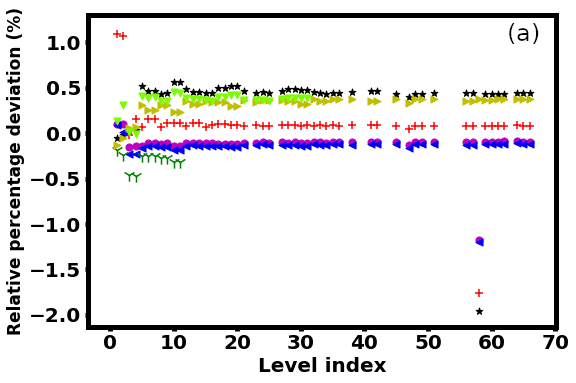} }}
    \qquad
    \subfloat{{\includegraphics[width=8cm,,height=5cm]{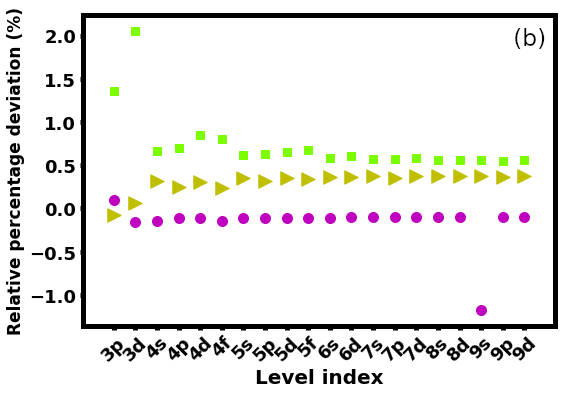} }}
    \caption{(a) Relative percentage deviation from the present AS\{9\}(star), AS\{10\}(plus), AS\{11\}(circle), AS\{11\}$\_$MCDF (triangle$\_$left), MBPT (triangle$\_$right) and previous work of Fischer et al.\cite{fischer2006relativistic} (tri$\_$down) and Liang et al.\cite{liang2009r} (triangle$\_$down) with respect to the NIST \cite{NIST} values. (b) Relative percentage deviation in energy (for multiplets) from the present AS\{11\}, MBPT and opacity database \cite{TOPbase} (square) with respect to the NIST \cite{NIST} values for Ar$^{7+}$.  }
    \label{fig:energy_ar}
\end{figure}

 We compare the present energy values for Ar$^{7+}$ in Table \ref{tab argon_energy} with the MCDF results for 12 levels reported by Fischer et al.\cite{fischer2006relativistic}, R-matrix calculations of Liang et al.\cite{liang2009r} for 32 levels and the corresponding results available at the NIST\cite{NIST} data base. 
For an enhanced analysis, the deviation of the present (AS\{9\} to AS\{11\}, AS\{11\}$\_$MCDF and MBPT) and previous \cite{fischer2006relativistic,liang2009r} results relative to the NIST values are shown in Fig. \ref{fig:energy_ar} (a). We find a fair convergence of our results with mean absolute deviations of 0.45\%, 0.16\%, 0.13\% and 0.16\% for AS\{9\}-AS\{11\} and AS\{11\}\_MCDF, respectively. The present AS\{11\} values show a maximum difference of 1.17\% for $9s$ state but differ by less than 0.15\% for all the other states. 
The results of Fischer et al.\cite{fischer2006relativistic} and Liang et al.\cite{liang2009r} have average variations of 0.30\% and 0.35\%, respectively, as compared to the NIST values. Further, the present MBPT results deviate by 0.32\% and 0.42\% relative to the NIST and the present AS\{11\} energies, respectively. 
Furthermore, the energies at opacity database \cite{TOPbase} were available for multiplets. Therefore, we converted our AS\{11\}, MBPT and NIST energies of the fine structure levels to their corresponding multiplets. To compare these results, in Fig.\ref{fig:energy_ar} (b) the relative difference of present AS\{11\}, MBPT and opacity \cite{TOPbase} results with respect to the NIST values is displayed. 
 Energies from \cite{TOPbase} show an absolute mean discrepancy of 0.74\% and a max difference of about 2\%. Therefore, our results are more precise, as evident from Fig.\ref{fig:energy_ar} (b). Large deviation in energies can lead to a fallacious prediction of stellar and plasma properties. Our results for almost all the levels are more accurate than that of \cite{fischer2006relativistic,liang2009r,TOPbase} and hence are of prime importance for stellar and plasma modeling. Moreover, we also provide energies of few fine structure levels which are missing in the NIST database. 
\setlength{\tabcolsep}{0.0001pt}

\begin{longtable}{|l|l|l|l|l|l|l|l|l|}
\caption{The present MCDF-RCI and MBPT energies as a function of increasing active space for lowest 71 levels of Kr$^{25+}$ and the values of NIST \cite{NIST}, Younis et al.\cite{younis2005energy} and Liang et al. \cite{liang2009r}.
\label{tab 2}} \\

\hline
\multicolumn{1}{|c|}{\textbf{Level }} & \multicolumn{5}{c|}{\textbf{Present}}  & \multicolumn{3}{c|}{\textbf{Others}}  \\
\cline{2-9}
 & \multicolumn{1}{l|}{\textbf{ AS\{9\} }} & \multicolumn{1}{l|}{\textbf{AS{10}}}& \multicolumn{1}{l|}{\textbf{AS\{11\}\_MCDF}}& \multicolumn{1}{l|}{\textbf{AS\{11\}}}& \multicolumn{1}{l|}{\textbf{MBPT }}& \multicolumn{1}{l|}{\textbf{\cite{NIST} }}& \multicolumn{1}{l|}{\textbf{\cite{younis2005energy} }}& \multicolumn{1}{l|}{\textbf{\cite{liang2009r} }}\\ \hline 
\endfirsthead
\multicolumn{9}{l}%
{{\bfseries \tablename\ \thetable{} -- continued from previous page}} \\
\hline
\multicolumn{1}{|c|}{\textbf{Level  }} & \multicolumn{5}{c|}{\textbf{Present}}  & \multicolumn{3}{c|}{\textbf{Others}}  \\
\cline{2-9}
\hline \multicolumn{1}{|l|}{\textbf{Level }} & \multicolumn{1}{l|}{\textbf{AS\{9\} }} & \multicolumn{1}{l|}{\textbf{AS\{10\}}}& \multicolumn{1}{l|}{\textbf{AS\{11\}\_MCDF}}& \multicolumn{1}{l|}{\textbf{AS\{11\}}}& \multicolumn{1}{l|}{\textbf{MBPT }}& \multicolumn{1}{l|}{\textbf{\cite{NIST}}}& \multicolumn{1}{l|}{\textbf{\cite{younis2005energy} }}&\multicolumn{1}{l|}{\textbf{\cite{liang2009r} }}\\ \hline 
\endhead
\hline \multicolumn{9}{|l|}{{Continued on next page}} \\ \hline
\endfoot
\hline \hline
\endlastfoot
$3s ~^2S_{1/2}$&        0.00 &        0.00 &        0.00 &        0.00 &        0.00 &        0.00 &        0.00 &        0.00 \\
$3p ~^2P_{1/2}$&   462926.54 &   455016.25 &   456109.56 &   454133.64 &   455455.14 &   454413 &   437051.89 &   450844\\
$3p ~^2P_{3/2}$&   590875.31 &   557232.95 &   562750.14 &   558960.15 &   559481.29 &   558678 &   566645.26 &   551051\\
$3d ~^2D_{3/2}$&  1161906.40 &  1164121.60 &  1168744.69 &  1163042.28 &  1165044.30 &  1164184 &  1164532.75 &  1162837\\
$3d ~^2D_{5/2}$&  1212723.01 &  1186972.69 &  1190420.82 &  1183462.06 &  1184506.94 &  1183991 &  1183746.63 &  1184156\\
$4s ~^2S_{1/2}$&  4517256.72 &  4488406.77 &  4497558.75 &  4491857.41 &  4490602.29 &  4493690 &  4492702.28 &  4499148\\
$4p ~^2P_{1/2}$&  4684748.80 &  4675059.10 &  4683313.68 &  4677700.31 &  4676217.17 &  4679430 &  4662023.14 &  4682602\\
$4p ~^2P_{3/2}$&  4734448.05 &  4715546.36 &  4725825.07 &  4719395.02 &  4717746.65 &  4720640 &  4729768.39 &  4720886\\
$4d ~^2D_{3/2}$&  4952830.10 &  4944727.15 &  4952521.76 &  4945424.13 &  4943817.37 &  4947290 &  4948873.72 &  4949529\\
$4d ~^2D_{5/2}$&  4934110.12 &  4950609.85 &  4961922.57 &  4954338.31 &  4952452.71 &  4955980 &  4954624.36 &  4958532\\
$4f ~^2F_{5/2}$&  5044831.15 &  5062725.91 &  5074141.49 &  5066254.53 &  5064743.41 &  5067310 &  5067562.10 &  5073401\\
$4f ~^2F_{7/2}$&  5056311.22 &  5066771.38 &  5077689.83 &  5069742.85 &  5068163.79 &  5070870 &  5070524.47 &  5076934\\
$5s ~^2S_{1/2}$&  6460371.21 &  6453044.59 &  6463764.26 &  6456747.41 &  6453990.26 &  6459150 &  6368810.01 &  6464707\\
$5p ~^2P_{1/2}$&  6530778.49 &  6545980.95 &  6556628.94 &  6549822.59 &  6546785.10 &  6551930 &  6560187.01 &  6556318\\
$5p ~^2P_{3/2}$&  6563195.06 &  6566654.86 &  6577808.64 &  6570565.96 &  6567458.69 &  6572690 &  6564191.52 &  6574752\\
$5d ~^2D_{3/2}$&  6664057.33 &  6676659.06 &  6688049.61 &  6680517.55 &  6677410.55 &  6682670 &  6689110.82 &  6685973\\
$5d ~^2D_{5/2}$&  7009147.40 &  6682973.26 &  6693373.28 &  6685603.70 &  6681897.29 &  6687210 &  6691182.87 &  6690542\\
$5f ~^2F_{5/2}$&  6716581.98 &  6736640.74 &  6748614.33 &  6740694.93 &  6737533.65 &  6741940 &  6741628.59 &  6747479\\
$5f ~^2F_{7/2}$&  6725210.36 &  6738442.99 &  6750440.55 &  6742476.50 &  6739282.07 &  6743720 &  6742818.77 &  6749291\\
$5g ~^2G_{7/2}$&  6728662.30 &  6746128.64 &  6757241.04 &  6749309.92 &  6746303.90 &  6751050 &             &  6756030\\
$5g ~^2G_{9/2}$&  6773329.25 &  6746523.96 &  6758199.64 &  6750267.36 &  6747365.37 &  6752110 &             &  6757100\\
$6s ~^2S_{1/2}$&  7480228.18 &  7487402.14 &  7498063.32 &  7490603.59 &  7485851.30 &  7492200 &             &  7498234\\
$6p ~^2P_{1/2}$&  7594159.27 &  7540049.47 &  7550636.30 &  7543326.19 &  7539379.54 &  7545010 &             &  7550313\\
$6p ~^2P_{3/2}$&  7547256.11 &  7551655.67 &  7562617.24 &  7555064.09 &  7551139.94 &  7556850 &             &  7560346\\
$6d ~^2D_{3/2}$&  7596211.16 &  7612939.41 &  7624610.44 &  7616897.53 &  7613010.37 &  7618560 &             &  7622921\\
$6d ~^2D_{5/2}$&  7815608.99 &  7621522.85 &  7628597.02 &  7620747.55 &  7615620.28 &  7621220 &             &  7625520\\
$6f ~^2F_{5/2}$&  7632780.13 &  7647261.41 &  7659155.51 &  7651211.98 &  7647270.14 &  7652900 &             &  7657878\\
$6f ~^2F_{7/2}$&  7685709.08 &  7648224.91 &  7660228.02 &  7652255.29 &  7648281.58 &  7653950 &             &  7658920\\
$6g ~^2G_{7/2}$&  7652923.44 &  7654018.54 &  7664562.06 &  7656625.13 &  7652685.12 &            &             &  7663302\\
$6g ~^2G_{9/2}$&  7689341.35 &  7653106.32 &  7665119.17 &  7657180.38 &  7653298.98 &            &             &  7663918\\
$6h ~^2H_{9/2}$&  7641026.76 &  7653532.77 &  7665589.99 &  7657612.70 &  7653793.98 &            &             &  7664129\\
$6h ~^2H_{11/2}$&  7633534.69 &  7653969.74 &  7665999.78 &  7658022.60 &  7654204.08 &            &             &  7664541\\
$7s ~^2S_{1/2}$&  8083247.23 &  8097794.35 &  8108615.92 &  8100961.80 &  8095864.09 &            &             &            \\
$7p ~^2P_{1/2}$&  8124133.76 &  8130989.62 &  8141332.25 &  8133773.04 &  8128925.40 &            &             &            \\
$7p ~^2P_{3/2}$&  8118966.86 &  8136912.06 &  8148350.18 &  8140624.84 &  8136225.52 &            &             &            \\
$7d ~^2D_{3/2}$&  8236640.99 &  8176998.98 &  8187156.47 &  8179356.59 &  8174518.39 &            &             &            \\
$7d ~^2D_{5/2}$&  8209942.20 &  8182561.07 &  8188403.03 &  8180519.07 &  8176162.55 &            &             &            \\
$7f ~^2F_{5/2}$&  8407338.31 &  8196268.95 &  8208261.19 &  8200303.26 &  8195889.18 &            &             &            \\
$7f ~^2F_{7/2}$&  8185730.11 &  8206219.58 &  8208889.12 &  8200910.98 &  8196526.03 &            &             &            \\
$7g ~^2G_{7/2}$&  8185710.22 &  8199710.14 &  8211638.22 &  8203697.37 &  8199412.29 &            &             &            \\
$7g ~^2G_{9/2}$&  8209781.25 &  8200439.44 &  8212124.28 &  8204181.88 &  8199798.30 &            &             &            \\
$7h ~^2H_{9/2}$&  8227746.48 &  8200375.93 &  8212444.77 &  8204464.00 &  8200125.52 &            &             &            \\
$7h ~^2H_{11/2}$&  8184621.96 &  8200689.13 &  8212760.89 &  8204780.04 &  8200384.10 &            &             &            \\
$7i ~^2I_{11/2}$&  8180538.23 &  8200633.81 &  8212661.23 &  8204719.11 &  8200534.12 &            &             &            \\
$7i ~^2I_{13/2}$&  8181279.92 &  8200866.78 &  8212844.61 &  8204902.45 &  8200718.50 &            &             &            \\
$8s ~^2S_{1/2}$&  8468163.87 &  8486966.34 &  8498353.69 &  8490600.14 &  8485550.36 &            &             &            \\
$8p ~^2P_{1/2}$&  8489657.07 &  8508577.92 &  8519992.25 &  8512287.73 &  8507494.53 &            &             &            \\
$8p ~^2P_{3/2}$&  8497545.38 &  8514645.46 &  8524985.18 &  8517175.81 &  8512341.12 &            &             &            \\
$8d ~^2D_{3/2}$&  8521583.95 &  8538753.08 &  8550260.49 &  8542409.84 &  8537688.38 &            &             &            \\
$8d ~^2D_{5/2}$&  8539475.55 &  8547568.25 &  8551294.59 &  8543388.53 &  8538790.37 &            &             &            \\
$8f ~^2F_{5/2}$&  8588679.12 &  8553446.47 &  8564645.20 &  8556678.33 &  8551914.79 &            &             &            \\
$8f ~^2F_{7/2}$&  8534460.74 &  8553858.83 &  8564943.84 &  8556962.51 &  8552341.29 &            &             &            \\
$8g ~^2G_{7/2}$&  8586581.88 &  8555658.00 &  8566918.80 &  8558975.25 &  8554321.46 &            &             &            \\
$8g ~^2G_{9/2}$&  8537883.99 &  8555472.37 &  8567193.70 &  8559248.99 &  8554580.05 &            &             &            \\
$8h ~^2H_{9/2}$&  8540622.64 &  8555308.69 &  8567382.26 &  8559399.11 &  8554796.36 &            &             &            \\
$8h ~^2H_{11/2}$&  8536092.84 &  8555623.70 &  8567687.63 &  8559704.45 &  8554969.53 &            &             &            \\
$8i ~^2I_{11/2}$&  8534866.86 &  8555571.30 &  8567572.06 &  8559627.59 &  8555078.25 &            &             &            \\
$8i ~^2I_{13/2}$&  8535058.44 &  8555649.98 &  8567668.90 &  8559724.38 &  8555201.90 &            &             &            \\
$9s ~^2S_{1/2}$&  8730098.55 &  8750493.60 &  8762306.00 &  8754502.12 &  8749676.28 &            &             &            \\
$9p ~^2P_{1/2}$&  8745473.14 &  8765783.25 &  8777603.13 &  8769805.97 &  8764978.31 &            &             &            \\
$9p ~^2P_{3/2}$&  8755470.81 &  8771465.21 &  8781390.18 &  8773532.28 &  8768357.45 &            &             &            \\
$9d ~^2D_{3/2}$&  8768681.04 &  8787013.28 &  8798713.37 &  8790833.12 &  8786005.10 &            &             &            \\
$9d ~^2D_{5/2}$&  8814175.14 &  8802075.16 &  8802311.11 &  8794390.05 &  8786778.75 &            &             &            \\
$9f ~^2F_{5/2}$&  8794609.35 &  8796954.81 &  8808818.54 &  8800845.37 &  8795951.77 &            &             &            \\
$9f ~^2F_{7/2}$&  8777370.86 &  8797124.60 &  8808960.72 &  8800977.21 &  8796251.17 &            &             &            \\
$9g ~^2G_{7/2}$&  8835088.12 &  8801596.05 &  8810530.81 &  8802584.49 &  8797664.25 &            &             &            \\
$9g ~^2G_{9/2}$&  8778751.27 &  8798511.79 &  8810571.09 &  8802625.68 &  8797845.48 &            &             &            \\
$9h ~^2H_{9/2}$&  8778388.44 &  8798752.23 &  8810828.66 &  8802843.89 &  8797992.35 &            &             &            \\
$9h ~^2H_{11/2}$&  8778448.57 &  8798959.06 &  8811033.86 &  8803049.08 &  8798113.98 &            &             &            \\
$9i ~^2I_{11/2}$&  8778162.21 &  8798898.96 &  8810916.99 &  8802970.90 &  8798191.82 &            &             &            \\
$9i ~^2I_{13/2}$&  8779976.51 &  8798982.46 &  8811002.08 &  8803055.94 &  8798278.68 &            &             &            \\
\end{longtable}

\subsubsection{Kr$^{25+}$}
\begin{figure}[H]
    \centering
    {{\includegraphics[width=7cm, height=6cm]{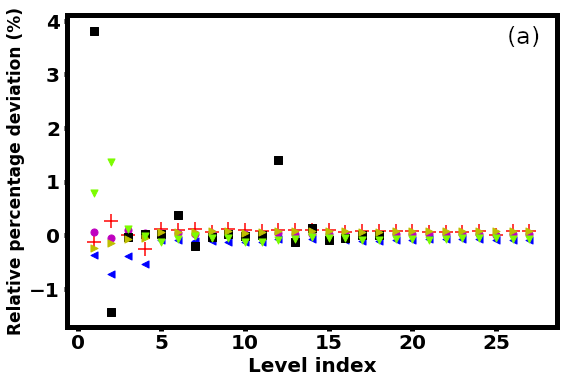} }}
    \qquad
    \subfloat{{\includegraphics[width=7.5cm, height=6cm]{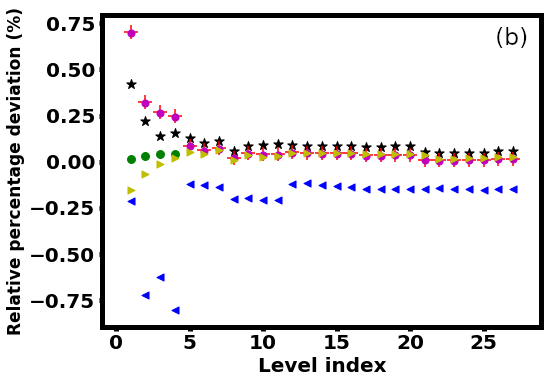} }}
    
    \caption{(a) Relative percentage deviation from the present AS\{10\}, AS\{11\}, AS\{11\}$\_$MCDF, MBPT and previous work of Younis et al.\cite{younis2005energy} (square) and Liang et al.\cite{liang2009r} (triangle$\_$down) with respect to the NIST \cite{NIST} values for Kr$^{25+}$. (b) Relative percentage deviation from the present AS\{9\}, AS\{10\}, AS\{11\}, AS\{11\}$\_$MCDF , MBPT and Vilkas et al.\cite{vilkas2008relativistic} (point) results with respect to the NIST \cite{NIST} values for Xe$^{43+}$.}
    \label{fig:energy_xe_kr}
\end{figure}

For sodium-like Kr$^{25+}$ results of only 28 levels (excluding the excited core levels) were available at the NIST, while Younis et al. \cite{younis2005energy} and Liang et al. \cite{liang2009r} have reported energies of the lowest 20 and 32 levels respectively. The present MCDF-RCI and MBPT results for each active space, the values available from the NIST, CIV3 \cite{younis2005energy} and R-matrix \cite{liang2009r} calculations are tabulated in Table. \ref{tab 2}. 
Further, the relative difference in energies of the present 
and previous \cite{younis2005energy,liang2009r} results with respect to the NIST values is displayed pictorially in Fig. \ref{fig:energy_xe_kr}(a). To distinguish the different results we did not include AS\{9\} values in Fig. \ref{fig:energy_xe_kr}(a) as the maximum deviation is found to be 6\% for AS\{9\}, although the mean deviation is 0.88\% only. It is noticeable from the figure that the present calculations converge well towards the NIST values with an increase in the active space and show  mean difference of 0.091\% and 0.032\% for AS\{10\} and AS\{11\}, respectively. 
The agreement with the NIST results improves from 0.15\%  AS\{11\}\_MCDF to 0.032\% as we move from AS\{11\}\_MCDF to AS\{11\} calculations. However, the RCI calculations introduce a remarkable rectification for the lowest four levels from 0.7\% to 0.05\%. This makes the importance of including the relativistic corrections for Kr$^{25+}$ abundantly clear. In our MBPT calculations, the mean discrepancy is only 0.079\% and 0.05\% relative to the NIST and the present AS\{11\} results, respectively. Such an excellent agreement between the two theories further validates the accuracy of our calculations. 
Younis et al. \cite{younis2005energy} and Liang et al.\cite{liang2009r} have mean deviations of 0.43\% and 0.14\%, respectively, which are more than our results. Moreover, energies reported in \cite{younis2005energy} for a few levels show relatively large differences.
For example, energy for $3p ~^2P_{1/2}$ state strays by 3.8\% in \cite{younis2005energy} whereas, the corresponding value deviates only by 0.06\%, 0.23\% and 1.37\% in the present AS\{11\}, MBPT and R-matrix \cite{liang2009r} calculations, respectively. 
%
%
An excellent agreement of our calculated energies for Kr$^{25+}$ ion with the NIST assists in establishing the reliability of the other electronic structure parameters determined in this work. Moreover, the results for $1s^22s^22p^6nl$ levels, where $n = 7-9$ and $l = 0-6$, are reported for the first time. 
%
%
\setlength{\tabcolsep}{0.5pt}
\begin{longtable}{|l|l|l|l|l|l|l|l|}
\caption{The present MCDF-RCI and MBPT energies as a function of increasing active space for lowest 71 levels of Xe$^{43+}$ and the values of NIST \cite{NIST} and Vilkas et al.\cite{vilkas2008relativistic} .\label{tab energy_xe}} \\
\hline
\multicolumn{1}{|c|}{\textbf{Level  }} & \multicolumn{5}{c|}{\textbf{Present}}  & \multicolumn{2}{c|}{\textbf{Others}}  \\
\cline{2-8}
 & \multicolumn{1}{l|}{\textbf{ AS\{9\} }} & \multicolumn{1}{l|}{\textbf{AS\{10\}}}& \multicolumn{1}{l|}{\textbf{AS\{11\}\_MCDF}}& \multicolumn{1}{l|}{\textbf{AS\{11\}}}& \multicolumn{1}{l|}{\textbf{MBPT }}& \multicolumn{1}{l|}{\textbf{\cite{NIST} }}&  \multicolumn{1}{l|}{\textbf{\cite{vilkas2008relativistic} }}\\ \hline 
\endfirsthead
\multicolumn{7}{l}%
{{\bfseries \tablename\ \thetable{} -- continued from previous page}} \\
\hline
\multicolumn{1}{|c|}{\textbf{Level  }} & \multicolumn{5}{c|}{\textbf{Present}}  & \multicolumn{2}{c|}{\textbf{Others}}  \\
\cline{2-8}
 & \multicolumn{1}{l|}{\textbf{AS\{9\} }} & \multicolumn{1}{l|}{\textbf{AS\{10\}}}& \multicolumn{1}{l|}{\textbf{AS\{11\}\_MCDF}}& \multicolumn{1}{l|}{\textbf{AS\{11\}}}& \multicolumn{1}{l|}{\textbf{MBPT }}& \multicolumn{1}{l|}{\textbf{\cite{NIST} }} &\multicolumn{1}{l|}{\textbf{\cite{vilkas2008relativistic} }}\\ \hline 
\endhead
\hline \multicolumn{7}{|l|}{{Continued on next page}} \\ \hline
\endfoot
\hline \hline
\endlastfoot
$3s  ~^2S _{1/2}$&         0.00 &         0.00 &         0.00 &         0.00 &         0.00 &         0.00 &        0.00 \\
$3p  ~^2P _{1/2}$&    803597.42 &    801321.00 &    808701.78 &    801366.80 &    808237.43 &    806985&   806861\\
$3p  ~^2P _{3/2}$&   1497989.67 &   1496410.97 &   1512125.74 &   1496481.71 &   1502291.61 &   1501276&  1500802\\
$3d  ~^2D _{3/2}$&   2520145.15 &   2516866.44 &   2539364.21 &   2516980.78 &   2523954.83 &   2523660&  2522571\\
$3d  ~^2D _{5/2}$&   2675240.99 &   2672720.77 &   2700952.76 &   2672886.16 &   2678868.12 &   2679380&  2678242\\
$4s  ~^2S _{1/2}$&  12247007.16 &  12252449.26 &  12277697.50 &  12252891.35 &  12256272.81 &  12263000&            \\
$4p  ~^2P _{1/2}$&  12583392.06 &  12587971.32 &  12611835.84 &  12588459.39 &  12590454.02 &  12596000&            \\
$4p  ~^2P _{3/2}$&  12865638.55 &  12870451.16 &  12897970.74 &  12870944.59 &  12872672.16 &  12880000&            \\
$4d  ~^2D _{3/2}$&  13252501.56 &  13257389.26 &  13287473.48 &  13257894.61 &  13259162.62 &  13260300&            \\
$4d  ~^2D _{5/2}$&  13319835.81 &  13324994.47 &  13357294.73 &  13325520.36 &  13326498.54 &  13331400&            \\
$4f  ~^2F _{5/2}$&  13523728.37 &  13530391.90 &  13564429.87 &  13531105.71 &  13532277.99 &  13535960&            \\
$4f  ~^2F _{7/2}$&  13552458.26 &  13559182.65 &  13593523.62 &  13559901.45 &  13560874.30 &  13565090&            \\
$5s  ~^2S _{1/2}$&  17691516.20 &  17698803.33 &  17729098.17 &  17699408.85 &  17699727.45 &  17707860&            \\
$5p  ~^2P _{1/2}$&  17861051.31 &  17867883.00 &  17897168.13 &  17868527.86 &  17867897.83 &  17876490&            \\
$5p  ~^2P _{3/2}$&  18002751.19 &  18009691.24 &  18040809.72 &  18010338.21 &  18009552.17 &  18018300&            \\
$5d  ~^2D _{3/2}$&  18191935.79 &  18199050.55 &  18231474.29 &  18199689.69 &  18198588.76 &  18207450&            \\
$5d  ~^2D _{5/2}$&  18226734.74 &  18233979.46 &  18267429.06 &  18234628.97 &  18233351.17 &  18242330&            \\
$5f  ~^2F _{5/2}$&  18327479.69 &  18335286.09 &  18369499.42 &  18336030.60 &  18334769.18 &  18342070&            \\
$5f  ~^2F _{7/2}$&  18342224.56 &  18350068.34 &  18384514.12 &  18350816.36 &  18349448.71 &  18356850&            \\
$5g  ~^2G _{7/2}$&  18355712.51 &  18364505.29 &  18399024.98 &  18365270.96 &  18363833.20 &  18371490&            \\
$5g  ~^2G _{9/2}$&  18364515.96 &  18373313.16 &  18407832.32 &  18374078.30 &  18372607.39 &  18380270&            \\
$6s  ~^2S _{1/2}$&  20577690.48 &  20585778.18 &  20618204.31 &  20586459.18 &  20581083.49 &  20588070&            \\
$6p  ~^2P _{1/2}$&  20674790.76 &  20682593.11 &  20714185.35 &  20683307.59 &  20681116.81 &  20684390&            \\
$6p  ~^2P _{3/2}$&  20755738.20 &  20763600.10 &  20796256.96 &  20764315.76 &  20762103.26 &  20765470&            \\
$6d  ~^2D _{3/2}$&  20862551.27 &  20870563.65 &  20903924.65 &  20871265.40 &  20868869.20 &  20872670&            \\
$6d  ~^2D _{5/2}$&  20882734.98 &  20890820.46 &  20924760.95 &  20891528.15 &  20889038.87 &  20892990&            \\
$6f  ~^2F _{5/2}$&  20939934.24 &  20948274.58 &  20982631.04 &  20949042.24 &  20946583.86 &  20951570&            \\
$6f  ~^2F _{7/2}$&  20948483.57 &  20956847.70 &  20991356.57 &  20957617.69 &  20955091.72 &  20960220&            \\
$6g  ~^2G _{7/2}$&  20957201.46 &  20966137.60 &  21000667.79 &  20966905.03 &  20964211.91 &             &            \\
$6g  ~^2G _{9/2}$&  20962293.87 &  20971233.34 &  21005769.67 &  20972000.14 &  20969289.17 &             &            \\
$6h  ~^2H _{9/2}$&  20962780.96 &  20971823.93 &  21006418.89 &  20972634.04 &  20970072.18 &             &            \\
$6h  ~^2H _{11/2}$&  20966173.73 &  20975216.88 &  21009809.62 &  20976027.11 &  20973455.91 &             &            \\
$7s  ~^2S _{1/2}$&  22289682.76 &  22298174.92 &  22331465.01 &  22298895.85 &  22295426.31 &             &            \\
$7p  ~^2P _{1/2}$&  22350274.97 &  22358567.04 &  22391303.54 &  22359317.14 &  22356355.04 &             &            \\
$7p  ~^2P _{3/2}$&  22400792.48 &  22409120.95 &  22442522.33 &  22409871.86 &  22406787.84 &             &            \\
$7d  ~^2D _{3/2}$&  22467016.38 &  22475467.69 &  22509295.57 &  22476202.68 &  22473056.20 &             &            \\
$7d  ~^2D _{5/2}$&  22479736.25 &  22488233.09 &  22522418.02 &  22488971.77 &  22485743.95 &             &            \\
$7f  ~^2F _{5/2}$&  22515349.40 &  22523970.78 &  22558410.61 &  22524752.80 &  22521565.77 &             &            \\
$7f  ~^2F _{7/2}$&  22520734.94 &  22529372.24 &  22563918.30 &  22530155.85 &  22526927.83 &             &            \\
$7g  ~^2G _{7/2}$&  22526538.65 &  22535556.65 &  22570100.73 &  22536330.22 &  22532941.82 &             &            \\
$7g  ~^2G _{9/2}$&  22529743.02 &  22538763.40 &  22573315.20 &  22539536.43 &  22536138.52 &             &            \\
$7h  ~^2H _{9/2}$&  22530137.77 &  22539205.73 &  22573805.73 &  22540015.53 &  22536692.30 &             &            \\
$7h  ~^2H _{11/2}$&  22532273.54 &  22541341.55 &  22575942.02 &  22542151.51 &  22538824.26 &             &            \\
$7i  ~^2I _{11/2}$&  22532258.73 &  22541423.95 &  22575998.39 &  22542218.41 &  22538981.14 &             &            \\
$7i  ~^2I _{13/2}$&  22533782.31 &  22542947.34 &  22577522.03 &  22543741.80 &  22540501.73 &             &            \\
$8s  ~^2S _{1/2}$&  23387642.46 &  23396360.97 &  23430082.45 &  23397105.03 &  23393244.27 &             &            \\
$8p  ~^2P _{1/2}$&  23427922.85 &  23436489.08 &  23469857.62 &  23437259.31 &  23433755.40 &             &            \\
$8p  ~^2P _{3/2}$&  23461538.67 &  23470128.92 &  23503938.27 &  23470899.74 &  23467309.68 &             &            \\
$8d  ~^2D _{3/2}$&  23505432.28 &  23514125.41 &  23548209.66 &  23514879.57 &  23511238.26 &             &            \\
$8d  ~^2D _{5/2}$&  23513952.82 &  23522676.18 &  23556996.91 &  23523432.80 &  23519737.41 &             &            \\
$8f  ~^2F _{5/2}$&  23537635.62 &  23546418.52 &  23580912.74 &  23547209.52 &  23543548.10 &             &            \\
$8f  ~^2F _{7/2}$&  23541246.05 &  23550040.09 &  23584607.56 &  23550832.21 &  23547141.38 &             &            \\
$8g  ~^2G _{7/2}$&  23545257.45 &  23554324.33 &  23588880.49 &  23555103.28 &  23551282.07 &             &            \\
$8g  ~^2G _{9/2}$&  23547403.61 &  23556472.20 &  23591034.48 &  23557250.73 &  23553422.75 &             &            \\
$8h  ~^2H _{9/2}$&  23547711.17 &  23556795.28 &  23591399.93 &  23557605.35 &  23553811.02 &             &            \\
$8h  ~^2H _{11/2}$&  23549142.17 &  23558226.27 &  23592831.42 &  23559036.49 &  23555239.51 &             &            \\
$8i  ~^2I _{11/2}$&  23549126.92 &  23558294.67 &  23592873.57 &  23559089.13 &  23555355.17 &             &            \\
$9i  ~^2I _{13/2}$&  23550147.92 &  23559315.52 &  23593894.62 &  23560109.99 &  23556374.33 &             &            \\
$9s  ~^2S _{1/2}$&  24133596.29 &  24142451.10 &  24176388.32 &  24143209.43 &  24139043.75 &             &            \\
$9p  ~^2P _{1/2}$&  24161673.84 &  24170405.25 &  24204150.29 &  24171187.66 &  24167329.13 &             &            \\
$9p  ~^2P _{3/2}$&  24185162.66 &  24193910.96 &  24227958.89 &  24194693.85 &  24190767.35 &             &            \\
$9d  ~^2D _{3/2}$&  24215749.05 &  24224586.56 &  24258824.98 &  24225352.49 &  24221384.48 &             &            \\
$9d  ~^2D _{5/2}$&  24221745.27 &  24230603.71 &  24264992.62 &  24231371.35 &  24227350.97 &             &            \\
$9f  ~^2F _{5/2}$&  24238285.60 &  24247167.80 &  24281696.89 &  24247964.58 &  24243982.52 &             &            \\
$9f  ~^2F _{7/2}$&  24741075.49 &  24750028.49 &  24784619.68 &  24750829.71 &  24246506.47 &             &            \\
$9g  ~^2G _{7/2}$&  24243697.85 &  24252795.43 &  24287360.47 &  24253578.42 &  24249466.93 &             &            \\
$9g  ~^2G _{9/2}$&  24245204.86 &  24254303.69 &  24288873.40 &  24255086.36 &  24250969.95 &             &            \\
$9h  ~^2H _{9/2}$&  24245448.20 &  24254542.91 &  24289151.01 &  24255353.39 &  24251247.00 &             &            \\
$9h  ~^2H _{11/2}$&  24246453.19 &  24255547.89 &  24290156.41 &  24256358.48 &  24252250.35 &             &            \\
$9i  ~^2I _{11/2}$&  24246436.90 &  24255606.51 &  24290188.49 &  24256400.98 &  24252334.23 &             &            \\
$9i  ~^2I _{13/2}$&  24247154.10 &  24256323.61 &  24290905.74 &  24257118.08 &  24253050.21 &             &            \\
\end{longtable}
\subsubsection{Xe$^{43+}$}
Table. \ref{tab energy_xe} displays comparison of our calculated energies for Na-like Xe$^{43+}$ ion with the previous results of the four levels reported by Vilkas et al.   \cite{vilkas2008relativistic} using the relativistic many-body møller–plesset perturbation theory and 28 levels available at the NIST database. Similar to Ar$^{7+}$ and Kr$^{25+}$, we show the relative percentage deviation of the two theoretical works with respect to NIST values in Fig. \ref{fig:energy_xe_kr} (b). Clearly, agreement with the NIST results improves with increment in the AS except for the first four levels where the deviation in AS\{9\} is the least. However, the absolute mean deviation is only 0.08\% in case of AS\{11\} compared to that obtained for AS\{9\} (0.1\%), AS\{10\} (0.09\%), and AS\{11\}\_MCDF (0.22\%). 
 Therefore, we find that the quality of the wavefunctions and hence, the electronic parameters get enhanced by including the relativistic corrections.
Although the mean error in MBPT results is slightly less than the AS\{11\} calculations, the two results digress only by 0.06\% and thus, are in excellent agreement.
%
%

Energies calculated by Vilkas et al.\cite{vilkas2008relativistic} for $3p$ levels show better agreement with the NIST values than the present calculations. However, for $3d$ states, our results better match the corresponding NIST energies.
Overall, we can say that the present calculations provide fairly precise energies along with the new results for 41 levels for Xe$^{43+}$.\\
Further, a thorough examination of Fig. \ref{fig:energy_ar} (a), \ref{fig:energy_xe_kr} (a) and \ref{fig:energy_xe_kr} (b), affirms the dominant role of relativistic effects in heavy ions. The relative difference in AS\{11\}\_MCDF and AS\{11\} increases from 0.03\% in Ar$^{7+}$ to 0.13\% and 0.26\% in Kr$^{25+}$ and Xe$^{43+}$, respectively. Therefore, including the relativistic effects in the atomic structure calculations becomes essential with increase in nuclear charge. 
\subsection{Transition parameters}
Using the MCDF-RCI method, we calculated the wavelengths, transition rates and oscillator strengths corresponding to E1, M1, E2 and M2 transitions among the $1s^22s^22p^6nl$ for $n = 3-9$ and $l = s,p,d,f,g,h,i$ levels of Na-like Ar$^{7+}$, Kr$^{25+}$ and Xe$^{43+}$ ions. The transition probabilities are obtained using the Coulomb and Babushkin gauges. We present the results through Tables 4 - 6 in the Babushkin gauge only, while Coulomb gauge results are used to determine the uncertainty parameter dT. It appraises the uncertainty in the obtained transition rates and is given by
\begin{eqnarray}
\label{dt}
 dT = \frac{|A_B - A_C|}{ \text{max} (A_B,A_C) },
\end{eqnarray}
where $A_B$ and $A_C$ are the transition rates in the Babushkin and Coulomb gauges, respectively.
We do not report the transition rates having values smaller than $ 10^{4}$ for E2, M1 and M2 transitions. Since E1 transitions are considered more important than the forbidden transitions, therefore, as an additional validation of the present results, we performed similar calculations using the MBPT technique. 
Detailed analysis of radiative parameters for each ion is as follows.
\begin{figure}[H]
    \centering
    \subfloat
    {{\includegraphics[width=8cm,height=6cm]{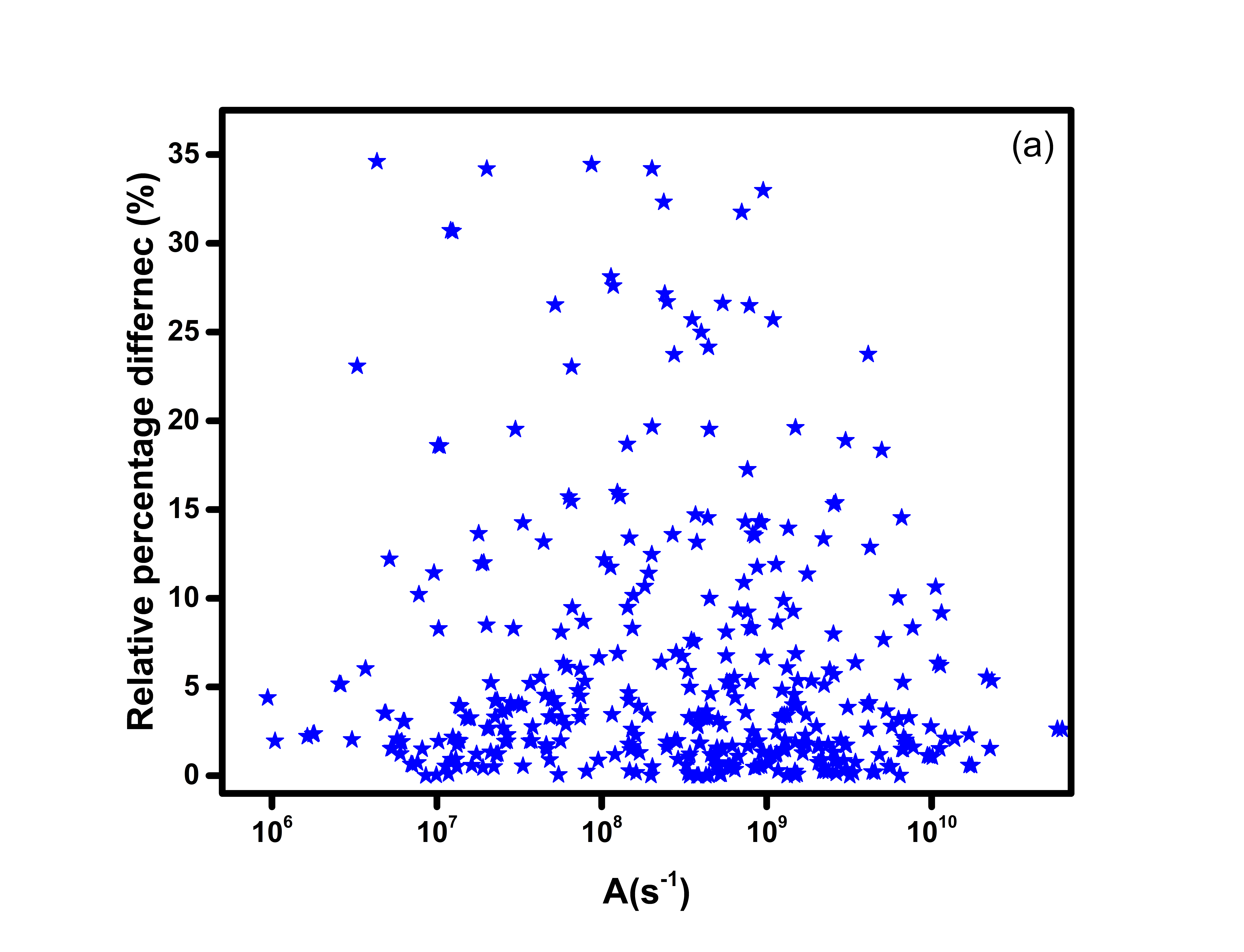} }}
    \quad
    \subfloat{{\includegraphics[width=8cm,height=6cm]{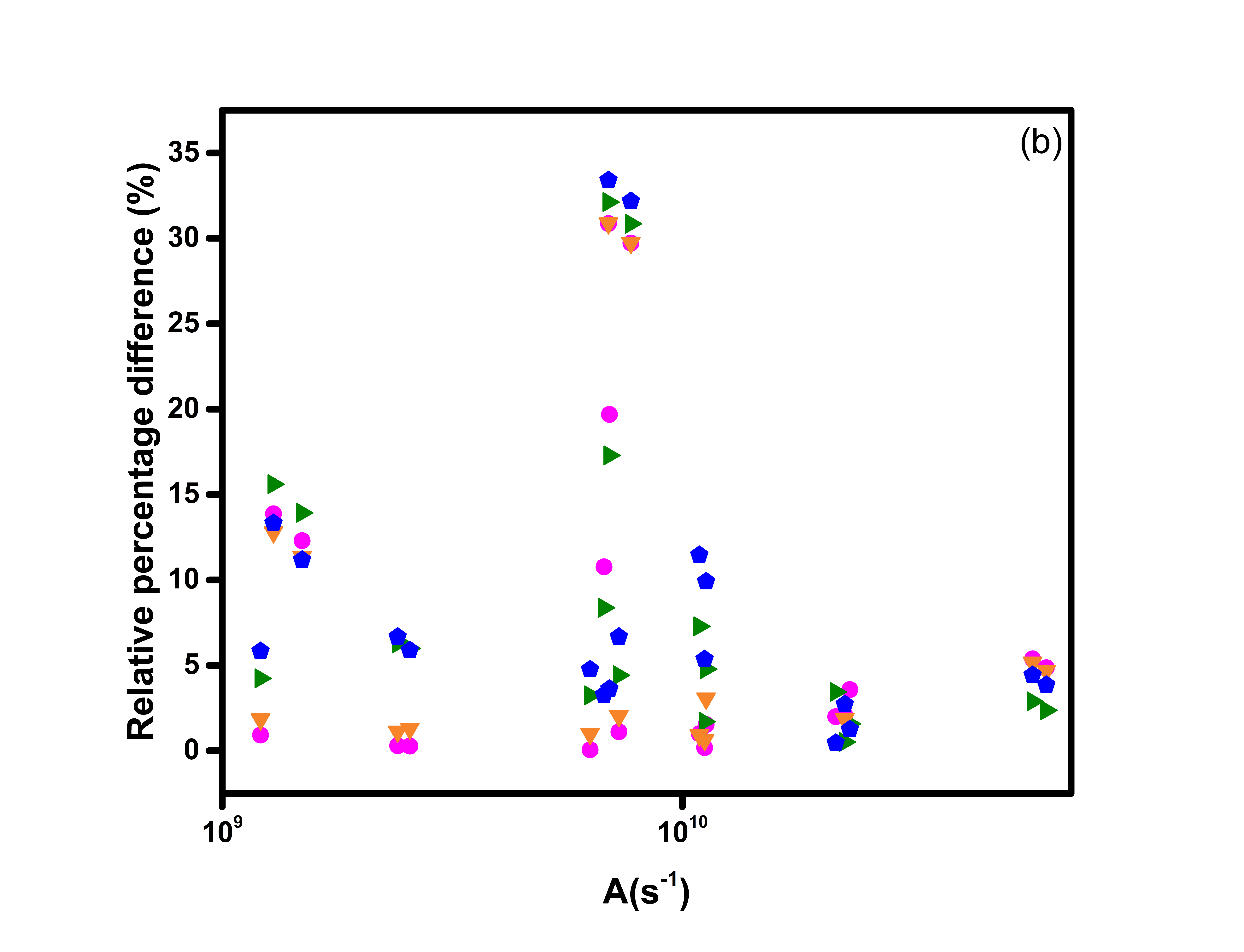} }}
    \quad
    \subfloat{{\includegraphics[width=8cm,height=6cm]{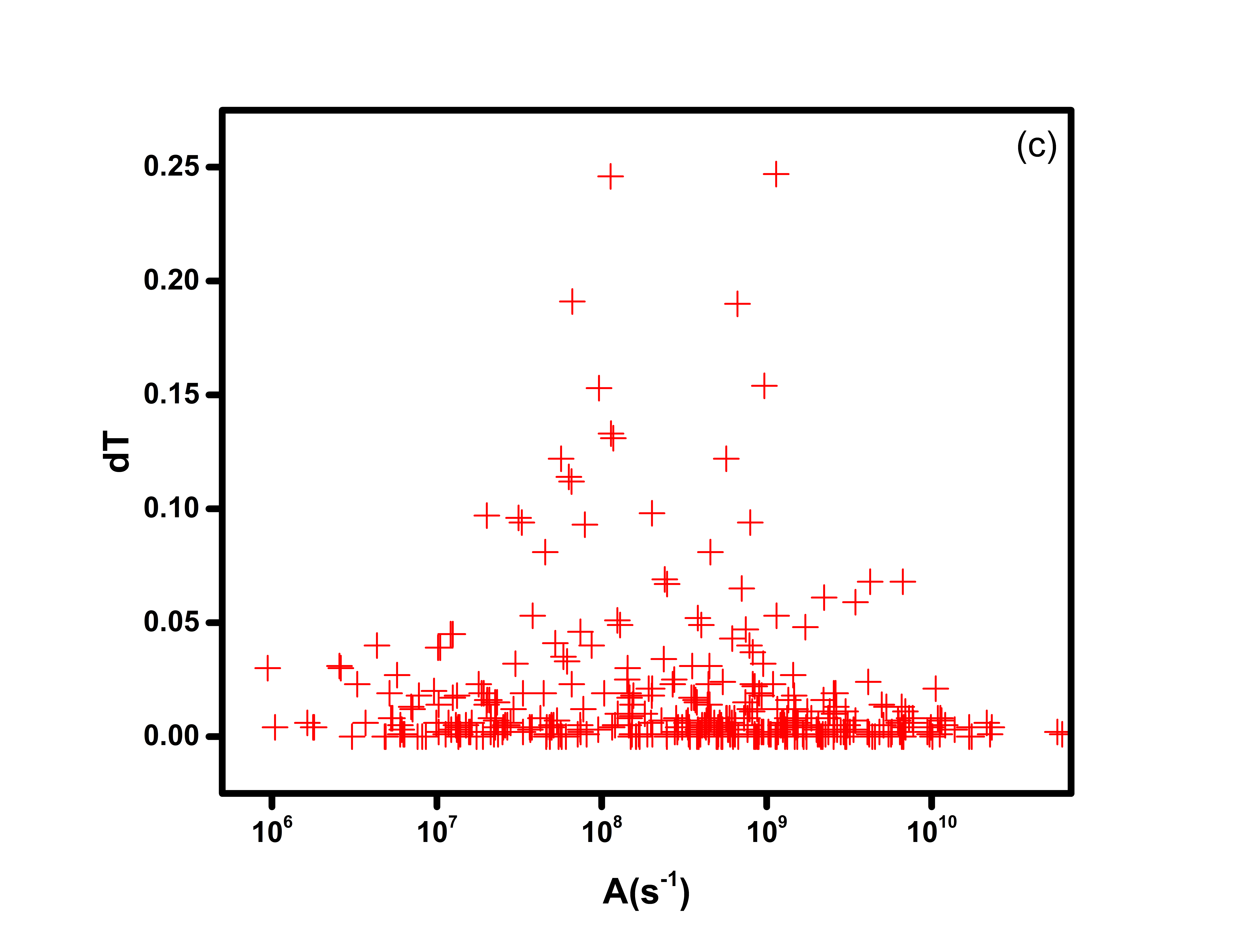} }}
    \quad
    \subfloat{{\includegraphics[width=8cm,height=6cm]{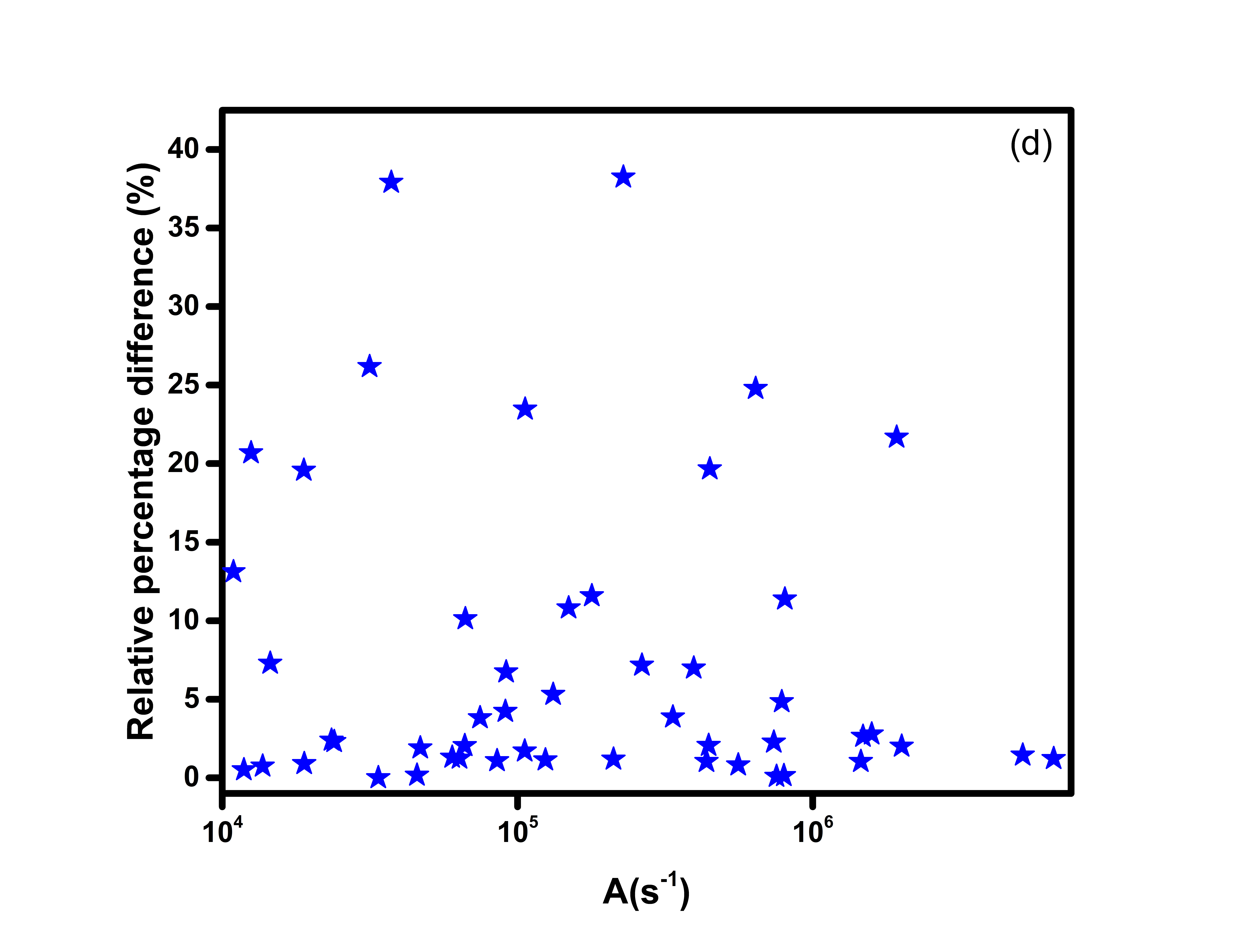}
    }}
    
    \caption{ The relative percentage deviation in (a) MBPT transition rates with respect to RCI values, (b) present RCI (circle), MBPT (triangle$\_$right), Liang et al. \cite{liang2009r} (triangle$\_$down), Fischer et al.\cite{fischer2006relativistic} (pentagon) transition rates with respect to NIST database \cite{NIST}. (c) The E1 dT parameter, (d) The relative percentage deviation in present RCI and Liang et al. \cite{liang2009r} values of E2 transition rates, for Ar$^{7+}$} 
    \label{fig:comp_nist_argon}
\end{figure}
\subsubsection{Ar$^{7+}$}
For Ar$^{7+}$, after neglecting the transition with $A$ $\leq 10^{4}$ s$^{-1}$, the radiative parameters of 377 and 250 lines are computed for E1 and E2 type transitions, respectively, and are presented in Tables S1-S2 (supplementary data). Since all the possible magnetic transitions among the states under consideration have $A$ smaller than 10$^{4}$ s$^{-1}$, these are not shown in Tables S1-S2. 
A detailed comparison of the present (RCI and MBPT) and previous results of Siegel et al. \cite{SIEGEL1998303}, Fischer et al.\cite{fischer2006relativistic}, Liang et al.\cite{liang2009r}, NIST \cite{NIST} and CAMBD \cite{CAMBD} for E1 lines  is shown in Table \ref{tab4}.
We observe that the present wavelengths are in excellent agreement with those of \cite{CAMBD} and show a mean deviation of only 0.16\%. Further, the gf values from the present work and Siegel et al. \cite{SIEGEL1998303} are also in excellent agreement.
Furthermore, in Fig.\ref{fig:comp_nist_argon}(a), we display the consistency between the transition rates calculated using the RCI and MBPT theories. The two theoretical results primarily lie within 20\%; however, more deviations are observed for a few weak lines with $A < 10^9$ s$^{-1}$.
For a more effective analysis, we demonstrate the comparison of the present transition rates and other previous results  \cite{fischer2006relativistic, liang2009r} with respect to the NIST values in Fig. \ref{fig:comp_nist_argon}(b). 
%
Our results are fairly consistent with the NIST values except for $3d ~^2D_{5/2}-4p ~^2p_{3/2}$ and $3d ~^2D_{3/2}-4p ~^2p_{1/2}$ lines. However, all the theoretical results match reasonably well for these transitions. 
%
%
The estimated uncertainty for the NIST A values lies within 18-25\%. In fact, for a few transitions, the accuracy range extends up to 50\%. Hence, our results are within the experimental uncertainties and can be considered reliable.
To further establish the accuracy of our calculations for E1 transitions, the dT parameter is shown in Fig. \ref{fig:comp_nist_argon} (c). Since its value lies below 0.05 for most cases, one can rely on the accuracy of the present results. 
Further, our calculated E2 transition probabilities are compared in Fig \ref{fig:comp_nist_argon} (d) with the corresponding values  available only for 83 lines from Liang et al \cite{liang2009r}. We display only 50 transitions here, for which the deviation is below 40\%. Except for a few transitions, the two theories show a reasonable agreement.  
Although Charro and Mart\'in \cite{charro2002relativistic} had also reported E2 transition probabilities, we found that their calculations deviate significantly from the present as well as previous \cite{liang2009r} results. Therefore, these results  \cite{charro2002relativistic} are not displayed in Fig \ref{fig:comp_nist_argon} (d).
Overall, most of the present radiative parameters are consistent with previous values in the literature. Furthermore, the radiative parameters of few allowed and forbidden lines of Ar$^{7+}$ are also reported for the first time.
%

\setlength\LTleft\fill          
\setlength\LTright\fill         
\begin{longtable}{|l|l|l|l|l|l|l|l|l|l|l|l|l|}
\caption{Comparison of the present E1 transition parameters with the values of Liang et al.\cite{liang2009r}, Fischer et al. \cite{fischer2006relativistic}, Siegel et al. \cite{SIEGEL1998303}, NIST \cite{NIST} and CAMBD atomic database\cite{CAMBD} for Ar$^{7+}$ .\label{tab4}} \\
\hline
 \multicolumn{2}{|c|}{\textbf{Levels  }} & \multicolumn{2}{c|}{\textbf{Wavelength (\AA)}}  & \multicolumn{5}{c|}{\textbf{A (s $^{-1}$)}}  & \multicolumn{2}{c|}{\textbf{gf}} \\
\hline \multicolumn{1}{|c|}{\textbf{Upper }} & \multicolumn{1}{c|}{\textbf{Lower }} & \multicolumn{1}{c|}{\textbf{RCI}}& \multicolumn{1}{c|}{\textbf{ \cite{CAMBD}}}& \multicolumn{1}{c|}{\textbf{RCI}}& \multicolumn{1}{c|}{\textbf{MBPT}}& \multicolumn{1}{c|}{\textbf{ \cite{NIST}}}& \multicolumn{1}{c|}{\textbf{ \cite{liang2009r} }}&
\multicolumn{1}{c|}{\textbf{ \cite{fischer2006relativistic}}}& \multicolumn{1}{c|}{\textbf{RCI}} & \multicolumn{1}{c|}{\textbf{ \cite{SIEGEL1998303} }}\\  \hline
\endfirsthead
\multicolumn{10}{c}%
{{\bfseries \tablename\ \thetable{} -- continued from previous page}} \\
\hline
 \multicolumn{2}{|c|}{\textbf{Levels  }} & \multicolumn{2}{c|}{\textbf{ Wavelength (\AA)}}  & \multicolumn{5}{c|}{\textbf{A (s $^{-1}$)}}  & \multicolumn{2}{c|}{\textbf{gf}} \\
\hline \multicolumn{1}{|c|}{\textbf{Upper }} & \multicolumn{1}{c|}{\textbf{Lower }} & \multicolumn{1}{c|}{\textbf{RCI}}& \multicolumn{1}{c|}{\textbf{ \cite{CAMBD}}}& \multicolumn{1}{c|}{\textbf{RCI}}& \multicolumn{1}{c|}{\textbf{MBPT}}& \multicolumn{1}{c|}{\textbf{ \cite{NIST}}}& \multicolumn{1}{c|}{\textbf{ \cite{liang2009r} }}&
\multicolumn{1}{c|}{\textbf{ \cite{fischer2006relativistic}}}& \multicolumn{1}{c|}{\textbf{RCI}} & \multicolumn{1}{c|}{\textbf{ \cite{SIEGEL1998303} }}\\  \hline 
\endhead
\hline \multicolumn{11}{|l|}{{Continued on next page}} \\ \hline
\endfoot
\hline \hline
\endlastfoot
$6p~^2P_{3/2}$&  $3s~^2S_{1/2}$&       106.55 &  106.66 &  4.08e+09 &  4.24e+09 &           &  3.29e+09 &           &   0.0277 &         \\
$6p~^2P_{1/2}$&  $3s~^2S_{1/2}$&       106.58 &         &  4.18e+09 &  4.35e+09 &           &  3.32e+09 &           &   0.0142 &         \\
$5p~^2P_{3/2}$&  $3s~^2S_{1/2}$&       119.96 &  120.09 &  6.76e+09 &  6.61e+09 &  6.10e+09 &  5.90e+09 &           &   0.0583 &  0.0568 \\
$5p~^2P_{1/2}$&  $3s~^2S_{1/2}$&       120.03 &  120.15 &  6.94e+09 &  6.80e+09 &  5.80e+09 &  6.01e+09 &           &   0.0300 &  0.0293 \\
$6d~^2D_{3/2}$&  $3p~^2P_{1/2}$&       122.45 &  122.62 &  6.24e+09 &  6.87e+09 &           &  5.60e+09 &           &   0.0561 &         \\
$6d~^2D_{5/2}$&  $3p~^2P_{3/2}$&       122.86 &  123.02 &  7.66e+09 &  8.30e+09 &           &  7.11e+09 &           &   0.1041 &         \\
$6d~^2D_{3/2}$&  $3p~^2P_{3/2}$&       122.86 &         &  1.26e+09 &  1.39e+09 &           &  1.18e+09 &           &   0.0114 &         \\
$6s~^2S_{1/2}$&  $3p~^2P_{1/2}$&       127.47 &  127.66 &  2.54e+09 &  2.74e+09 &           &  2.16e+09 &           &   0.0123 &         \\
$6s~^2S_{1/2}$&  $3p~^2P_{3/2}$&       127.91 &  128.09 &  5.10e+09 &  5.50e+09 &           &  4.52e+09 &           &   0.0250 &         \\
$5d~^2D_{3/2}$&  $3p~^2P_{1/2}$&       137.69 &  137.90 &  9.89e+09 &  1.02e+10 &           &  9.31e+09 &           &   0.1124 &  0.1118 \\
$5d~^2D_{5/2}$&  $3p~^2P_{3/2}$&       138.20 &  138.41 &  1.21e+10 &  1.23e+10 &           &  1.14e+10 &           &   0.2073 &  0.2052 \\
$5d~^2D_{3/2}$&  $3p~^2P_{3/2}$&       138.21 &         &  2.00e+09 &  2.06e+09 &           &  1.90e+09 &           &   0.0229 &  0.0228 \\
$5s~^2S_{1/2}$&  $3p~^2P_{1/2}$&       149.64 &  148.73 &  4.82e+09 &  4.87e+09 &           &  4.36e+09 &           &   0.0323 &  0.0322 \\
$5s~^2S_{1/2}$&  $3p~^2P_{3/2}$&       150.25 &  149.33 &  9.70e+09 &  9.81e+09 &           &  8.90e+09 &           &   0.0656 &  0.0654 \\
$4p~^2P_{3/2}$&  $3s~^2S_{1/2}$&       158.75 &  158.92 &  1.09e+10 &  1.02e+10 &  1.10e+10 &  9.74e+09 &  1.11e+10 &   0.1646 &  0.1595 \\
$6f~^2F_{5/2}$&  $3d~^2D_{3/2}$&       158.89 &         &  1.06e+10 &  1.17e+10 &           &  1.10e+10 &           &   0.2400 &         \\
$6f~^2F_{7/2}$&  $3d~^2D_{5/2}$&       158.92 &  159.05 &  1.15e+10 &  1.25e+10 &           &  1.18e+10 &           &   0.3473 &         \\
$6f~^2F_{5/2}$&  $3d~^2D_{5/2}$&       158.92 &         &  7.64e+08 &  8.35e+08 &           &  7.88e+08 &           &   0.0174 &         \\
$4p~^2P_{1/2}$&  $3s~^2S_{1/2}$&       159.00 &  159.18 &  1.13e+10 &  1.06e+10 &  1.11e+10 &  1.00e+10 &  1.14e+10 &   0.0854 &  0.0830 \\
$6p~^2P_{3/2}$&  $3d~^2D_{3/2}$&       165.18 &         &  1.43e+08 &  1.57e+08 &           &  1.32e+08 &           &   0.0023 &         \\
$6p~^2P_{3/2}$&  $3d~^2D_{5/2}$&       165.21 &         &  1.33e+09 &  1.41e+09 &           &  1.19e+09 &           &   0.0217 &         \\
$6p~^2P_{1/2}$&  $3d~^2D_{3/2}$&       165.26 &         &  1.44e+09 &  1.58e+09 &           &  1.32e+09 &           &   0.0118 &         \\
$4d~^2D_{3/2}$&  $3p~^2P_{1/2}$&       179.09 &  179.39 &  1.38e+10 &  1.35e+10 &           &  1.29e+10 &  1.40e+10 &   0.2650 &  0.2620 \\
$4d~^2D_{5/2}$&  $3p~^2P_{3/2}$&       179.94 &  180.25 &  1.69e+10 &  1.65e+10 &           &  1.58e+10 &  1.69e+10 &   0.4915 &  0.4848 \\
$4d~^2D_{3/2}$&  $3p~^2P_{3/2}$&       179.96 &         &  2.81e+09 &  2.75e+09 &           &  2.63e+09 &  2.82e+09 &   0.0546 &  0.0540 \\
$5f~^2F_{5/2}$&  $3d~^2D_{3/2}$&       184.09 &  184.26 &  2.16e+10 &  2.28e+10 &  2.20e+10 &  2.21e+10 &           &   0.6573 &         \\
$5f~^2F_{7/2}$&  $3d~^2D_{5/2}$&       184.12 &  184.30 &  2.31e+10 &  2.44e+10 &  2.40e+10 &  2.37e+10 &           &   0.9408 &         \\
$5f~^2F_{5/2}$&  $3d~^2D_{5/2}$&       184.13 &         &  1.54e+09 &  1.62e+09 &           &  1.58e+09 &           &   0.0470 &         \\
$5p~^2P_{3/2}$&  $3d~^2D_{3/2}$&       199.80 &  200.13 &  2.90e+08 &  2.93e+08 &           &  2.71e+08 &           &   0.0070 &  0.0069 \\
$5p~^2P_{3/2}$&  $3d~^2D_{5/2}$&       199.84 &  200.00 &  2.64e+09 &  2.64e+09 &           &  2.45e+09 &           &   0.0632 &  0.0625 \\
$5p~^2P_{1/2}$&  $3d~^2D_{3/2}$&       199.99 &  200.14 &  2.92e+09 &  2.95e+09 &           &  2.73e+09 &           &   0.0351 &  0.0350 \\
$4s~^2S_{1/2}$&  $3p~^2P_{1/2}$&       228.92 &  229.43 &  1.12e+10 &  1.10e+10 &  1.12e+10 &  1.06e+10 &  1.13e+10 &   0.1757 &  0.1752 \\
$4s~^2S_{1/2}$&  $3p~^2P_{3/2}$&       230.36 &  230.87 &  2.26e+10 &  2.22e+10 &  2.21e+10 &  2.15e+10 &  2.25e+10 &   0.3589 &  0.3601 \\
$4f~^2F_{5/2}$&  $3d~^2D_{3/2}$&       259.91 &  260.25 &  5.77e+10 &  5.92e+10 &  6.10e+10 &  5.83e+10 &  5.78e+10 &   3.5070 &         \\
$4f~^2F_{7/2}$&  $3d~^2D_{5/2}$&       259.96 &  260.33 &  6.18e+10 &  6.35e+10 &  6.50e+10 &  6.25e+10 &  6.20e+10 &   5.0120 &         \\
$4f~^2F_{5/2}$&  $3d~^2D_{5/2}$&       259.98 &         &  4.12e+09 &  4.23e+09 &           &  4.16e+09 &  4.13e+09 &   0.2506 &         \\
$6p~^2P_{3/2}$&  $4s~^2S_{1/2}$&       276.44 &         &  1.24e+09 &  1.28e+09 &           &  1.11e+09 &           &   0.0567 &         \\
$6p~^2P_{1/2}$&  $4s~^2S_{1/2}$&       276.65 &         &  1.28e+09 &  1.32e+09 &           &  1.13e+09 &           &   0.0293 &         \\
$6d~^2D_{3/2}$&  $4p~^2P_{1/2}$&       305.17 &         &  1.50e+09 &  1.60e+09 &           &  1.39e+09 &           &   0.0838 &         \\
$6d~^2D_{5/2}$&  $4p~^2P_{3/2}$&       306.10 &         &  1.86e+09 &  1.96e+09 &           &  1.76e+09 &           &   0.1569 &         \\
$6d~^2D_{3/2}$&  $4p~^2P_{3/2}$&       306.12 &         &  3.07e+08 &  3.28e+08 &           &  2.93e+08 &           &   0.0173 &         \\
$4p~^2P_{3/2}$&  $3d~^2D_{3/2}$&       336.92 &         &  7.67e+08 &  7.54e+08 &           &  7.40e+08 &  7.67e+08 &   0.0522 &  0.0518 \\
$4p~^2P_{3/2}$&  $3d~^2D_{5/2}$&       337.05 &  337.26 &  6.91e+09 &  6.79e+09 &  1.00e+10 &  6.66e+09 &  6.91e+09 &   0.4710 &  0.4666 \\
$4p~^2P_{1/2}$&  $3d~^2D_{3/2}$&       338.07 &  338.22 &  7.73e+09 &  7.61e+09 &  1.10e+10 &  7.46e+09 &  7.73e+09 &   0.2649 &  0.2630 \\
$6s~^2S_{1/2}$&  $4p~^2P_{1/2}$&       338.37 &         &  1.54e+09 &  1.60e+09 &           &  1.46e+09 &           &   0.0529 &         \\
$6s~^2S_{1/2}$&  $4p~^2P_{3/2}$&       339.53 &         &  3.10e+09 &  3.22e+09 &           &  2.97e+09 &           &   0.1071 &         \\
$6f~^2F_{5/2}$&  $4d~^2D_{3/2}$&       378.59 &         &  5.31e+09 &  5.50e+09 &           &  5.34e+09 &           &   0.6845 &         \\
$6f~^2F_{7/2}$&  $4d~^2D_{5/2}$&       378.66 &         &  5.74e+09 &  5.90e+09 &           &  5.73e+09 &           &   0.9864 &         \\
$6f~^2F_{5/2}$&  $4d~^2D_{5/2}$&       378.67 &         &  3.82e+08 &  3.93e+08 &           &  3.82e+08 &           &   0.0493 &         \\
$5p~^2P_{3/2}$&  $4s~^2S_{1/2}$&       389.31 &         &  1.67e+09 &  1.64e+09 &           &  1.58e+09 &           &   0.1520 &  0.1496 \\
$5p~^2P_{1/2}$&  $4s~^2S_{1/2}$&       390.04 &         &  1.74e+09 &  1.71e+09 &           &  1.63e+09 &           &   0.0794 &  0.0783 \\
$6g~^2G_{7/2}$&  $4f~^2F_{5/2}$&       408.05 &         &  5.47e+09 &  5.44e+09 &           &  5.49e+09 &           &   1.0930 &         \\
$6g~^2G_{9/2}$&  $4f~^2F_{7/2}$&       408.09 &         &  5.67e+09 &  5.64e+09 &           &  5.69e+09 &           &   1.4160 &         \\
$6g~^2G_{7/2}$&  $4f~^2F_{7/2}$&       408.10 &         &  2.02e+08 &  2.01e+08 &           &  2.03e+08 &           &   0.0405 &         \\
$6p~^2P_{3/2}$&  $4d~^2D_{3/2}$&       416.36 &         &  1.46e+08 &  1.52e+08 &           &  1.43e+08 &           &   0.0151 &         \\
$6p~^2P_{3/2}$&  $4d~^2D_{5/2}$&       416.46 &         &  1.32e+09 &  1.37e+09 &           &  1.29e+09 &           &   0.1378 &         \\
$6p~^2P_{1/2}$&  $4d~^2D_{3/2}$&       416.83 &         &  1.46e+09 &  1.53e+09 &           &  1.42e+09 &           &   0.0763 &         \\
$6d~^2D_{5/2}$&  $4f~^2F_{5/2}$&       418.86 &         &  9.62e+06 &  1.07e+07 &           &  9.19e+06 &           &   0.0015 &         \\
$6d~^2D_{3/2}$&  $4f~^2F_{5/2}$&       418.89 &         &  2.01e+08 &  2.26e+08 &           &  1.93e+08 &           &   0.0211 &         \\
$6d~^2D_{5/2}$&  $4f~^2F_{7/2}$&       418.92 &         &  1.93e+08 &  2.15e+08 &           &  1.84e+08 &           &   0.0304 &         \\
$5d~^2D_{3/2}$&  $4p~^2P_{1/2}$&       421.42 &         &  1.65e+09 &  1.67e+09 &           &  1.58e+09 &           &   0.1760 &  0.1782 \\
$5d~^2D_{5/2}$&  $4p~^2P_{3/2}$&       423.17 &         &  2.05e+09 &  2.07e+09 &           &  1.96e+09 &           &   0.3308 &  0.3335 \\
$5d~^2D_{3/2}$&  $4p~^2P_{3/2}$&       423.22 &         &  3.42e+08 &  3.46e+08 &           &  3.29e+08 &           &   0.0367 &  0.0372 \\
$3d~^2D_{3/2}$&  $3p~^2P_{1/2}$&       517.65 &  519.43 &  6.30e+09 &  6.50e+09 &  6.30e+09 &  6.60e+09 &  6.36e+09 &   1.0130 &  1.0194 \\
$3d~^2D_{5/2}$&  $3p~^2P_{3/2}$&       524.72 &  526.46 &  7.28e+09 &  7.52e+09 &  7.20e+09 &  7.68e+09 &  7.35e+09 &   1.8030 &  1.8156 \\
$3d~^2D_{3/2}$&  $3p~^2P_{3/2}$&       525.03 &  526.87 &  1.21e+09 &  1.25e+09 &  1.20e+09 &  1.27e+09 &  1.22e+09 &   0.2002 &  0.2015 \\
$5s~^2S_{1/2}$&  $4p~^2P_{1/2}$&       557.72 &         &  3.19e+09 &  3.19e+09 &           &  3.13e+09 &           &   0.2973 &  0.2990 \\
$5s~^2S_{1/2}$&  $4p~^2P_{3/2}$&       560.88 &         &  6.43e+09 &  6.42e+09 &           &  6.31e+09 &           &   0.6062 &  0.6100 \\
$5f~^2F_{5/2}$&  $4d~^2D_{3/2}$&       561.79 &  562.46 &  9.41e+09 &  9.51e+09 &           &  9.44e+09 &           &   2.6710 &  2.6868 \\
$5f~^2F_{7/2}$&  $4d~^2D_{5/2}$&       561.92 &  562.61 &  1.01e+10 &  1.02e+10 &           &  1.01e+10 &           &   3.8200 &  3.8424 \\
$5f~^2F_{5/2}$&  $4d~^2D_{5/2}$&       561.96 &         &  6.73e+08 &  6.80e+08 &           &  6.75e+08 &           &   0.1911 &  0.1922 \\
$5g~^2G_{7/2}$&  $4f~^2F_{5/2}$&       628.17 &         &  1.69e+10 &  1.68e+10 &           &  1.69e+10 &           &   7.9800 &         \\
$5g~^2G_{9/2}$&  $4f~^2F_{7/2}$&       628.25 &         &  1.75e+10 &  1.74e+10 &           &  1.75e+10 &           &  10.3500 &         \\
$5g~^2G_{7/2}$&  $4f~^2F_{7/2}$&       628.29 &         &  6.24e+08 &  6.21e+08 &           &  6.25e+08 &           &   0.2956 &         \\
$5d~^2D_{5/2}$&  $4f~^2F_{5/2}$&       674.01 &         &  2.29e+07 &  2.26e+07 &           &  2.21e+07 &           &   0.0093 &         \\
$5d~^2D_{3/2}$&  $4f~^2F_{5/2}$&       674.14 &         &  4.80e+08 &  4.76e+08 &           &  4.65e+08 &           &   0.1309 &         \\
$5d~^2D_{5/2}$&  $4f~^2F_{7/2}$&       674.14 &         &  4.58e+08 &  4.52e+08 &           &  4.43e+08 &           &   0.1871 &         \\
$3p~^2P_{3/2}$&  $3s~^2S_{1/2}$&       700.94 &  700.24 &  2.56e+09 &  2.70e+09 &  2.55e+09 &  2.70e+09 &  2.58e+09 &   0.7533 &  0.7622 \\
$3p~^2P_{1/2}$&  $3s~^2S_{1/2}$&       714.53 &  713.81 &  2.41e+09 &  2.55e+09 &  2.40e+09 &  2.56e+09 &  2.43e+09 &   0.3685 &  0.3728 \\
$5p~^2P_{3/2}$&  $4d~^2D_{3/2}$&       739.08 &         &  3.33e+08 &  3.32e+08 &           &  3.29e+08 &           &   0.1091 &  0.1088 \\
$5p~^2P_{3/2}$&  $4d~^2D_{5/2}$&       739.39 &  739.80 &  3.00e+09 &  2.99e+09 &           &  2.96e+09 &           &   0.9845 &  0.9816 \\
$5p~^2P_{1/2}$&  $4d~^2D_{3/2}$&       741.74 &  742.50 &  3.36e+09 &  3.35e+09 &           &  3.31e+09 &           &   0.5536 &  0.5528 \\
$6p~^2P_{3/2}$&  $5s~^2S_{1/2}$&       767.54 &         &  4.28e+08 &  4.28e+08 &           &  4.03e+08 &           &   0.1511 &         \\
$6p~^2P_{1/2}$&  $5s~^2S_{1/2}$&       769.13 &         &  4.47e+08 &  4.47e+08 &           &  4.17e+08 &           &   0.0793 &         \\
$6d~^2D_{3/2}$&  $5p~^2P_{1/2}$&       810.03 &         &  3.42e+08 &  3.59e+08 &           &  3.25e+08 &           &   0.1346 &         \\
$6d~^2D_{5/2}$&  $5p~^2P_{3/2}$&       813.12 &         &  4.32e+08 &  4.48e+08 &           &  4.14e+08 &           &   0.2569 &         \\
$6d~^2D_{3/2}$&  $5p~^2P_{3/2}$&       813.22 &         &  7.15e+07 &  7.50e+07 &           &  6.93e+07 &           &   0.0284 &         \\
$6f~^2F_{5/2}$&  $5d~^2D_{3/2}$&      1038.89 &         &  2.36e+09 &  2.40e+09 &           &  2.38e+09 &           &   2.2910 &         \\
$6f~^2F_{7/2}$&  $5d~^2D_{5/2}$&      1039.11 &         &  2.54e+09 &  2.57e+09 &           &  2.55e+09 &           &   3.2910 &         \\
$6f~^2F_{5/2}$&  $5d~^2D_{5/2}$&      1039.20 &         &  1.70e+08 &  1.72e+08 &           &  1.70e+08 &           &   0.1646 &         \\
$6s~^2S_{1/2}$&  $5p~^2P_{1/2}$&      1095.28 &         &  1.17e+09 &  1.18e+09 &           &  1.15e+09 &           &   0.4217 &         \\
$6s~^2S_{1/2}$&  $5p~^2P_{3/2}$&      1101.12 &         &  2.36e+09 &  2.37e+09 &           &  2.33e+09 &           &   0.8588 &         \\
$6g~^2G_{7/2}$&  $5f~^2F_{5/2}$&      1154.62 &         &  4.37e+09 &  4.36e+09 &           &  4.37e+09 &           &   6.9830 &         \\
$6g~^2G_{9/2}$&  $5f~^2F_{7/2}$&      1154.75 &         &  4.53e+09 &  4.52e+09 &           &  4.53e+09 &           &   9.0540 &         \\
$6g~^2G_{7/2}$&  $5f~^2F_{7/2}$&      1154.82 &         &  1.62e+08 &  1.61e+08 &           &  1.62e+08 &           &   0.2587 &         \\
$6h~^2H_{9/2}$&  $5g~^2G_{7/2}$&      1163.76 &         &  6.59e+09 &  6.49e+09 &           &  6.59e+09 &           &  13.3800 &         \\
$6h~^2H_{11/2}$&  $5g~^2G_{9/2}$&      1163.84 &         &  6.74e+09 &  6.64e+09 &           &  6.74e+09 &           &  16.4200 &         \\
$6h~^2H_{9/2}$&  $5g~^2G_{9/2}$&      1163.89 &         &  1.50e+08 &  1.48e+08 &           &  1.50e+08 &           &   0.3041 &         \\
$6f~^2F_{5/2}$&  $5g~^2G_{7/2}$&      1170.93 &         &  5.10e+07 &  4.88e+07 &           &  4.96e+07 &           &   0.0629 &         \\
$6f~^2F_{7/2}$&  $5g~^2G_{9/2}$&      1170.95 &         &  4.95e+07 &  4.74e+07 &           &  4.82e+07 &           &   0.0814 &         \\
$6d~^2D_{5/2}$&  $5f~^2F_{5/2}$&      1245.59 &         &  1.63e+07 &  1.62e+07 &           &  1.58e+07 &           &   0.0227 &         \\
$6d~^2D_{5/2}$&  $5f~^2F_{7/2}$&      1245.82 &         &  3.26e+08 &  3.24e+08 &           &  3.17e+08 &           &   0.4549 &         \\
$6d~^2D_{3/2}$&  $5f~^2F_{5/2}$&      1245.82 &         &  3.41e+08 &  3.41e+08 &           &  3.33e+08 &           &   0.3171 &         \\
$6p~^2P_{3/2}$&  $5d~^2D_{3/2}$&      1383.23 &         &  1.47e+08 &  1.48e+08 &           &  1.46e+08 &           &   0.1689 &         \\
$6p~^2P_{3/2}$&  $5d~^2D_{5/2}$&      1383.79 &         &  1.33e+09 &  1.33e+09 &           &  1.32e+09 &           &   1.5260 &         \\
$6p~^2P_{1/2}$&  $5d~^2D_{3/2}$&      1388.40 &         &  1.48e+09 &  1.49e+09 &           &  1.47e+09 &           &   0.8570 &         \\
$4d~^2D_{3/2}$&  $4p~^2P_{1/2}$&      1440.66 &         &  1.29e+09 &  1.27e+09 &  1.50e+09 &  1.30e+09 &  1.31e+09 &   1.6080 &  1.6038 \\
$4d~^2D_{5/2}$&  $4p~^2P_{3/2}$&      1460.76 &         &  1.49e+09 &  1.46e+09 &  1.70e+09 &  1.51e+09 &  1.51e+09 &   2.8620 &  2.8564 \\
$4d~^2D_{3/2}$&  $4p~^2P_{3/2}$&      1461.94 &         &  2.48e+08 &  2.43e+08 &           &  2.51e+08 &  2.50e+08 &   0.3179 &  0.3170 \\
$4p~^2P_{3/2}$&  $4s~^2S_{1/2}$&      1881.09 &         &  5.38e+08 &  5.54e+08 &           &  5.48e+08 &  5.48e+08 &   1.1430 &  1.1510 \\
$4p~^2P_{1/2}$&  $4s~^2S_{1/2}$&      1917.54 &         &  5.08e+08 &  5.24e+08 &           &  5.20e+08 &  5.17e+08 &   0.5600 &  0.5642 \\
$5d~^2D_{3/2}$&  $5p~^2P_{1/2}$&      3024.45 &         &  3.86e+08 &  3.72e+08 &           &  3.86e+08 &           &   2.1150 &  2.1040 \\
$5d~^2D_{5/2}$&  $5p~^2P_{3/2}$&      3066.67 &         &  4.45e+08 &  4.30e+08 &           &  4.48e+08 &           &   3.7630 &  3.8556 \\
$5d~^2D_{3/2}$&  $5p~^2P_{3/2}$&      3069.39 &         &  7.40e+07 &  7.16e+07 &           &  7.43e+07 &           &   0.4180 &  0.4160 \\
$5p~^2P_{3/2}$&  $5s~^2S_{1/2}$&      3935.02 &         &  1.61e+08 &  1.65e+08 &           &  1.63e+08 &           &   1.4960 &  1.5044 \\
$5p~^2P_{1/2}$&  $5s~^2S_{1/2}$&      4011.45 &         &  1.52e+08 &  1.56e+08 &           &  1.55e+08 &           &   0.7339 &  0.7380 \\
$4f~^2F_{5/2}$&  $4d~^2D_{3/2}$&      5116.30 &         &  2.50e+07 &  2.59e+07 &           &  2.16e+07 &           &   0.5894 &         \\
$4f~^2F_{7/2}$&  $4d~^2D_{5/2}$&      5123.02 &         &  2.67e+07 &  2.77e+07 &           &  2.29e+07 &           &   0.8408 &         \\
$6d~^2D_{3/2}$&  $6p~^2P_{1/2}$&      5445.62 &         &  1.46e+08 &  1.39e+08 &           &  1.46e+08 &           &   2.5910 &         \\
$6d~^2D_{5/2}$&  $6p~^2P_{3/2}$&      5521.98 &         &  1.68e+08 &  1.61e+08 &           &  1.69e+08 &           &   4.6060 &         \\
$6d~^2D_{3/2}$&  $6p~^2P_{3/2}$&      5526.63 &         &  2.80e+07 &  2.68e+07 &           &  2.80e+07 &           &   0.5122 &         \\
$6p~^2P_{3/2}$&  $6s~^2S_{1/2}$&      7113.62 &         &  6.06e+07 &  6.23e+07 &           &  6.13e+07 &           &   1.8370 &         \\
$6p~^2P_{1/2}$&  $6s~^2S_{1/2}$&      7252.49 &         &  5.72e+07 &  5.90e+07 &           &  5.83e+07 &           &   0.9015 &         \\
$5f~^2F_{5/2}$&  $5d~^2D_{3/2}$&      9880.92 &         &  1.22e+07 &  1.21e+07 &           &  1.08e+07 &           &   1.0730 &         \\
$5f~^2F_{7/2}$&  $5d~^2D_{5/2}$&      9894.52 &         &  1.30e+07 &  1.29e+07 &           &  1.14e+07 &           &   1.5310 &         \\
$6f~^2F_{5/2}$&  $6d~^2D_{3/2}$&     17042.45 &         &  5.75e+06 &  5.63e+06 &           &  5.12e+06 &           &   1.5010 &         \\
$6f~^2F_{7/2}$&  $6d~^2D_{5/2}$&     17061.76 &         &  6.12e+06 &  6.00e+06 &           &  5.42e+06 &           &   2.1360 &         \\
\end{longtable}
%
%
%
%
%
%
%
\subsubsection{Kr$^{25+}$}
For Kr$^{25+}$, the number of E1, M1, E2 and M2 transitions among the presently considered levels are 481, 36, 524 and 102, respectively. The radiative parameters for these lines are given in Tables S3-S4 of the supplementary data.

Table \ref{tab5}, shows a thorough analysis of our calculated and previously reported E1 transition parameters \cite{liang2009r, SAMPSON1990209,NIST,CAMBD}.
%
The present wavelengths agree extremely well with results of 26 and 41 lines available at the NIST \cite{NIST} and CAMBD\cite{CAMBD} databases with an average deviation of  0.049\% and 0.045\%, respectively. 
%
%
Our weighted oscillator strengths have a mean deviation of only 2.71\% relative to the theoretical results of Sampson et al. \cite{SAMPSON1990209}, with only six transitions exhibiting a variation above 5\%. Liang et al. \cite{liang2009r} also reported gf values and mentioned that their results agree within 20\% with \cite{SAMPSON1990209}. Since our results also show a similar comparison with these R-matrix calculations \cite{liang2009r} for gf, the latter are not included in Table \ref{tab5}.   
%

\begin{figure}[H]
    \centering
    \subfloat
    {{\includegraphics[width=5cm,height=4.5cm]{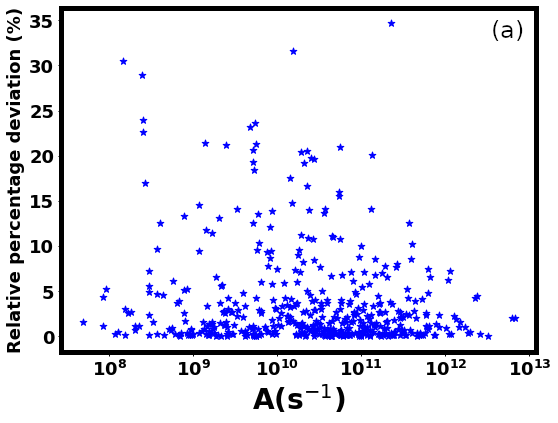} }}
    \quad
    \subfloat{{\includegraphics[width=5cm,height=4.5cm]{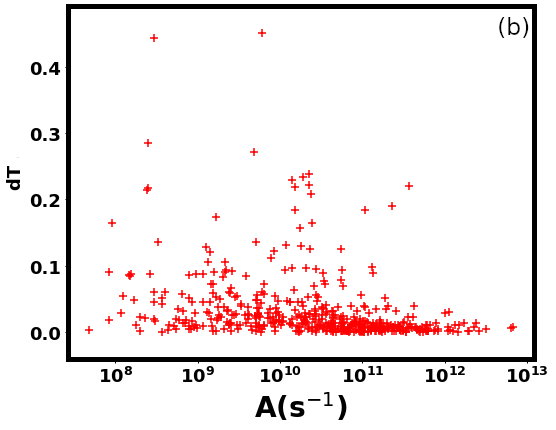} }}
    \quad
    \subfloat{{\includegraphics[width=4.9cm,height=4.5cm]{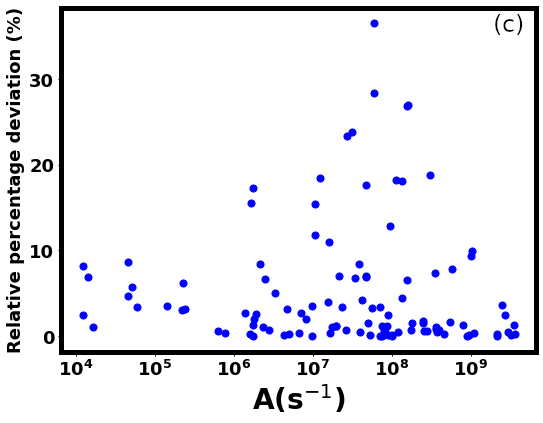} }}

    \caption{(a) The relative absolute percentage deviation in the present MBPT values with respect to RCI results, (b) The E1 dT parameter, (c) The relative absolute percentage deviation in the present RCI and Liang et al. \cite{liang2009r} values of E2 transition rates, for  Kr$^{25+}$ }
    \label{fig:kr_trans}
\end{figure}

The present A values for E1 transitions lie primarily within 10\% of those reported by Liang et al. \cite{liang2009r} with a few exceptions.
However, significant discrepancies are encountered in comparing the transition rates of Younis et al. \cite{younis2005energy} with both present, and the R-matrix \cite{liang2009r} values. Therefore, we do not report a full comparison with \cite{younis2005energy}. A good agreement of our results with \cite{liang2009r} supports the accuracy of our calculations.
%
%
Further, to clarify the ambiguity, we calculated the E1 transition probability by MBPT method using FAC and found a good agreement with the RCI values as displayed in Fig.\ref{fig:kr_trans} (a).
Moreover, in Fig \ref{fig:kr_trans} (b), we evaluated the uncertainty in the E1 A values from length (Coulomb) and velocity (Babushkin) gauges.
It is evident that a profusion of transitions has low dT values (less than 0.1); hence, brace the accuracy of our calculations.
%
%
%
%
%
%
%
%
%
%
\setlength{\tabcolsep}{1pt}
\begin{longtable}{|l|l|l|l|l|l|l|l|l|l|l|l|}
\caption{Comparison of the present E1 transition parameters with the values of Liang et al.\cite{liang2009r}, Sampson et al. \cite{SAMPSON1990209}, NIST \cite{NIST} and CAMBD \cite{CAMBD} atomic database for Kr$^{25+}$ .\label{tab5}} \\
\hline
 \multicolumn{2}{|c|}{\textbf{Levels  }} & \multicolumn{3}{c|}{\textbf{Wavelength (\AA)}}  & \multicolumn{3}{c|}{\textbf{A (s $^{-1}$)}}  & \multicolumn{2}{c|}{\textbf{gf}} \\
\hline \multicolumn{1}{|c|}{\textbf{Upper }} & \multicolumn{1}{c|}{\textbf{Lower }} & \multicolumn{1}{c|}{\textbf{RCI}}& \multicolumn{1}{c|}{\textbf{\cite{NIST}}}& \multicolumn{1}{c|}{\textbf{\cite{CAMBD}}}& \multicolumn{1}{c|}{\textbf{RCI}}& \multicolumn{1}{c|}{\textbf{MBPT}}& \multicolumn{1}{c|}{\textbf{\cite{liang2009r} }}&
\multicolumn{1}{c|}{\textbf{RCI}}& \multicolumn{1}{c|}{\textbf{ \cite{SAMPSON1990209} }}\\  \hline
\endfirsthead
\multicolumn{10}{c}%
{{\bfseries \tablename\ \thetable{} -- continued from previous page}} \\
\hline
 \multicolumn{2}{|c|}{\textbf{Levels  }} & \multicolumn{3}{c|}{\textbf{Wavelength (\AA)}}  & \multicolumn{3}{c|}{\textbf{A (s $^{-1}$)}}  & \multicolumn{2}{c|}{\textbf{gf}} \\
\hline \multicolumn{1}{|c|}{\textbf{Upper }} & \multicolumn{1}{c|}{\textbf{Lower }} & \multicolumn{1}{c|}{\textbf{RCI}}& \multicolumn{1}{c|}{\textbf{\cite{NIST}}}& \multicolumn{1}{c|}{\textbf{\cite{CAMBD}}}&  \multicolumn{1}{c|}{\textbf{RCI}}& \multicolumn{1}{c|}{\textbf{MBPT}}&  \multicolumn{1}{c|}{\textbf{ \cite{liang2009r} }}&
\multicolumn{1}{c|}{\textbf{RCI}}& \multicolumn{1}{c|}{\textbf{\cite{SAMPSON1990209}  }}\\ \hline 
\endhead
\hline \multicolumn{10}{|l|}{{Continued on next page}} \\ \hline
\endfoot
\hline \hline
\endlastfoot
$5p~^2P_{3/2}$&  $3s~^2S_{1/2}$&      15.22 &   15.21 &   15.21 &  7.74e+11 &  7.64e+11 &  6.94e+11 &   0.1075 &  0.1076 \\
$5d~^2D_{3/2}$&  $3p~^2P_{1/2}$&      16.06 &   16.07 &   16.06 &  1.43e+12 &  1.45e+12 &  1.38e+12 &   0.2216 &  0.2222 \\
$5d~^2D_{5/2}$&  $3p~^2P_{3/2}$&      16.32 &   16.34 &   16.32 &  1.72e+12 &  1.74e+12 &  1.70e+12 &   0.4124 &  0.4152 \\
$5f~^2F_{5/2}$&  $3d~^2D_{3/2}$&      17.93 &   17.94 &   17.93 &  2.23e+12 &  2.33e+12 &  2.31e+12 &   0.6450 &  0.6700 \\
$5f~^2F_{7/2}$&  $3d~^2D_{5/2}$&      17.99 &   17.99 &   17.99 &  2.37e+12 &  2.47e+12 &  2.47e+12 &   0.9188 &  0.9576 \\
$4p~^2P_{3/2}$&  $3s~^2S_{1/2}$&      21.19 &   21.19 &   21.18 &  1.34e+12 &  1.32e+12 &  1.30e+12 &   0.3619 &  0.3640 \\
$4p~^2P_{1/2}$&  $3s~^2S_{1/2}$&      21.38 &   21.37 &   21.35 &  1.48e+12 &  1.46e+12 &  1.42e+12 &   0.2034 &  0.2054 \\
$4d~^2D_{3/2}$&  $3p~^2P_{1/2}$&      22.27 &   22.26 &   22.25 &  2.60e+12 &  2.59e+12 &  2.59e+12 &   0.7726 &  0.7618 \\
$4d~^2D_{5/2}$&  $3p~^2P_{3/2}$&      22.75 &   22.74 &   22.73 &  3.17e+12 &  3.17e+12 &  3.16e+12 &   1.4760 &  1.4612 \\
$4s~^2S_{1/2}$&  $3p~^2P_{1/2}$&      24.77 &   24.77 &   24.76 &  5.83e+11 &  5.70e+11 &  5.36e+11 &   0.1073 &  0.1000 \\
$4s~^2S_{1/2}$&  $3p~^2P_{3/2}$&      25.43 &   25.42 &   25.41 &  1.26e+12 &  1.23e+12 &  1.18e+12 &   0.2437 &  0.2292 \\
$4f~^2F_{5/2}$&  $3d~^2D_{3/2}$&      25.62 &   25.62 &   25.61 &  6.28e+12 &  6.40e+12 &  6.41e+12 &   3.7090 &  3.7092 \\
$4f~^2F_{7/2}$&  $3d~^2D_{5/2}$&      25.73 &   25.73 &   25.71 &  6.70e+12 &  6.83e+12 &  6.84e+12 &   5.3170 &  5.3316 \\
$5p~^2P_{3/2}$&  $4s~^2S_{1/2}$&      48.11 &   48.11 &   48.11 &  2.76e+11 &  2.74e+11 &  2.67e+11 &   0.3830 &         \\
$5p~^2P_{1/2}$&  $4s~^2S_{1/2}$&      48.59 &   48.59 &   48.59 &  3.08e+11 &  3.06e+11 &  2.91e+11 &   0.2180 &         \\
$5d~^2D_{3/2}$&  $4p~^2P_{1/2}$&      49.93 &   49.93 &   49.92 &  4.63e+11 &  4.61e+11 &  4.52e+11 &   0.6919 &         \\
$5d~^2D_{5/2}$&  $4p~^2P_{3/2}$&      50.86 &   50.86 &   50.85 &  5.78e+11 &  5.76e+11 &  5.69e+11 &   1.3440 &         \\
$5f~^2F_{5/2}$&  $4d~^2D_{3/2}$&      55.70 &   55.71 &   55.71 &  1.09e+12 &  1.09e+12 &  1.09e+12 &   3.0410 &         \\
$5f~^2F_{7/2}$&  $4d~^2D_{5/2}$&      55.92 &   55.93 &   55.93 &  1.17e+12 &  1.17e+12 &  1.17e+12 &   4.3800 &         \\
$5g~^2G_{7/2}$&  $4f~^2F_{5/2}$&      59.42 &   59.38 &   59.39 &  1.88e+12 &  1.89e+12 &  1.89e+12 &   7.9670 &         \\
$5g~^2G_{9/2}$&  $4f~^2F_{7/2}$&      59.51 &   59.46 &   59.45 &  1.95e+12 &  1.95e+12 &  1.96e+12 &  10.3300 &         \\
$3d~^2D_{3/2}$&  $3p~^2P_{1/2}$&     141.06 &  140.89 &  140.87 &  3.30e+10 &  3.40e+10 &  3.40e+10 &   0.3939 &  0.4030 \\
$3d~^2D_{5/2}$&  $3p~^2P_{3/2}$&     160.13 &  159.92 &  159.87 &  2.72e+10 &  2.80e+10 &  2.90e+10 &   0.6280 &  0.6432 \\
$3d~^2D_{3/2}$&  $3p~^2P_{3/2}$&     165.54 &  165.16 &  165.12 &  4.08e+09 &  4.22e+09 &  4.35e+09 &   0.0671 &  0.0688 \\
$3p~^2P_{3/2}$&  $3s~^2S_{1/2}$&     178.90 &  178.99 &  178.97 &  2.17e+10 &  2.23e+10 &  2.13e+10 &   0.4169 &  0.4266 \\
$3p~^2P_{1/2}$&  $3s~^2S_{1/2}$&     220.20 &  220.06 &  220.03 &  1.15e+10 &  1.18e+10 &  1.15e+10 &   0.1667 &  0.1706 \\
$6p~^2P_{3/2}$&  $3s~^2S_{1/2}$&      13.24 &          &   13.23 &  4.60e+11 &  4.60e+11 &  3.17e+11 &   0.0483 &         \\
$6p~^2P_{1/2}$&  $3s~^2S_{1/2}$&      13.26 &          &   13.25 &  4.92e+11 &  4.94e+11 &  3.19e+11 &   0.0259 &         \\
$6d~^2D_{3/2}$&  $3p~^2P_{1/2}$&      13.96 &          &   13.96 &  8.32e+11 &  8.51e+11 &  6.91e+11 &   0.0972 &         \\
$6d~^2D_{5/2}$&  $3p~^2P_{3/2}$&      14.16 &          &   14.16 &  1.01e+12 &  1.02e+12 &  9.29e+11 &   0.1815 &         \\
$5p~^2P_{1/2}$&  $3s~^2S_{1/2}$&      15.27 &          &   15.26 &  8.35e+11 &  8.27e+11 &  7.23e+11 &   0.0584 &  0.0586 \\
$6f~^2F_{5/2}$&  $3d~^2D_{3/2}$&      15.41 &          &   15.41 &  1.08e+12 &  1.15e+12 &  1.10e+12 &   0.2310 &         \\
$6f~^2F_{7/2}$&  $3d~^2D_{5/2}$&      15.46 &          &   15.46 &  1.14e+12 &  1.22e+12 &  1.19e+12 &   0.3260 &         \\
$5s~^2S_{1/2}$&  $3p~^2P_{1/2}$&      16.66 &          &   16.65 &  2.74e+11 &  2.67e+11 &  2.11e+11 &   0.0228 &  0.0212 \\
$5s~^2S_{1/2}$&  $3p~^2P_{3/2}$&      16.96 &          &   16.95 &  5.84e+11 &  5.70e+11 &  5.05e+11 &   0.0504 &  0.0472 \\
$4d~^2D_{3/2}$&  $3p~^2P_{3/2}$&      22.80 &          &   22.79 &  5.35e+11 &  5.35e+11 &  5.31e+11 &   0.1668 &  0.1652 \\
$4p~^2P_{3/2}$&  $3d~^2D_{5/2}$&      28.28 &          &   28.28 &  3.55e+11 &  3.40e+11 &  3.40e+11 &   0.1702 &  0.1632 \\
$4p~^2P_{1/2}$&  $3d~^2D_{3/2}$&      28.45 &          &   28.45 &  4.29e+11 &  4.12e+11 &  4.10e+11 &   0.1040 &  0.1000 \\
$6f~^2F_{5/2}$&  $4d~^2D_{3/2}$&      36.96 &          &   36.96 &  5.86e+11 &  5.93e+11 &  5.76e+11 &   0.7204 &         \\
$6f~^2F_{7/2}$&  $4d~^2D_{5/2}$&      37.07 &          &   37.06 &  6.28e+11 &  6.33e+11 &  6.21e+11 &   1.0340 &         \\
$5s~^2S_{1/2}$&  $4p~^2P_{3/2}$&      57.56 &          &   57.52 &  4.00e+11 &  3.99e+11 &  3.80e+11 &   0.3976 &         \\
$4p~^2P_{3/2}$&  $3d~^2D_{3/2}$&      28.12 &          &         &  3.86e+10 &  3.71e+10 &  3.66e+10 &   0.0183 &  0.0176 \\
$4f~^2F_{5/2}$&  $3d~^2D_{5/2}$&      25.75 &          &         &  4.46e+11 &  4.55e+11 &  4.55e+11 &   0.2660 &  0.2658 \\
$5d~^2D_{3/2}$&  $3p~^2P_{3/2}$&      16.34 &          &         &  2.88e+11 &  2.92e+11 &  2.82e+11 &   0.0461 &  0.0464 \\
$5p~^2P_{1/2}$&  $3d~^2D_{3/2}$&      18.56 &          &         &  1.78e+11 &  1.66e+11 &  1.53e+11 &   0.0184 &  0.0172 \\
$5p~^2P_{3/2}$&  $3d~^2D_{3/2}$&      18.49 &          &         &  1.62e+10 &  1.50e+10 &  1.37e+10 &   0.0033 &  0.0032 \\
$5p~^2P_{3/2}$&  $3d~^2D_{5/2}$&      18.56 &          &         &  1.47e+11 &  1.37e+11 &  1.31e+11 &   0.0303 &  0.0288 \\
$5f~^2F_{5/2}$&  $3d~^2D_{5/2}$&      17.99 &          &         &  1.57e+11 &  1.64e+11 &  1.63e+11 &   0.0458 &  0.0474 \\
$6d~^2D_{3/2}$&  $3p~^2P_{3/2}$&      14.17 &          &         &  1.66e+11 &  1.70e+11 &  1.53e+11 &   0.0200 &         \\
$6s~^2S_{1/2}$&  $3p~^2P_{1/2}$&      14.21 &          &         &  1.51e+11 &  1.49e+11 &  7.19e+10 &   0.0091 &         \\
$6s~^2S_{1/2}$&  $3p~^2P_{3/2}$&      14.43 &          &         &  3.18e+11 &  3.16e+11 &  2.30e+11 &   0.0198 &         \\
$6f~^2F_{5/2}$&  $3d~^2D_{5/2}$&      15.46 &          &         &  7.54e+10 &  8.08e+10 &  7.84e+10 &   0.0162 &         \\
$6p~^2P_{3/2}$&  $3d~^2D_{3/2}$&      15.64 &          &         &  8.45e+09 &  7.65e+09 &  5.45e+09 &   0.0012 &         \\
$6p~^2P_{1/2}$&  $3d~^2D_{3/2}$&      15.67 &          &         &  9.28e+10 &  8.46e+10 &  6.10e+10 &   0.0068 &         \\
$6p~^2P_{3/2}$&  $3d~^2D_{5/2}$&      15.69 &          &         &  7.25e+10 &  6.97e+10 &  5.86e+10 &   0.0107 &         \\
$6p~^2P_{3/2}$&  $4s~^2S_{1/2}$&      32.65 &          &         &  1.82e+11 &  1.82e+11 &  1.50e+11 &   0.1163 &         \\
$6p~^2P_{1/2}$&  $4s~^2S_{1/2}$&      32.77 &          &         &  1.98e+11 &  1.99e+11 &  1.56e+11 &   0.0639 &         \\
$6d~^2D_{3/2}$&  $4p~^2P_{1/2}$&      34.02 &          &         &  3.09e+11 &  3.10e+11 &  2.73e+11 &   0.2144 &         \\
$6d~^2D_{5/2}$&  $4p~^2P_{3/2}$&      34.47 &          &         &  3.82e+11 &  3.80e+11 &  3.58e+11 &   0.4087 &         \\
$6d~^2D_{3/2}$&  $4p~^2P_{3/2}$&      34.51 &          &         &  6.38e+10 &  6.40e+10 &  5.96e+10 &   0.0456 &         \\
$6s~^2S_{1/2}$&  $4p~^2P_{1/2}$&      35.55 &          &         &  9.82e+10 &  9.84e+10 &  7.29e+10 &   0.0372 &         \\
$6s~^2S_{1/2}$&  $4p~^2P_{3/2}$&      36.09 &          &         &  2.09e+11 &  2.09e+11 &  1.77e+11 &   0.0815 &         \\
$6f~^2F_{5/2}$&  $4d~^2D_{5/2}$&      37.08 &          &         &  4.18e+10 &  4.22e+10 &  4.12e+10 &   0.0517 &         \\
$6p~^2P_{3/2}$&  $4d~^2D_{3/2}$&      38.32 &          &         &  8.83e+09 &  8.73e+09 &  7.57e+09 &   0.0078 &         \\
$6p~^2P_{3/2}$&  $4d~^2D_{5/2}$&      38.45 &          &         &  8.19e+10 &  7.97e+10 &  7.29e+10 &   0.0726 &         \\
$6p~^2P_{1/2}$&  $4d~^2D_{3/2}$&      38.49 &          &         &  9.65e+10 &  9.55e+10 &  8.00e+10 &   0.0429 &         \\
$6g~^2G_{7/2}$&  $4f~^2F_{5/2}$&      38.60 &          &         &  6.11e+11 &  6.18e+11 &  6.14e+11 &   1.0920 &         \\
$6g~^2G_{9/2}$&  $4f~^2F_{7/2}$&      38.65 &          &         &  6.31e+11 &  6.38e+11 &  6.36e+11 &   1.4130 &         \\
$6g~^2G_{7/2}$&  $4f~^2F_{7/2}$&      38.66 &          &         &  2.25e+10 &  2.27e+10 &  2.26e+10 &   0.0403 &         \\
$6d~^2D_{5/2}$&  $4f~^2F_{5/2}$&      39.15 &          &         &  7.93e+08 &  8.06e+08 &  6.86e+08 &   0.0011 &         \\
$6d~^2D_{5/2}$&  $4f~^2F_{7/2}$&      39.20 &          &         &  1.61e+10 &  1.63e+10 &  1.45e+10 &   0.0222 &         \\
$6d~^2D_{3/2}$&  $4f~^2F_{5/2}$&      39.21 &          &         &  1.69e+10 &  1.75e+10 &  1.48e+10 &   0.0156 &         \\
$5d~^2D_{3/2}$&  $4p~^2P_{3/2}$&      50.99 &          &         &  9.79e+10 &  9.77e+10 &  9.61e+10 &   0.1527 &         \\
$5f~^2F_{5/2}$&  $4d~^2D_{5/2}$&      55.98 &          &         &  7.80e+10 &  7.82e+10 &  7.78e+10 &   0.2199 &         \\
$5s~^2S_{1/2}$&  $4p~^2P_{1/2}$&      56.21 &          &         &  1.86e+11 &  1.86e+11 &  1.74e+11 &   0.1767 &         \\
$5g~^2G_{7/2}$&  $4f~^2F_{7/2}$&      59.54 &          &         &  6.95e+10 &  6.97e+10 &  6.98e+10 &   0.2953 &         \\
$5p~^2P_{3/2}$&  $4d~^2D_{3/2}$&      61.53 &          &         &  1.83e+10 &  1.82e+10 &  1.78e+10 &   0.0416 &         \\
$5d~^2D_{5/2}$&  $4f~^2F_{5/2}$&      61.75 &          &         &  1.84e+09 &  1.84e+09 &  1.80e+09 &   0.0063 &         \\
$5p~^2P_{3/2}$&  $4d~^2D_{5/2}$&      61.87 &          &         &  1.68e+11 &  1.66e+11 &  1.65e+11 &   0.3867 &         \\
$5d~^2D_{5/2}$&  $4f~^2F_{7/2}$&      61.89 &          &         &  3.72e+10 &  3.71e+10 &  3.67e+10 &   0.1280 &         \\
$5d~^2D_{3/2}$&  $4f~^2F_{5/2}$&      61.95 &          &         &  3.98e+10 &  3.99e+10 &  3.91e+10 &   0.0916 &         \\
$5p~^2P_{1/2}$&  $4d~^2D_{3/2}$&      62.33 &          &         &  2.02e+11 &  2.00e+11 &  1.93e+11 &   0.2348 &         \\
$6p~^2P_{3/2}$&  $5s~^2S_{1/2}$&      91.05 &          &         &  8.32e+10 &  8.29e+10 &  7.73e+10 &   0.4135 &         \\
$6p~^2P_{1/2}$&  $5s~^2S_{1/2}$&      92.03 &          &         &  9.34e+10 &  9.31e+10 &  8.39e+10 &   0.2371 &         \\
$6d~^2D_{3/2}$&  $5p~^2P_{1/2}$&      93.71 &          &         &  1.26e+11 &  1.26e+11 &  1.19e+11 &   0.6649 &         \\
$6d~^2D_{5/2}$&  $5p~^2P_{3/2}$&      95.22 &          &         &  1.60e+11 &  1.60e+11 &  1.55e+11 &   1.3060 &         \\
$6d~^2D_{3/2}$&  $5p~^2P_{3/2}$&      95.57 &          &         &  2.72e+10 &  2.72e+10 &  2.62e+10 &   0.1488 &         \\
$6f~^2F_{5/2}$&  $5d~^2D_{3/2}$&     103.02 &          &         &  2.89e+11 &  2.89e+11 &  2.85e+11 &   2.7600 &         \\
$6f~^2F_{7/2}$&  $5d~^2D_{5/2}$&     103.45 &          &         &  3.12e+11 &  3.11e+11 &  3.07e+11 &   4.0040 &         \\
$6f~^2F_{5/2}$&  $5d~^2D_{5/2}$&     103.56 &          &         &  2.09e+10 &  2.08e+10 &  2.05e+10 &   0.2014 &         \\
$6s~^2S_{1/2}$&  $5p~^2P_{1/2}$&     106.29 &          &         &  7.35e+10 &  7.34e+10 &  6.65e+10 &   0.2489 &         \\
$6s~^2S_{1/2}$&  $5p~^2P_{3/2}$&     108.69 &          &         &  1.57e+11 &  1.57e+11 &  1.47e+11 &   0.5575 &         \\
$6g~^2G_{7/2}$&  $5f~^2F_{5/2}$&     109.18 &          &         &  4.88e+11 &  4.88e+11 &  4.87e+11 &   6.9710 &         \\
$6g~^2G_{9/2}$&  $5f~^2F_{7/2}$&     109.33 &          &         &  5.05e+11 &  5.06e+11 &  5.05e+11 &   9.0500 &         \\
$6g~^2G_{7/2}$&  $5f~^2F_{7/2}$&     109.39 &          &         &  1.80e+10 &  1.81e+10 &  1.80e+10 &   0.2590 &         \\
$6h~^2H_{9/2}$&  $5g~^2G_{7/2}$&     110.10 &          &         &  7.36e+11 &  7.35e+11 &  7.36e+11 &  13.3700 &         \\
$6h~^2H_{11/2}$&  $5g~^2G_{9/2}$&     110.16 &          &         &  7.52e+11 &  7.51e+11 &  7.52e+11 &  16.4100 &         \\
$6h~^2H_{9/2}$&  $5g~^2G_{9/2}$&     110.21 &          &         &  1.67e+10 &  1.67e+10 &  1.67e+10 &   0.3038 &         \\
$6f~^2F_{7/2}$&  $5g~^2G_{9/2}$&     110.87 &          &         &  5.55e+09 &  5.55e+09 &  5.42e+09 &   0.0819 &         \\
$6f~^2F_{5/2}$&  $5g~^2G_{7/2}$&     110.88 &          &         &  5.80e+09 &  5.76e+09 &  5.55e+09 &   0.0641 &         \\
$6d~^2D_{5/2}$&  $5f~^2F_{5/2}$&     113.63 &          &         &  1.33e+09 &  1.34e+09 &  1.29e+09 &   0.0154 &         \\
$6d~^2D_{5/2}$&  $5f~^2F_{7/2}$&     113.86 &          &         &  2.69e+10 &  2.71e+10 &  2.64e+10 &   0.3133 &         \\
$6d~^2D_{3/2}$&  $5f~^2F_{5/2}$&     114.13 &          &         &  2.88e+10 &  2.90e+10 &  2.79e+10 &   0.2252 &         \\
$6p~^2P_{3/2}$&  $5d~^2D_{3/2}$&     114.34 &          &         &  8.66e+09 &  8.64e+09 &  8.32e+09 &   0.0679 &         \\
$6p~^2P_{3/2}$&  $5d~^2D_{5/2}$&     115.01 &          &         &  8.05e+10 &  7.92e+10 &  7.72e+10 &   0.6383 &         \\
$6p~^2P_{1/2}$&  $5d~^2D_{3/2}$&     115.90 &          &         &  9.46e+10 &  9.44e+10 &  8.86e+10 &   0.3811 &         \\
$4d~^2D_{3/2}$&  $4p~^2P_{1/2}$&     373.52 &          &         &  7.83e+09 &  7.86e+09 &  7.83e+09 &   0.6548 &         \\
$4d~^2D_{5/2}$&  $4p~^2P_{3/2}$&     425.63 &          &         &  6.38e+09 &  6.40e+09 &  6.65e+09 &   1.0400 &         \\
$4p~^2P_{3/2}$&  $4s~^2S_{1/2}$&     439.49 &          &         &  5.30e+09 &  5.30e+09 &  4.96e+09 &   0.6139 &         \\
$4d~^2D_{3/2}$&  $4p~^2P_{3/2}$&     442.42 &          &         &  9.43e+08 &  9.49e+08 &  9.84e+08 &   0.1107 &         \\
$4p~^2P_{1/2}$&  $4s~^2S_{1/2}$&     538.09 &          &         &  2.86e+09 &  2.86e+09 &  2.80e+09 &   0.2482 &         \\
$5d~^2D_{3/2}$&  $5p~^2P_{1/2}$&     765.14 &          &         &  2.51e+09 &  2.51e+09 &  2.48e+09 &   0.8823 &         \\
$5d~^2D_{5/2}$&  $5p~^2P_{3/2}$&     869.28 &          &         &  2.07e+09 &  2.04e+09 &  2.12e+09 &   1.4050 &         \\
$5p~^2P_{3/2}$&  $5s~^2S_{1/2}$&     878.59 &          &         &  1.72e+09 &  1.71e+09 &  1.57e+09 &   0.7972 &         \\
$5d~^2D_{3/2}$&  $5p~^2P_{3/2}$&     909.49 &          &         &  3.00e+08 &  3.00e+08 &  3.12e+08 &   0.1488 &         \\
$5p~^2P_{1/2}$&  $5s~^2S_{1/2}$&    1074.40 &          &         &  9.36e+08 &  9.28e+08 &  9.07e+08 &   0.3239 &         \\
$6d~^2D_{3/2}$&  $6p~^2P_{1/2}$&    1359.22 &          &         &  9.88e+08 &  9.91e+08 &  9.74e+08 &   1.0950 &         \\
$6d~^2D_{5/2}$&  $6p~^2P_{3/2}$&    1522.45 &          &         &  8.47e+08 &  8.03e+08 &  8.38e+08 &   1.7650 &         \\
$6p~^2P_{3/2}$&  $6s~^2S_{1/2}$&    1551.34 &          &         &  6.72e+08 &  6.98e+08 &  6.18e+08 &   0.9698 &         \\
$6p~^2P_{1/2}$&  $6s~^2S_{1/2}$&    1896.72 &          &         &  3.66e+08 &  3.83e+08 &  3.66e+08 &   0.3950 &         \\
\end{longtable}
The present E2 transition probabilities show a good agreement with Liang et al. \cite{liang2009r} and are displayed in Fig \ref{fig:kr_trans} (c). We found large discrepancies in the RQDO A values of Charro and Mart\'in \cite{charro2002relativistic} relative to both the present and previous \cite{liang2009r} results and hence, did not include those for comparison. Similar behavior was seen for Ar$^{7+}$. This could be due to the non-inclusion of indirect relativistic effects in RQDO method as mentioned in \cite{charro2002relativistic}.

\begin{figure}[H]
    \centering
    {{\includegraphics[width=6cm,height=5cm]{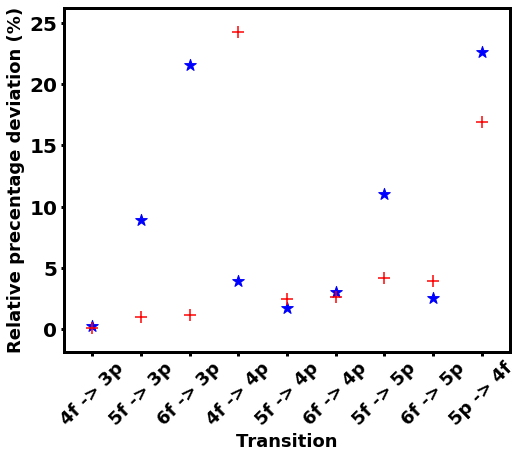} }}
    \caption{The relative percentage deviation between the present and Charro and Mart\'in \cite{charro2002relativistic} values of E2 oscillator strengths for Kr$^{25+}$ (star), Xe$^{43+}$ (plus).}
   \label{fig:e2_comp_charo}
\end{figure}

Further, we compared the present E2 oscillator strengths with the multiplets results from another work of Charro and Mart\'in \cite{charro2002systematic} based on the RQDO procedure. The relative percentage deviation between the two results is displayed in Fig\ref{fig:e2_comp_charo} and a fair consistency (within 22\%) is evident.
%

%
%
%
%
\subsubsection{ Xe$^{43+}$}
The transition parameters of 462, 92, 534 and 270 lines for  E1, M1, E2 and M2 transitions, respectively, of Xe$^{43+}$, are tabulated in Tables S5-S6 (Supplementary data). 

\begin{figure}[H]

    \centering
    \subfloat
    {{\includegraphics[width=8cm,height=6cm]{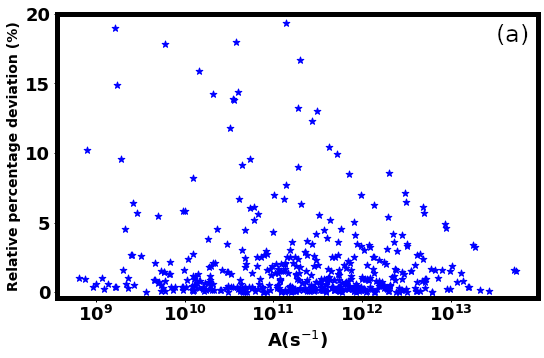} }}
    \qquad
    \subfloat{{\includegraphics[width=8cm,height=6cm]{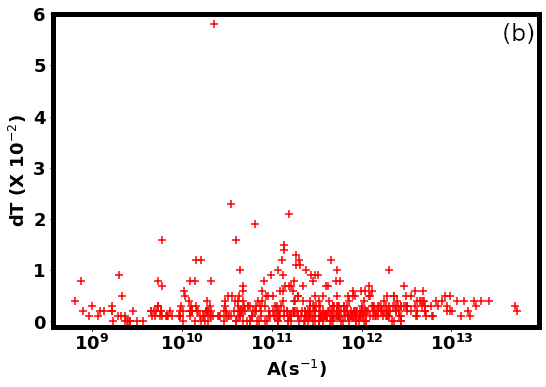} }}

    \caption{(a) The relative absolute percentage deviation in the present MBPT values with respect to RCI results, (b) The E1 dT parameter, for Xe$^{43+}$}
    \label{fig:xe_trans}
\end{figure}

Table \ref{tab6}, presents comprehensive comparison of the present and previous \cite{NIST,Spectra_W3,SAMPSON1990209,johnson1996transition,vilkas2008relativistic} results for wavelengths, E1 transition probabilities and corresponding weighted oscillator strengths.
It is evident that our results are in good agreement with the wavelengths of 4 transitions listed at NIST and have a maximum difference of 0.866 $\AA$ for $3p~^2P_{1/2}$ - $3s~^2S_{1/2}$ line.
The wavelengths of 34 allowed transitions from the earlier work \cite{matsushima1991spectra} available at the SPECTRA-W3 database \cite{Spectra_W3} shows an excellent match with the present results within a marginal difference of 0.009 \AA.
%
%
The present gf values also agree with the theoretical calculations of \cite{SAMPSON1990209}. The most considerable difference of 0.058 between the two results occurs for $4f~^2F_{7/2}$ - $3d~^2D_{5/2}$ line, whereas the mean difference is found to be only 0.007. Further, the mean deviation in our transition rates are within 2.0\% and 0.8\% on comparing with the results reported by Vilkas et al. \cite{vilkas2008relativistic} for five lines and Johnson et al. \cite{johnson1996transition} for four lines, respectively. 
%
%
The relative deviation in the MBPT to RCI transition rates is shown in Fig. \ref{fig:xe_trans}(a). The average difference between the present two results is only 1.96\%. However, a few transitions digress by more than 10\%. 
The transition $3d~^2D_{5/2}$ - $9p~^2P_{3/2}$ deviated the most by 19\%. This inconsistency could be due to the difference in the configuration interaction effects considered in the two physical models.
%
Further, the uncertainty dT in E1 transition rates is illustrated in Fig \ref{fig:xe_trans}(b). The majority of E1 transitions are below the dT value of 0.01. Hence, our results of transition rates from the length and velocity gauges are in excellent agreement with a mean uncertainty (dT) of 0.003\% for E1 transitions.
%
%
Comparison of the present E2 oscillator strengths with the multiplet f values from Charro and Mart\'in \cite{charro2002systematic} is shown 
in Fig \ref{fig:e2_comp_charo}. Our results agree well for almost all the transitions, except for $4f - 4p$ and $5p - 4f$ lines that stray relatively larger by 24\% and 16\% from the present results. This is possibly due to the small value ($~10^{-06}$) of oscillator strengths corresponding to these lines. 
Overall, the agreement among all the results is pretty satisfactory. To our knowledge, no other comprehensive results exist for Na-like Xe$^{43+}$ ion; hence, most radiative parameters are new.
\setlength{\tabcolsep}{1pt}
\begin{longtable}{|l|l|l|l|l|l|l|l|l|l|l|}
\caption{Comparison of the present E1 transition parameters with the values of Johnson et al. \cite{johnson1996transition}, Sampson et al. \cite{SAMPSON1990209}, Vilkas et al. \cite{vilkas2008relativistic}, NIST \cite{NIST} and Spectra W3 \cite{Spectra_W3} atomic database  for Xe$^{43+}$ .\label{tab6}} \\
\hline
 \multicolumn{2}{|c|}{\textbf{Levels  }} & \multicolumn{3}{c|}{\textbf{Wavelength (\AA)}}  & \multicolumn{3}{c|}{\textbf{A (s $^{-1}$)}}  & \multicolumn{2}{c|}{\textbf{gf}} \\
\hline \multicolumn{1}{|c|}{\textbf{Upper }} & \multicolumn{1}{c|}{\textbf{Lower }} & \multicolumn{1}{c|}{\textbf{RCI}}& \multicolumn{1}{c|}{\textbf{\cite{NIST}}}& \multicolumn{1}{c|}{\textbf{\cite{Spectra_W3}}}& \multicolumn{1}{c|}{\textbf{RCI}}& \multicolumn{1}{c|}{\textbf{MBPT}}& \multicolumn{1}{c|}{\textbf{Others  }}& \multicolumn{1}{c|}{\textbf{RCI  }}& \multicolumn{1}{c|}{\textbf{ \cite{SAMPSON1990209} }}\\  \hline
\endfirsthead
\multicolumn{10}{c}%
{{\bfseries \tablename\ \thetable{} -- continued from previous page}} \\
\hline \multicolumn{1}{|c|}{\textbf{Upper }} & \multicolumn{1}{c|}{\textbf{Lower }} & \multicolumn{1}{c|}{\textbf{RCI}}& \multicolumn{1}{c|}{\textbf{\cite{NIST}}}& \multicolumn{1}{c|}{\textbf{\cite{Spectra_W3}}}& \multicolumn{1}{c|}{\textbf{RCI}}& \multicolumn{1}{c|}{\textbf{MBPT}}& \multicolumn{1}{c|}{\textbf{Others  }}& \multicolumn{1}{c|}{\textbf{RCI  }}& \multicolumn{1}{c|}{\textbf{  \cite{SAMPSON1990209} }}\\  \hline
\endhead
\hline \multicolumn{10}{|l|}{{Continued on next page}} \\ \hline
\endfoot
\hline \hline
\endlastfoot
$6p~^2P_{3/2}$&  $3s~^2S_{1/2}$&       4.82 &         &   4.81 &  3.56e+12 &  3.51e+12 &           &   0.0496 &         \\
$6p~^2P_{1/2}$&  $3s~^2S_{1/2}$&       4.83 &         &   4.83 &  4.13e+12 &  4.10e+12 &           &   0.0290 &         \\
$6d~^2D_{3/2}$&  $3p~^2P_{1/2}$&       4.98 &         &   4.98 &  6.68e+12 &  6.78e+12 &           &   0.0994 &         \\
$6d~^2D_{5/2}$&  $3p~^2P_{3/2}$&       5.16 &         &   5.16 &  7.82e+12 &  7.95e+12 &           &   0.1871 &         \\
$6f~^2F_{5/2}$&  $3d~^2D_{3/2}$&       5.43 &         &   5.43 &  8.48e+12 &  8.89e+12 &           &   0.2245 &         \\
$6f~^2F_{7/2}$&  $3d~^2D_{5/2}$&       5.47 &         &   5.47 &  8.87e+12 &  9.28e+12 &           &   0.3183 &         \\
$5p~^2P_{3/2}$&  $3s~^2S_{1/2}$&       5.55 &         &   5.55 &  5.94e+12 &  5.84e+12 &           &   0.1097 &  0.1096 \\
$5p~^2P_{1/2}$&  $3s~^2S_{1/2}$&       5.60 &         &   5.59 &  7.04e+12 &  6.97e+12 &           &   0.0661 &  0.0660 \\
$5d~^2D_{3/2}$&  $3p~^2P_{1/2}$&       5.75 &         &   5.75 &  1.16e+13 &  1.17e+13 &           &   0.2292 &  0.2298 \\
$5s~^2S_{1/2}$&  $3p~^2P_{1/2}$&       5.92 &         &   5.92 &  1.90e+12 &  1.84e+12 &           &   0.0200 &  0.0188 \\
$5d~^2D_{5/2}$&  $3p~^2P_{3/2}$&       5.97 &         &   5.97 &  1.38e+13 &  1.39e+13 &           &   0.4418 &  0.4468 \\
$5d~^2D_{3/2}$&  $3p~^2P_{3/2}$&       5.99 &         &        &  2.32e+12 &  2.34e+12 &           &   0.0499 &  0.0500 \\
$5s~^2S_{1/2}$&  $3p~^2P_{3/2}$&       6.17 &         &   6.17 &  4.51e+12 &  4.39e+12 &           &   0.0515 &  0.0492 \\
$5f~^2F_{5/2}$&  $3d~^2D_{3/2}$&       6.32 &         &   6.32 &  1.77e+13 &  1.83e+13 &           &   0.6347 &  0.6552 \\
$5f~^2F_{7/2}$&  $3d~^2D_{5/2}$&       6.38 &         &   6.38 &  1.85e+13 &  1.91e+13 &           &   0.9043 &  0.9360 \\
$5f~^2F_{5/2}$&  $3d~^2D_{5/2}$&       6.38 &         &        &  1.22e+12 &  1.26e+12 &           &   0.0447 &  0.0456 \\
$5p~^2P_{3/2}$&  $3d~^2D_{3/2}$&       6.45 &         &        &  1.03e+11 &  9.56e+10 &           &   0.0026 &  0.0024 \\
$5p~^2P_{1/2}$&  $3d~^2D_{3/2}$&       6.51 &         &        &  1.35e+12 &  1.26e+12 &           &   0.0171 &  0.0164 \\
$5p~^2P_{3/2}$&  $3d~^2D_{5/2}$&       6.52 &         &        &  9.79e+11 &  9.11e+11 &           &   0.0250 &  0.0240 \\
$4p~^2P_{3/2}$&  $3s~^2S_{1/2}$&       7.77 &         &   7.76 &  1.02e+13 &  1.00e+13 &           &   0.3701 &  0.3710 \\
$4p~^2P_{1/2}$&  $3s~^2S_{1/2}$&       7.94 &         &   7.94 &  1.28e+13 &  1.26e+13 &           &   0.2423 &  0.2440 \\
$4d~^2D_{3/2}$&  $3p~^2P_{1/2}$&       8.03 &         &   8.03 &  2.14e+13 &  2.14e+13 &           &   0.8269 &  0.8190 \\
$4d~^2D_{5/2}$&  $3p~^2P_{3/2}$&       8.45 &         &   8.45 &  2.64e+13 &  2.64e+13 &           &   1.6990 &  1.6940 \\
$4d~^2D_{3/2}$&  $3p~^2P_{3/2}$&       8.50 &         &   8.50 &  4.53e+12 &  4.53e+12 &           &   0.1963 &  0.1952 \\
$4s~^2S_{1/2}$&  $3p~^2P_{1/2}$&       8.73 &         &   8.73 &  3.97e+12 &  3.89e+12 &3.97e+12\cite{johnson1996transition}   &   0.0907 &  0.0860 \\
$4f~^2F_{5/2}$&  $3d~^2D_{3/2}$&       9.08 &         &   9.08 &  5.08e+13 &  5.16e+13 &           &   3.7670 &  3.7908 \\
$4f~^2F_{7/2}$&  $3d~^2D_{5/2}$&       9.19 &         &   9.19 &  5.37e+13 &  5.45e+13 &           &   5.4350 &  5.4930 \\
$4f~^2F_{5/2}$&  $3d~^2D_{5/2}$&       9.21 &         &        &  3.57e+12 &  3.62e+12 &           &   0.2723 &  0.2724 \\
$4s~^2S_{1/2}$&  $3p~^2P_{3/2}$&       9.30 &         &   9.29 &  9.61e+12 &  9.46e+12 &    9.61e+12\cite{johnson1996transition}   &   0.2491 &  0.2400 \\
$4p~^2P_{3/2}$&  $3d~^2D_{3/2}$&       9.66 &         &        &  2.41e+11 &  2.32e+11 &           &   0.0135 &  0.0132 \\
$4p~^2P_{3/2}$&  $3d~^2D_{5/2}$&       9.81 &         &   9.80 &  2.31e+12 &  2.22e+12 &           &   0.1329 &  0.1296 \\
$4p~^2P_{1/2}$&  $3d~^2D_{3/2}$&       9.93 &         &   9.93 &  3.20e+12 &  3.09e+12 &           &   0.0945 &  0.0924 \\
$6f~^2F_{5/2}$&  $4d~^2D_{3/2}$&      13.00 &         &  13.01 &  4.76e+12 &  4.80e+12 &           &   0.7239 &         \\
$6f~^2F_{7/2}$&  $4d~^2D_{5/2}$&      13.10 &         &  13.11 &  5.05e+12 &  5.08e+12 &           &   1.0390 &         \\
$5d~^2D_{3/2}$&  $4p~^2P_{1/2}$&      17.82 &         &  17.82 &  3.92e+12 &  3.91e+12 &           &   0.7460 &         \\
$5d~^2D_{5/2}$&  $4p~^2P_{3/2}$&      18.64 &         &  18.64 &  5.08e+12 &  5.07e+12 &           &   1.5880 &         \\
$5f~^2F_{5/2}$&  $4d~^2D_{3/2}$&      19.69 &         &  19.70 &  9.02e+12 &  9.04e+12 &           &   3.1450 &         \\
$5f~^2F_{7/2}$&  $4d~^2D_{5/2}$&      19.90 &         &  19.91 &  9.64e+12 &  9.66e+12 &           &   4.5770 &         \\
$5g~^2G_{7/2}$&  $4f~^2F_{5/2}$&      20.69 &         &  20.68 &  1.55e+13 &  1.55e+13 &           &   7.9420 &         \\
$5s~^2S_{1/2}$&  $4p~^2P_{3/2}$&      20.71 &         &  20.71 &  3.12e+12 &  3.11e+12 &           &   0.4009 &         \\
$5g~^2G_{9/2}$&  $4f~^2F_{7/2}$&      20.77 &         &  20.77 &  1.60e+13 &  1.60e+13 &           &  10.3200 &         \\
$3d~^2D_{3/2}$&  $3p~^2P_{1/2}$&      58.29 &   59.10 &        &  1.66e+11 &  1.69e+11 &  1.66e+11\cite{vilkas2008relativistic} &   0.3375 &  0.3430 \\
$3p~^2P_{3/2}$ &  $3s~^2S_{1/2}$ &    66.8234 &   66.58 &          &  1.556e+11 &  1.603e+11 &  1.57e+11 \cite{vilkas2008relativistic} &   0.4168 &         \\
& & & & & & & 1.57e+11 \cite{johnson1996transition} & & \\
$3d~^2D_{5/2}$&  $3p~^2P_{3/2}$&      85.00 &   84.80 &        &  6.49e+10 &  6.61e+10 &  6.52e+10\cite{vilkas2008relativistic} &   0.4220 &  0.4296 \\
$3d~^2D_{3/2}$&  $3p~^2P_{3/2}$&      97.99 &         &        &  6.94e+09 &  7.09e+09 &  6.96e+09\cite{vilkas2008relativistic} &   0.0399 &  0.0408 \\
$3p~^2P_{1/2}$&  $3s~^2S_{1/2}$&     124.79 &  123.92 &        &  2.29e+10 &  2.40e+10 &  2.34e+10\cite{vilkas2008relativistic} &   0.1071 &  0.1094 \\
&  &   &   &    &   & &  2.34e+10 \cite{johnson1996transition} &   & \\
  \hline
\end{longtable}
%
%
%
%

%
%
%
%

%
\subsection{Lifetime}
The lifetime ($\tau$) of a level $l$ is given as,
\begin{equation}
    \tau_l = \frac{1}{\Sigma_u A_{ul}} ,
\end{equation}
where the summation is over the transition probabilities of E1, E2, M1 and M2 transitions. The lifetimes calculated for the presently considered levels of Na-like Ar$^{7+}$, Kr$^{25+}$ and Xe$^{43+}$ are tabulated in Table S7 (supplementary data). For a few levels, theoretical \cite{theodosiou1988accurate, vilkas2008relativistic} and experimental  \cite{kink1997lifetime, trabert1994experimental} results are available for comparison. Therefore, our calculated lifetimes are compared with these studies in Table. \ref{Tab:lifetime}. Our results for Xe$^{43+}$ have an excellent match with the corresponding values of Vilkas et al. \cite{vilkas2008relativistic}. However, these are within 1.4\%, 0.8\% and  3\%, for Ar$^{7+}$, Kr$^{25+}$ and Xe$^{43+}$, respectively, in comparison to the results of Theodosiou and Curtis \cite{theodosiou1988accurate}. 
Further, the present lifetimes for the levels of Kr$^{25+}$ and Xe$^{43+}$  also show a good agreement with the measurements of Kink et al.\cite{kink1997lifetime} and Träbert et al. \cite{trabert1994experimental} and lie within the experimental uncertainties.
\begin{figure}[H]

    \centering
    \subfloat
    {{\includegraphics[width=5cm,height=4cm]{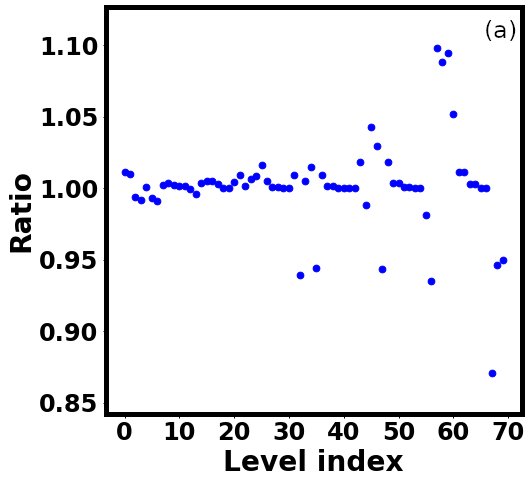} }}
    \quad
    \subfloat{{\includegraphics[width=5cm,height=4cm]{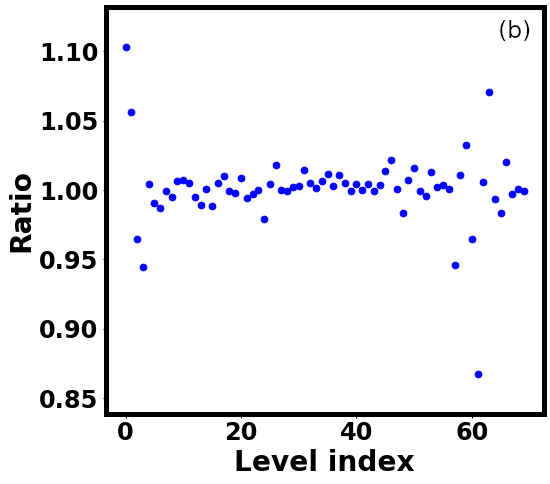} }}
    \quad
    \subfloat{{\includegraphics[width=5cm,height=4cm]{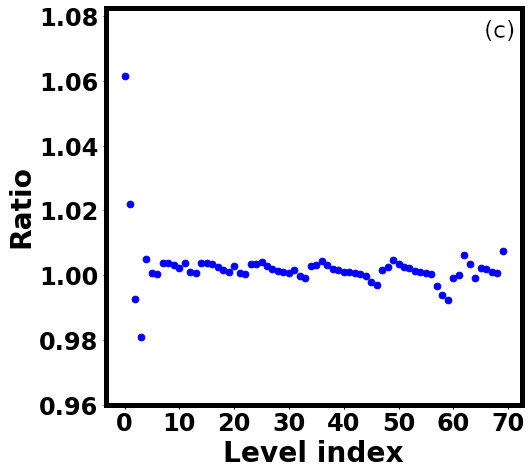} }}

    \caption{The ratio of lifetimes in length and velocity gauge for, (a) Ar$^{7+}$, (b) Kr$^{25+}$, (c) Xe$^{43+}$. }
    \label{fig:Lifetime}
\end{figure}
\begin{table}[H]
\caption{The lifetimes (unit: $10^{-10} sec$) from the present and previous results of Theodosiou and Curtis\cite{theodosiou1988accurate}, Träbert et al.\cite{trabert1994experimental} and Kink et al.\cite{kink1997lifetime} for Na-like Ar$^{7+}$, Kr$^{25+}$ and Xe$^{43+}$ ions.}
\centering
 \begin{tabular}{|c |c |c |c |c |}
 
  \hline
  Ion & Level & RCI & Other theories   & Measurements \\ 
  \hline
 Ar$^{7+}$ &  &  & & \\
  & $3p~^2P_{1/2}$  & 4.154 & 4.094\cite{theodosiou1988accurate} & 4.17 $\pm$ 0.1 \cite{reistad1986oscillator}\\
  & $3p~^2P_{3/2}$  &  3.911 & 3.856\cite{theodosiou1988accurate} & 3.89 $\pm$ 0.1 \cite{reistad1986oscillator}\\
  & $3p~^2D_{3/2}$  &  1.336 & 1.3375\cite{theodosiou1988accurate} & 1.70 $\pm$ 0.1 \cite{reistad1986oscillator}\\
  & $3d~^2D_{5/2}$  & 1.374 & 1.3799\cite{theodosiou1988accurate} & 1.66 $\pm$ 0.08 \cite{reistad1986oscillator}\\
  \hline
   Kr$^{25+}$ &  &  &  &\\
  & $3p~^2P_{1/2}$  & 0.8723 & 0.8664\cite{theodosiou1988accurate} &\\
  & $3p~^2P_{3/2}$  &  0.4604 & 0.4586\cite{theodosiou1988accurate} & 0.452 $\pm$ 0.02\cite{kink1997lifetime} \\
  & $3p~^2D_{3/2}$  &  0.2696 & 0.2674\cite{theodosiou1988accurate} &\\
  & $3d~^2D_{5/2}$  & 0.3673 & 0.3644\cite{theodosiou1988accurate} &\\
  \hline
   Xe$^{43+}$ & & & &\\
  & $3p~^2P_{1/2}$  & 0.4360 & 0.4198\cite{theodosiou1988accurate} & 0.42 $\pm$ 0.03\cite{trabert1994experimental} \\
  & & & 0.4270\cite{vilkas2008relativistic}& \\
  & $3p~^2P_{3/2}$  &  0.0646 & 0.0627\cite{theodosiou1988accurate} & 0.068 $\pm$ 0.006\cite{trabert1994experimental}\\
  & & & 0.0636\cite{vilkas2008relativistic}& \\
  & $3p~^2D_{3/2}$  &  0.0579 & 0.0562\cite{theodosiou1988accurate}  & 0.06 $\pm$ 0.01\cite{trabert1994experimental}\\
  & & & 0.0579\cite{vilkas2008relativistic} & \\
  & $3d~^2D_{5/2}$  & 0.1540 & 0.1493\cite{theodosiou1988accurate} & 0.14 $\pm$ 0.04 \cite{trabert1994experimental}\\
  & & & 0.1530\cite{vilkas2008relativistic} & \\
  \hline
 \end{tabular}
 \label{Tab:lifetime}
\end{table}
%
%
%
Further, we perform an accuracy check by plotting the ratio of lifetimes ($\tau_b/\tau_c$) for all the 71 levels in the length and velocity gauges in Figs. \ref{fig:Lifetime} (a), (b) and (c), for Ar$^{7+}$, Kr$^{25+}$ and Xe$^{43+}$, respectively. 
For Xe$^{43+}$, the ratio is close to one for most of the levels and the largest value is 1.06 for $3p ~^2P_{1/2}$ state. Similarly, for Kr$^{25+}$ and Ar$^{7+}$, the majority of levels have a ratio near to one. The maximum differences of 0.86 and 0.87 occur for the $9d ~^2D_{5/2}$ and $9s~^2S_{1/2}$ states, respectively. Overall we find a decent consistency between the lifetime results calculated from the two gauges.
\subsection{Hyperfine Interaction constants and Land\'e g$_J$ Factors}
Using the GRASP2018, we computed the hyperfine constants (A$_J$ and B$_J$), and Land\'e g$_J$ factors for the 71 levels of the presently studied Na-like inert gas ions. To provide generalized results, we assumed the nuclear magnetic and quadrupole moments to equal one. Table S8 (supplementary data) represents our RCI results of A$_J/\mu, $B$_J$/Q, and Land\'e g$_J$ factors for all the there ions. The comparison of the present values with the only available RCC calculations of Dutta and Majumder \cite{dutta2013electron} for Ar$^{7+}$ is shown in Table \ref{Tab:hfs}. A$_J$ and B$_J$/Q have been reported for only 3s, 3p and 3d states in \cite{dutta2013electron}. The maximum absolute difference between the present and RCC values is observed for the $3p~^2P_{3/2}$ state where A$_J/\mu_I$ and B$_J/$/Q stray by 40.5 (MHz/units of $\mu_I$)  and  28.9 (Mhz/b), respectively. 
Further, Dutta and Majumder \cite{dutta2013electron} also calculated the HFS for the Dirac-Fock (DF) reference states without considering the correlations and Gaunt corrections. Hence, we carried out similar MCDF calculations of HFS for the MR set AS\{9\} labelled as MCDF$\_$AS\{9\}. The comparison of these two results is also given in Table.\ref{Tab:hfs}. An excellent agreement with a maximum relative deviation of 0.1\% exists between the present MCDF$\_$AS\{9\} and previous DF B$_J$/Q values. However, our RCI A$_J/\mu_I$ values exhibit a better match with the RCC calculations \cite{dutta2013electron} as compared to the agreement between the MCDF$\_$AS\{9\} and DF results \cite{dutta2013electron}. Therefore, the slight variation between the RCI and RCC results could be due to the correlation and relativistic corrections considered in these two calculations. 
%
%
%
%
\begin{table}[H]
\caption{The present calculated hyperfine constants A$_J/\mu_I$ (MHz per units of $\mu_I$), B$_J/$Q (MHz/barn) and the corresponding results of Dutta and Majumder \cite{dutta2013electron} for Ar$^{7+}$}
\centering
 \begin{tabular}{ |c |c |c |c |c|c |c|c|c| }
 
  \hline
  \multicolumn{1}{|c|}{}  & \multicolumn{4}{c|}{A$_J$/$\mu_I$ (MHz/units of $\mu_I$ )} & \multicolumn{4}{c|}{B$_J$/Q (MHz/barn)} \\
  \cline{2-9}
Level   & RCI  & RCC \cite{dutta2013electron} & MCDF$\_$AS\{9\} & DF \cite{dutta2013electron} & RCI & RCC \cite{dutta2013electron} &   MCDF$\_$AS\{9\} & DF \cite{dutta2013electron} \\ 
  \hline
  $3s~^2S_{1/2}$  & 5638.00 & 5669.86 & 5200.00 & 5080.69 &  -& - &  -& -  \\  
  $3p~^2P_{1/2}$  &  1529.00 & 1518.29 &1379.00 & 1346.27& - & -  &  - & - \\
  $3p~^2P_{3/2}$  &  302.00 & 261.54 & 268.70 & 262.34 & 1953.00& 1924.15 &  1744.00 &1742.25  \\
  $3d~^2D_{3/2}$  & 87.27 & 84.15 &  85.00& 83.10 & 182.70& 180.22  &183.70 &183.68 \\
  $3d~^2D_{5/2}$  &  9.69 & 10.59 & 36.36 & 35.54 & 260.30 &256.92 & 261.34 & 261.30 \
  \\
  \hline
 \end{tabular}
 \label{Tab:hfs}
\end{table}

\begin{figure}[H]

    \centering
    \subfloat
    {{\includegraphics[width=8cm,height=6cm]{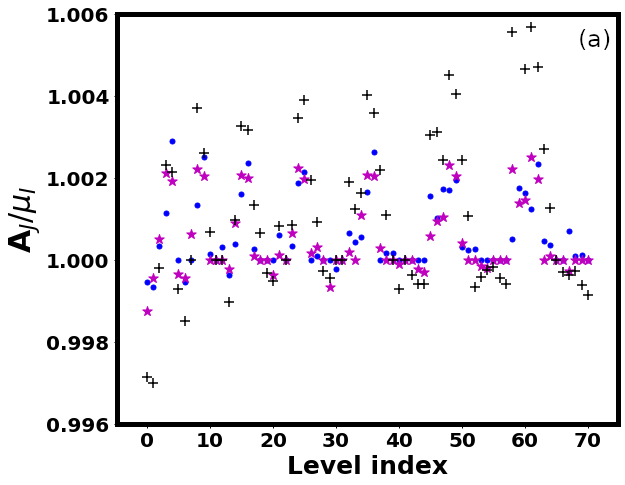} }}
    \qquad
    \subfloat{{\includegraphics[width=8cm,height=6cm]{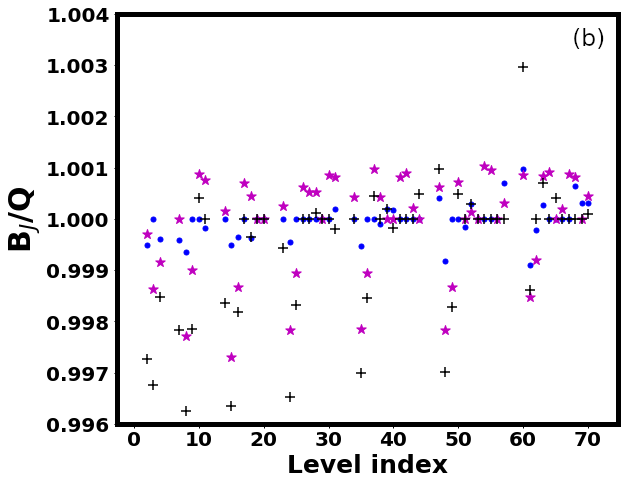} }}
    \caption{The ratio of the present RCI and MCDF HFS constants (a) A$_J/\mu_I$ and (b) B$_J/Q$, for Ar$^{7+}$ (point), Kr$^{25+}$(star), Xe$^{43+}$(plus).}
    \label{fig:hfs}
\end{figure}

Since the study of hyperfine interactions is a valuable probe to the QED effects; therefore, to analyze this, we present the ratio of HFS constants calculated with and without including 
QED corrections in Fig. \ref{fig:hfs} for all three ions. The ratio is maximum for Xe$^{43+}$ and slightly comparable for Ar$^{7+}$ and Kr$^{25+}$. The observation supports the fact that heavy ions are more sensitive to QED effects.

Further, to the best of our knowledge, there are no other studies involving the estimation of HFS and Land\'e g$_J$ factors of higher levels of Na-like ions. Therefore we can say that most parameters are the first report. Since hyperfine constants are crucial for precise abundance estimation, and line modeling and Land\'e g$_J$ factors are requisite for exploring magnetic fields in hot stars, we believe our results will benefit such studies and be helpful for comparisons for future works.
\subsection{Isotope Shifts}
\begin{figure}[H]

    \centering
    \subfloat
    {{\includegraphics[width=5cm,height=4cm]{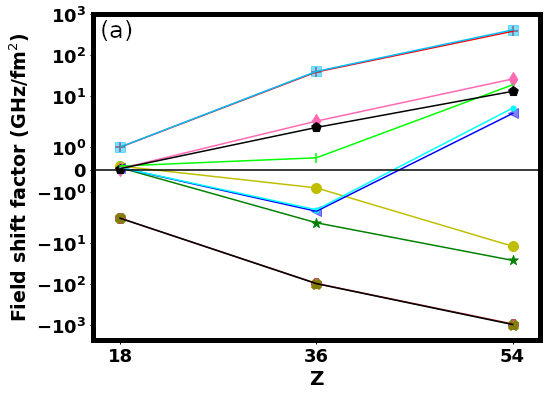} }}
    \qquad
    \subfloat{{\includegraphics[width=5cm,height=4cm]{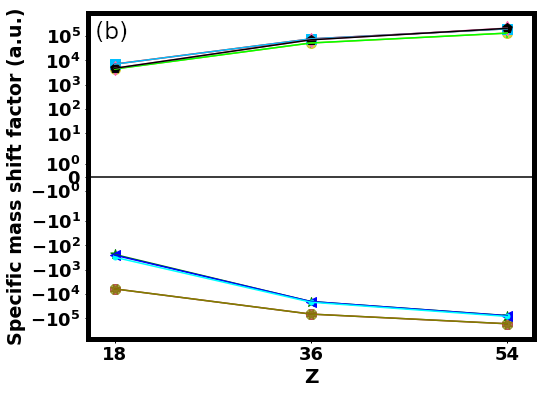} }}
    \qquad
    \subfloat{{\includegraphics[width=5cm,height=4cm]{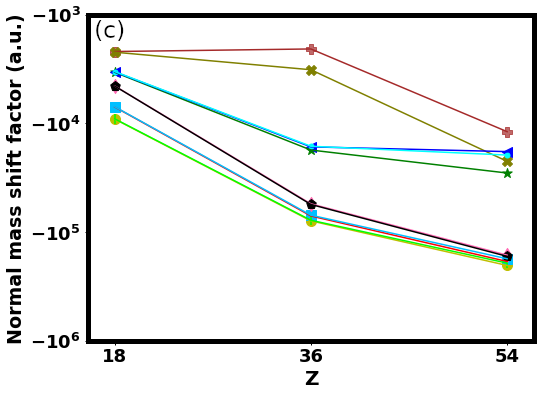} }}

    \caption{The (a) FS, (b) SMS and (c) NMS factors for $3p~^2P_{1/2} \rightarrow 3s~^2S_{1/2}$ (plus (filled)), $3p~^2P_{3/2} \rightarrow 3s~^2S_{1/2}$ (x (filled)), $3d~^2D_{3/2} \rightarrow 3p~^2P_{1/2}$ (star), $4s~^2S_{1/2} \rightarrow 3p~^2P_{1/2}$ (plus), $4d~^2D_{3/2} \rightarrow 3p~^2P_{1/2}$ (circle), $3d~^2D_{3/2} \rightarrow 3p~^2P_{3/2}$ ( triangle$\_$left), $3d~^2D_{5/2} \rightarrow 3p~^2P_{3/2}$ (point), $4s~^2S_{1/2} \rightarrow 3p~^2P_{3/2}$ (square), $4d~^2D_{5/2} \rightarrow 3p~^2P_{3/2}$ (vline), $4p~^2P_{1/2} \rightarrow 3d~^2D_{3/2}$ (diamond), $4p~^2P_{3/2} \rightarrow 3d~^2D_{3/2}$ (pentagon) of Ar$^{7+}$, Kr$^{25+}$ and Xe$^{43+}$}
    \label{fig:IS}
\end{figure}

Using the RIS4 module, we computed the electronic factors corresponding to normal mass, specific mass, and field shifts for the 71 levels of Na-like Ar$^{7+}$, Kr$^{25+}$ and Xe$^{43+}$ ions. We listed them in Table S9 (supplementary data).
Previous results of IS factors are available from the MBPT calculations of Safronova and Johnson \cite{PhysRevA.64.052501} for Ar$^{7+}$, CIDF results of Tupitsyn et al. \cite{tupitsyn2003relativistic} for Ar$^{7+}$ and Xe$^{43+}$, and RMBPT and GRASP values of Silwal et al. \cite{silwal2018measuring} for Xe$^{43+}$. Table. \ref{Tab:IS} displays comparison of the present and aforementioned studies \cite{PhysRevA.64.052501,tupitsyn2003relativistic,silwal2018measuring}.
We observed that our calculated FS values show an excellent agreement with \cite{tupitsyn2003relativistic}, but the SMS results have a relative error of 5.7\% and 2.4\% in Ar$^{7+}$ and Xe$^{43+}$, respectively. 
A minimum and maximum deviation of 0.3\% and 2\% is noticed in present FS results of Ar$^{7+}$ when compared with Safranova and Johnson \cite{PhysRevA.64.052501} values. In contrast, for SMS parameters, the difference goes up to 15\%. 
%
Further, our IS value for the D1 line of Xe$^{43+}$ deviates only by 0.5\% and 0.4\% with the GRASP2K and RMBPT calculations from Silwal et al. \cite{silwal2018measuring}, respectively. Altogether, a reasonably good agreement exists between the present and previous results. 
\begin{table}[H]
\caption{The normal mass shift (NMS), specific mass shift (SMS) and filed shift (FS) values from present and the results of Safronova and Johnson \cite{PhysRevA.64.052501}, Tupitsyn et al. \cite{tupitsyn2003relativistic} and GRASP2K (\cite{silwal2018measuring}$^a$) and RMBPT (\cite{silwal2018measuring}$^b$) of Silwal et al. \cite{silwal2018measuring}}
\centering
\footnotesize
\begin{tabular}{|c |c |c |c |c |c|c |c| }
\hline
\multicolumn{1}{|c|}{\textbf{Level  }} & \multicolumn{1}{|c|}{\textbf{  Transition}} &\multicolumn{2}{c|}{\textbf{NMS (a.u.)}}  & \multicolumn{2}{c|}{\textbf{SMS (a.u.)}} &\multicolumn{2}{c|}{\textbf{FS (GHz/fm$^2$)}}  \\
\cline{3-8}

  & & Present & Others  & Present & Others & Present & Others \\ 
  \hline
 Ar$^{7+}$ & $3p~^2P_{1/2} - 3s~^2S_{1/2}$  &  -2184.098 &-  & -6132.980& -6504.1\cite{tupitsyn2003relativistic} & 2417.500 & 2417.5\cite{tupitsyn2003relativistic} \\
  & & & & &-6363\cite{PhysRevA.64.052501}& &2426.46 \cite{PhysRevA.64.052501}\\
  &$3p~^2P_{3/2} - 3s~^2S_{1/2}$  &  -2215.140 & - & -6117.170& -6328\cite{PhysRevA.64.052501}& 2421.70 &2428.53\cite{PhysRevA.64.052501}\\
  &$3p~^2D_{3/2} - 3s~^2S_{1/2}$  & -5571.958 &  -& -6366.946& -7499 \cite{PhysRevA.64.052501} & 2334.40 &2282.975\cite{PhysRevA.64.052501}\\
  &$3d~^2D_{5/2} - 3s~^2S_{1/2}$  &  -5552.825 & -& -6422.460 & -7537\cite{PhysRevA.64.052501} & 2334.20 & 2282.582\cite{PhysRevA.64.052501}\\
  \hline
  Xe$^{43+}$ & $3p~^2P_{1/2} - 3s~^2S_{1/2}$  & -11922.120 &  & -167581.049 & -171701.00\cite{tupitsyn2003relativistic} & 958270.0 & 965738.00\cite{tupitsyn2003relativistic} \\
  & & &-13144.574 \cite{silwal2018measuring}$^a$  & & -170605.61 \cite{silwal2018measuring}$^a$ & & 958984.59 \cite{silwal2018measuring}$^a$\\
  & & &-13144.574 \cite{silwal2018measuring}$^b$ & &-170331.77 \cite{silwal2018measuring}$^b$&  & 965738.00 \cite{silwal2018measuring}$^b$\\
  &$3p~^2P_{3/2} - 3s~^2S_{1/2}$  &  -22360.743 &  -& -166657.021& -& 988260.0 &- \\
  &$3d~^2D_{3/2} - 3s~^2S_{1/2}$  & -40548.925 & - & -244603.794& - &984560.0 &-\\
  &$3d~^2D_{5/2} - 3s~^2S_{1/2}$  &  -41931.357 & -& -250234.586 &- & 983250.0 & -\\
  \hline
 \end{tabular}
 \label{Tab:IS}
\end{table}
Further in Fig.\ref{fig:IS} (a), (b) and (c), we have analysed the trend in FS, SMS and NMS factors of the three Na-like ions for few allowed transitions among the $n = 3$ and $n = 4$ levels. It can be clearly observed that the IS factors for Xe$^{43+}$ are the order of magnitudes more than those of Ar$^{7+}$ and Kr$^{25+}$.
Moreover, a difference in the sign of FS factor is seen in Kr$^{25+}$ for $3d ~^2D_{3/2} - 3p ~^2P_{3/2}$ and $ 3d ~^2D_{5/2} - 3p ~^2P_{3/2}$ transitions. For these lines, the FS is positive for Ar$^{7+}$ and Xe$^{43+}$, whereas it is negative for Kr$^{25+}$. Somewhat similar trend is observed in $3d ~^2D_{3/2} - 3p ~^2P_{1/2}$ and $ 4d ~^2D_{3/2} - 3p ~^2P_{1/2} $ lines where the FS value changes from positive in Ar$^{7+}$ to negative in Kr$^{25+}$ and Xe$^{43+}$. The possible reason for these features could be the J dependence of these transitions and can be subjected to further investigations in future studies of Na-like iso-electronic series. The plots for SMS and NMS show no such trends. 
Most of the present results are arrived at first in the literature for all three ions. Also, we are confident that our calculated IS parameters will be significant in stellar modeling, heavy-ion storage experiments, and the investigation of nuclear charge radii.
\section{Conclusions}
We performed a systematic computation of level energies, lifetimes, multipole radiative rates, transition wavelengths, hyperfine constants, Land\'e g$_J$ and isotope shift factors for 71 states of the $nl$ $n\leq9; l\leq6$ configurations of Na-like Ar$^{7+}$, Kr$^{25+}$ and Xe$^{43+}$ ions.
The MCDF-RCI method is implemented to carry out these calculations using the GRASP2018 and its module RIS4. 
A large section of the present results of the three ions is reported for the first time. The gross absence of previous theoretical or experimental results led us to carry out similar independent calculations using the MBPT theory to establish our MCDF-RCI results' accuracy.
The average deviations between the present MCDF-RCI and MBPT level energies are 0.427\%, 0.05\% and 0.06\% for Ar$^{7+}$, Kr$^{25+}$, and Xe$^{43+}$, respectively. The transition rates computed using the MCDF-RCI and MBPT approach are within 10\% agreement for most of the lines, which confirms the reliability of the present MCDF-RCI and MBPT results. Further, a good agreement is obtained in comparing our results with the corresponding values from NIST, Spectra W3, CAMBD atomic database, and other previous theoretical and experimental studies, wherever available. The absolute mean difference of only 0.13\%, 0.03\% and 0.08\% in the present MCDF-RCI energies is observed with reference to the NIST values for Ar$^{7+}$, Kr$^{25+}$ and Xe$^{43+}$, respectively. 
Our calculated lifetimes in the length and velocity gauges are highly consistent and agree well with the other results. There is a good match between our HFS and IS results with the earlier theoretical and experimental works. We believe that the present calculations are accurate and comprehensive. These results will fulfill the demand for atomic parameters for HCI and will also be beneficial in studying chemical abundance in space, spectra analysis, and diagnosis of plasma.











\bibliographystyle{ieeetr}
\bibliography{mybib.bib}
\end{document}